\documentclass[11pt,onecolumn, oneside]{dansthesis2}
\usepackage{graphicx}
\usepackage{amsmath}
\usepackage{amsfonts}
\usepackage{amssymb}

\title{Foundations of quantum theory and quantum information applications}
\author{Ernesto Fagundes Galv{\~a}o} \college{Wolfson College}
\degree{Doctor of Philosophy, University of Oxford}
\degreedate{Trinity Term 2002}

\begin{document}
\baselineskip=12pt plus1pt
\setcounter{secnumdepth}{3}
\setcounter{tocdepth}{3}
\maketitle

\begin{romanpages}
\addcontentsline{toc}{chapter}{Abstract}
\begin{abstract}
\begin{center}
\emph{\Large{Foundations of quantum theory  and \\quantum information
applications}}

\vspace{3 mm}
{Thesis submitted for the Degree of Doctor of Philosophy.}
\vspace{3 mm}

\emph{\large{Ernesto Fagundes Galv\~{a}o}\\Wolfson College, University
 of Oxford}

Trinity Term 2002

\vspace{5 mm}
\end{center}

This thesis establishes a number of connections between foundational
issues in quantum theory, and some quantum information
applications. It starts with a review of quantum contextuality and
non-locality, multipartite entanglement characterisation, and of a few
quantum information protocols.

Quantum non-locality and contextuality are shown to be essential for
different implementations of quantum information protocols known as quantum random access codes and quantum communication
complexity protocols. The simplest versions of these protocols are
shown to be equivalent to tests of two- and multi-party contextuality
and non-locality. From them I derive sufficient experimental
conditions for tests of these quantum properties.

I also discuss how the distribution of quantum information through
quantum cloning processes can be useful in quantum
computing. Regarding entanglement characterisation, some results are
obtained relating two problems, that of additivity of the relative
entropy of entanglement, and that of identifying different types of tripartite
entanglement in the asymptotic regime of manipulations of many copies of a given state.

The thesis ends with a description of an information processing task
in which a single qubit substitutes for an arbitrarily
large amount of classical communication. This result is interpreted in
different ways: as a gap between quantum and classical computation
space complexity; as a
bound on the amount of classical communication necessary to simulate
entanglement; and as a basic result on hidden-variable theories for
quantum mechanics. In this case, the resource behind the quantum
advantage is simply the real nature of the set of quantum pure
states. I also show that the advantage of quantum
over classical communication can be established in a feasible
experiment.

\end{abstract}

\addcontentsline{toc}{chapter}{Acknowledgements}
\begin{acknowledgements}

My warmest thanks go to my supervisor Lucien Hardy, who managed to provide intellectual stimulation and guidance while giving
me freedom to find my own way in research. His sharp insights and
enthusiasm for quantum theory were a constant source of inspiration
for my work.

I am also thankful to Artur Ekert for his support, and for creating a
great research atmosphere at the Centre for Quantum
Computation. Over four years at CQC I ended up meeting most people in
the research community, and learned much in enjoyable
conversations with many of them. A partial list includes Sougato Bose, Dagmar Bru\ss, Mark
Bowdrey, Garry Bowen, Holly Cummins, Hilary Greaves, Patrick Hayden,
Leah Henderson, John Howell, Hitoshi Inamori, Daniel Jonathan, Antia
Lamas Linares, Jan-{\AA}ke Larsson, P\'{e}rola Milman, Daniel Oi, Tony Short, Vlatko Vedral and Jon Walgate.

I had the chance to do research and publish with two other co-authors: Martin Plenio and Shashank Virmani. I am very thankful for
their warm hospitality during many lunch times and afternoon
discussions at Imperial College.

I thank all of my friends, and specially Joana and Z\'{e} for their
love and support. A big `thank you' to my family, whose love, support
and encouragement made it possible for me to choose an academic
career.

I was supported financially by a scholarship from the Brazilian agency
Coordena\c{c}\~{a}o de Aperfei\c{c}oamento de Pessoal de N\'{\i}vel Superior (CAPES);
by Universities UK through an Overseas Research Students Awards Scheme
(ORS) award; and by Wolfson College, which provided some funding for
academic travel.

\end{acknowledgements}

\tableofcontents
\listoffigures
\end{romanpages}

\chapter{Introduction}

The conceptual revolution brought to science by quantum theory is
now almost a century old. Despite this grand old age, it still seems
that the theory's full significance has not yet been appreciated
outside a limited circle of physicists and philosophers
of science.

It is true that terms like `quantum leap' and `uncertainty
principle' have been incorporated by everyday language. They are often used
to convey concepts which are only remotely related to their
physical meaning, but at least this shows that a wider public has taken note of some aspects of the interpretation of quantum
mechanics. This timid acknowledgement, however, is
completely overshadowed by the penetration of general relativity into
mainstream culture, a theory of comparable age and importance.

This acceptance of relativity's ideas by the general public shows up
in popular science magazines and newspapers. Terms like black holes
and space-time are presented as the theoretical constructs they are, and
not just as misinformed analogies, as usually happens with
quantum theory. I think this can be attributed at least to two factors.

First, quantum theory's rupture with the previous theoretical paradigm
was more radical than relativity's. A non-scientist may not have a good understanding of what space-time and black holes are, but at least he or she
can be comforted in the knowledge that professional physicists
apparently do. This is definitely \textit{not} the case with
quantum theory, whose interpretation has always been a motive for
argument even among experts. This seems to arise from the overturning
of concepts preserved even by relativity, such as locality and perhaps
determinism.  This is why it seems quite appropriate to group the
other great physical theories and name them `classical' (relativity
included), while quantum theory remains in a class of its own.

Another factor which keeps the discussion of the foundations of
quantum theory away from a wider audience is the lack of practical
applications. Relativity also has few widespread applications, but
this gap in public perception seems to have been filled by astronomical
observations and thought experiments involving spaceships
and black holes, for example. In
comparison, quantum theory's applications seem much less convincing:
the stability of matter and quantum tunnelling are two examples that
come to my mind. Working applications of quantum theory's most
baffling characteristics are not yet part of day-to-day life.

In the last few years, the new field of quantum information has been
changing that. Practical quantum information applications are just
around the corner, with prototypes of quantum cryptographic setups and small-scale quantum computers already working in laboratories. These
applications illustrate some of the counter-intuitive features of
quantum theory in a particularly compelling way. A sign of this is the
increasing use of these applications in introductory quantum mechanics
courses. In a broader context, we see the public being engaged
by the excitement of research in quantum physics, through television,
general lectures and numerous popular science articles that have been
appearing in the general press.

By bringing the information-theoretic structure of quantum mechanics to the
fore, these applications are very helpful to the researcher as
well. Quantum information applications offer simple examples and
procedures that illustrate important characteristics of quantum
mechanics. As we will see in this thesis, these applications
often suggest what is essential and what is accessory
in quantum theory, highlighting features which may be of practical use
and theoretical importance.

My research has been directed at examining these applications, and
working out what they can tell us about foundational issues in quantum
theory, such as contextuality and non-locality. It appears that taking
into account the role of information in quantum theory will be
unavoidable for major further developments. This is hinted at by the
existence of information-based results (such as black hole entropy)
which will need to be accounted for by some future unification of
gravity and quantum mechanics.

Of course, a deeper theoretical understanding of the
information-theoretic aspects of the theory should also lead to new or
improved applications. Another research goal of mine is to turn the
insights coming from the foundations of quantum theory into useful
quantum information applications that can hopefully be implemented in
the near future. Given the theoretical promises and the current rate of
experimental progress, quantum computing and other quantum
technologies may assume a great economic importance in the next
decades.

Having stated two of my main research motivations, let me now give a
preview of the work I have been doing along these lines in the last
years, and which constitute this thesis.

\textbf{Chapters 2, 3 and 4} present introductory material
respectively on the foundations of quantum theory, quantum
entanglement, and a few quantum information applications. Most of the
results here are drawn from the recent literature, but the choice of
topics and the presentation is made so as to prepare terrain for
chapters 5 to 9. The original material in the introductory chapters
includes an analogy used to present Hardy's non-locality proof in section \ref{sec
duel}, and a curious fidelity balance result for cloning machines in
section \ref{sec curfbr}.

In \textbf{chapter 5} I 
investigate the characterisation of tripartite entanglement, which is
a much harder mathematical problem than the bipartite case. This
problem is important as multi-particle entanglement shows up in natural
systems, as well as in quantum information protocols such as the one I
discuss in chapter \ref{chap feasible}. This chapter results from
collaboration with Martin Plenio and Shashank Virmani from Imperial
College (London). The work I report is mostly my own, while in the
published version of these results \cite{GalvaoPV00} there are also analytical results obtained mostly thanks to my collaborators'
considerable mathematical skills.  Section \ref{sec wvsghz} results
from numerical calculations performed in joint work with Leah
Henderson from the University of Bristol. I also published a second
paper on multipartite entanglement in joint work with Lucien Hardy
\cite{GalvaoH00}, but as its results did not fit in naturally with the rest of this thesis, I decided not to include them here.

Encoding information in quantum systems involves some
counter-intuitive constraints.  One of these is the
\textit{no-cloning theorem}, which states that it is impossible to
make perfect copies of an arbitrary unknown quantum state. Non-perfect
but good-quality copying is feasible, but the question is: what are
these \textit{quantum cloning machines} good for? In \textbf{chapter
6} I show that quantum cloning can be used to distribute quantum
information in an efficient manner during quantum computations. This
chapter resulted from collaboration with my supervisor Lucien Hardy,
having been published in \cite{GalvaoH00b}.

In \textbf{chapter 7} I turn to a quantum information application
known as \textit{quantum random access codes}. We will see that some
simple versions of these codes had already cropped up under different
guises previously, without being noticed. I recognise  the equivalence
between these different protocols and show that their higher-than-classical performance
derives from either quantum contextuality or non-locality. I also
analyse the feasibility of implementing these codes experimentally,
showing that a successful implementation would be equivalent to a
contextuality or non-locality test. The quantum random
access code using a qutrit in section \ref{sec 21trit} was found in
joint work with Daniel Oi.

In theory, quantum communication is better than classical
communication for many tasks. However, given the difficulty of
implementing quantum protocols experimentally, usually it is the
classical protocols that outperform the quantum versions in practical setups. In \textbf{chapter 8} I
present solutions to a communication complexity problem which rely
either on entanglement or on quantum communication. The analysis of
this problem leads to sufficient conditions for testing multipartite non-locality and contextuality. Moreover, I show it is possible
to demonstrate experimentally the advantage of  quantum over classical
communication in a feasible experiment. Some of these results appeared
in preprint format \cite{Galvao00}, while others were published \cite{Galvao02}.

In \textbf{chapter 9} I present a generalisation of the communication
complexity problem discussed in chapter 8. I rigorously prove that a
single qubit of quantum communication can be better than any finite number of classical bits for a particular task. The result lends
itself to different interpretations. First, I  show it means that
entanglement is very hard to simulate with classical
communication. I also demonstrate that quantum
computers can have an unbounded advantage in memory size with respect
to classical computers. Interestingly, all these results can be shown
to stem from a single axiom that distinguishes quantum theory from
classical probability theory. This chapter represents joint work with
my supervisor, and some of it has appeared in preprint format \cite{GalvaoH01}. Finally, in \textbf{chapter 10} I offer
some concluding remarks, and point at some open problems for further research.

I hope the reader can recognise my two main research motivations in
the treatment I give to the topics addressed in this thesis. In
working on them I have acquired a more intuitive grasp of what
quantum theory can offer for information processing, and
why. Fortunately, the source of surprises seems to be endless.

\chapter{On the foundations of quantum theory \label{chap found}}

Since the development of the mathematical formalism of quantum
mechanics in the 1920's, the theory has established itself as the most
thoroughly tested and successful physical theory ever. It provides
solutions to many fundamental problems in physics, ranging from the
stability of matter to the nuclear fusion reaction generating heat in our Sun.

While its mathematical structure and predictive power can be precisely
stated, quantum theory's conceptual foundations have been a
motive for discussion for decades. This is understandable, given the
many fundamental differences between quantum theory and the great
classical theories, such as thermodynamics and general
relativity. In this chapter I will touch upon some of the known
counter-intuitive features of quantum mechanics, reviewing some of the
work done to clarify the conceptual foundations of the theory. Understanding these foundational issues is important in
appreciating the beauty and subtlety of Nature itself, but also
sheds light on why and how quantum information applications can be
better than their classical counterparts.

Maybe the first thing a student notices when introduced to quantum
theory is that, unlike classical theories, quantum mechanical
predictions are in general intrinsically probabilistic. Some of the
founding fathers of quantum mechanics (notably Einstein) were very
sceptical about this. They suspected that quantum theory was just an
approximation to a more fundamental, deterministic theory, which they
hoped would be developed in the future. Interestingly, after
the work of de Broglie and Bohm, we know that a deterministic
description of quantum phenomena is \textit{possible}. This
deterministic (or causal) interpretation of quantum theory will be briefly sketched in section \ref{sec dhvt}.

The discovery of a wholly deterministic interpretation of quantum
theory was a herald of other important developments. Starting in the mid-1950's, some researchers have been able to understand better some of the unavoidable conceptual characteristics of any
theory compatible with quantum mechanics. This remarkable achievement
pointed out some surprising, highly counter-intuitive features of quantum
theory. In this chapter we will briefly review some of what has been found
so far by investigating the necessary limitations of more general
\textit{hidden-variable theories} compatible with quantum mechanics.

In section \ref{sec contextuality} I will present some results showing
that quantum mechanics is contextual. We will see this means that
in general, the outcomes of a given observable $A$ cannot be assigned
pre-defined values, independently of which other commuting observables
we choose to measure together with $A$. More than a mere curiosity of
quantum theory, we will see in chapters \ref{chap qcc} and \ref{chap
feasible} that the higher-than-classical performance of some quantum
information protocols can be attributed to this quantum property.

In section \ref{sec nonlocality} we will see how quantum non-locality
arises naturally from contextuality. We will examine different
types of proof of quantum non-locality, and present some material
which will be relevant to the discussion of the uses of non-locality
we will analyse in the remaining chapters of this thesis.

In chapters \ref{chap qcc}, \ref{chap feasible} and \ref{chap qubit} we
will identify the fundamental quantum characteristics behind higher-than-classical performances in a number of information
processing applications. We will see that the power of quantum systems
with respect to classical systems can be traced back to fundamental
quantum traits that appear in axiomatic approaches to quantum theory. In
sections \ref{sec hardyaxioms} and \ref{sec invinfo} I will
sketch some aspects of the work of Hardy, and Brukner and Zeilinger on the foundations
of quantum theory. We will see that their approaches are directly
relevant to understanding some quantum information processing
protocols I will present later.

In order to flesh out some of the most surprising of quantum theory's
characteristics, we must start by presenting the hidden-variable
theory research programme, and what it can tell us us about physical
reality.

\section{Why hidden-variable theories? \label{sec whvt}}

Many characteristics of quantum theory run against the physical
intuition we developed in a mostly classical world. Quantum theory's description of
Nature agrees more precisely with experiments than any previous
physical theory, but at the same time its predictions are
intrinsically probabilistic.

A great deal of research has been done to try to find out whether this is a sign of
quantum theory's incompleteness, or whether this non-determinism is a
fundamental trait of Nature. \textit{Hidden-variable theories} (HVT's) were built as
attempts to find alternatives compatible with quantum mechanics, but
with different conceptual foundations. In particular, many attempts
have been made to build hidden-variable theories which could predict
quantum phenomena while retaining the determinism and locality of
classical physics.

HVT's are important for many reasons. It is productive for
theoreticians to view a problem through more than one alternative
conceptual structure, provided by different HVT's. This helps one to
develop an intuition about quantum phenomena. This is particularly helpful
because of the apparent inadequacy of classical intuition to account
for many important quantum characteristics. Alternative HVT's put
familiar problems in a new light, and may suggest new or improved
applications in quantum information theory, for example.

They also help us to identify the need (or not) for certain conceptual
characteristics in any possible acceptable future extension of quantum
mechanics. This is an indirect way of making explicit some possibly
subtle traits of Nature, which may then be brought to the fore in future
theories superseding quantum mechanics.

In the next section we will briefly comment on the success of creating
a deterministic HVT for quantum mechanics (Bohmian mechanics). This
shows that an inherently probabilistic description is not necessary.

By exploring hidden-variable theories, we will see in subsequent
sections that there are some  characteristics which are unavoidable in
any HVT compatible with quantum mechanics, for example contextuality
and non-locality. In following chapters we will see how these
characteristics can be explored in quantum information processing
tasks.

\subsection{A deterministic hidden-variable theory is possible \label{sec dhvt}}

In his influential book \textit{Mathematical foundations of quantum
mechanics} \cite{vonNeumann32}, John von Neumann claimed to have
proven that quantum mechanics could not be replaced by any alternative
deterministic HVT. For many years this seems to have blocked progress
in this area.

In 1966 Bell explicitly built a toy deterministic HVT describing
any projective measurement on a single qubit \cite{Bell66}. Bell
refuted von Neumann's claims in a
simple but bold manner, showing
that von Neumann's theorem only applied to a very restricted class of
HVT's, which automatically ruled out the sort of successful HVT Bell came up
with. Despite its conceptual importance,  Bell's result was
quite restricted: it did not account for Positive Operator Valued
Measure (POVM)\footnote{For a description of the concept of POVM's
see the excellent book by Peres \cite[section 9.5]{Peres93}. } measurements, nor for the
qubit's dynamics. The question of whether it was possible to devise a
deterministic HVT for general quantum mechanics (in arbitrary
dimensions) remained open.

Or rather, it did not: as Bell himself pointed out, this question had
already been closed long before Bell's 1966 paper. Bell brought to
attention the work of Bohm \cite{Bohm52, Bohm52b} that proposed such a
a theory in 1952, based on earlier work by de Broglie in the late
1920's \cite{deBroglie30, deBroglie30b}. For a long time Bohm's papers
\cite{Bohm52, Bohm52b} were largely ignored by the physics community,
for a conjunction of historical factors.

In Bohm's causal (i.e. deterministic) interpretation of quantum theory, the
whole experimental set-up establishes a field which can be calculated
from the wavefunction $\psi(t)$. Preparation of a quantum state, in this
interpretation, consists of preparing an ensemble of states with
different values of the hidden variable $\lambda$. This
hidden-variable is just the real position of the particle, and is
distributed according to the initial probability distribution for
finding the particle at position $x$, so it is proportional to
$|\psi|^2$. As time passes, the particles in this ensemble follow
well-specified trajectories, `guided' by the field which can be
calculated from the time-evolved $\psi(t)$.

Bohm showed that many measurements
(spin measurements, for example) can be reduced to position
measurements. In a Stern-Gerlach experiment, for example, each spin-$\tfrac{1}{2}$ particle follows a well-defined trajectory through the
magnet, impinging on the end screen in one of two possible
spots. The indeterminacy we see in the experiments arises from our
lack of control over the preparation procedure: we always
create an ensemble of states distributed according to the initial
probability distribution. In the case of the particles in the
Stern-Gerlach experiment, half of them follow trajectories leading to
the upper spot, and half to the lower one. But each particle follows a well-defined
trajectory, the indeterminacy arising from our lack of knowledge and
control over the ensemble we prepare. This approach puts the quantum
indeterminacy on the same footing as the indeterminacy in classical
statistical mechanics ensembles.

Bohmian mechanics has been further developed in recent years. For
example, Bohmian trajectories have been calculated for the double-slit
experiment \cite{PhilippidisDH79}; the hydrogen atom \cite{Holland93};
and to barrier-tunneling \cite{DewdneyH82}.

By adopting Bohm's interpretation of quantum mechanics, we recover a
deterministic theory. There is, however, a conceptual price to pay. We
need to introduce hidden-variables which we are unable to manipulate,
but which completely determine all the particles properties, through
their trajectories. If these hidden-variables have a physical
existence, one is allowed to think they will eventually become
accessible in the future, after some technological advances. This may be an impossible dream, however;
there has been research suggesting that manipulation of Bohm's hidden
variables would result in problematic possibilities such as
faster-than-light signalling \cite{Valentini01, Valentini01b}.

Independently of the interpretation we give for quantum theory, it
retains counter-intuitive conceptual characteristics, such as contextuality and
non-locality. In the next sections we will briefly explore these two
fundamental characteristics of quantum mechanics. We will see that
they can be harnessed to obtain computation and communication gains in
quantum information applications.

\section{Quantum contextuality \label{sec contextuality}}

Let us consider a quantum system in a pure
state $\left|\psi\right\rangle$. Mathematically, $\left|\psi\right\rangle$ can be
fully characterised by a complete set $S$ of
mutually commuting operators (called observables) of which
$\left|\psi\right\rangle$ is a simultaneous eigenvector, together with the
eigenvalues that will be found if each of those is measured. The usual
physical interpretation is that each observable corresponds to a
physical property being measured, such as angular momentum, position, and so on. Being a simultaneous eigenstate of the set $S$ of
observables, whenever $\left|\psi\right\rangle$ is measured for these properties the results
will be predictable and the same, corresponding to one particular eigenvalue of each observable $O_i \in S$.

By analogy with classical physics, when we physically implement the
measurement of observable $O_i$ with a suitable physical apparatus, we
say we are measuring the system's property which is defined by
$O_i$. For example, we say that we are measuring the electron's intrinsic angular momentum along the $z$ direction when we send through an
electron through a Stern-Gerlach magnet oriented along $z$.

A set $S$ of observables completely characterises a
pure state when the outcomes of all the observables in
$S$ are predictable. We tend to say that all the properties of the
system are known, and also tend to think that these properties exist in
principle independently from each other. This intuition is
corroborated by performing different experiments on multiple copies of
the same state. Each experiment can measure a different subset of the
observables in $S$; we still
obtain the same, predictable outcomes, independently on the
experimental context (i.e. independently on the particular sub-set of
observables $O_i \in S$ chosen to be measured in each experiment).

But now suppose that we are given an unknown pure state, prepared by
another party. Now we have no guarantee that it will be an eigenstate
of any particular set of observables that we decide to measure. Still,
our classical intuition leads us to expect that even so, the result we will
obtain when measuring observable $O_k$ is independent on the
experimental \textit{context}, i.e. on which other commuting
observables I decide to simultaneously measure together with
observable $O_k$. This intuition fits with our attribution of
pre-defined properties to the system, which are independent of each
other and which can be measured selectively at will.

This \textit{non-contextuality hypothesis} can be shown to be
\textit{false} for quantum mechanics. Let us now review some proofs of
quantum contextuality, and briefly discuss the significance of
this result.

\subsection{Two simple contextuality proofs}

When we consider the typical situation relevant to Heisenberg's
principle, one produces a state and tries to measure observables which
do not commute. Then we find that measurement results seem to indicate
that there is no predefined value for the outcomes of non-commuting observables, prior to the measurement.

This non-existence of pre-defined properties, however, may be just an
enticing suggestion of the usual Copenhagen interpretation of quantum
theory. Instead of just accepting this suggestion unquestionably, it turns
out that we can actually prove that these values cannot be
pre-defined non-contextually. In this
section we will review two simple contextuality proofs, due to Peres
\cite{Peres90} and Mermin \cite{Mermin90b}. We start with Peres'
proof.

\subsubsection{Peres' contextuality proof}

Let us consider a particle in the singlet state:

\begin{equation}
\left|\phi^{-} \right \rangle = \frac{1}{\sqrt{2}} \left(\left|0_A1_B
\right\rangle - \left|1_A0_B \right\rangle \right),\label{firstphiminus}
\end{equation}
where $\left|0\right\rangle_j$ and $\left|1\right\rangle_j$ represent
two orthogonal states of party $j$'s subsystem. This is an example of
a canonical maximally entangled pair of qubits, also called an
Einstein- Podolsky- Rosen (EPR) pair, or alternatively a \textit{Bell
state}.

This state is a simultaneous eigenvector of observables which can be
conveniently written using the three Pauli matrices:
\begin{equation}
\sigma_x= \left(\begin{array}{c c}0 & 1 \\ 1 & 0\\
\end{array}
\right), \hspace{4 mm}
\sigma_y= \left(\begin{array}{c c}0 & -i \\ i & 0\\
\end{array}
\right), \hspace{4 mm}
\sigma_z= \left(\begin{array}{c c}1 & 0 \\ 0 & -1\\
\end{array}
\right).
\end{equation}

State (\ref{firstphiminus}) is an eigenvector (with eigenvalue -1) of the observables $\sigma_{x1} \otimes \sigma_{x2}$,
$\sigma_{y1} \otimes \sigma_{y2}$, and $\sigma_{z1} \otimes
\sigma_{z2}$, where the sub-indices indicate which particle (1 or 2) we
are referring to. The non-contextuality assumption
tells us that the outcomes for each one-particle observable are pre-defined, and do
not depend on which other commuting operators we decide to
measure. For example, the outcome $x_1=\pm 1$ that we obtain by measuring
$\sigma_{x1}$ must not depend on whether we decide to measure
$\sigma_{x2}$, $\sigma_{y2}$ or $\sigma_{z2}$, since these three
operators commute with $\sigma_{x1}$. By giving these hypothetical
measurement outcomes definite values $x_1, x_2, y_1, y_2, z_1, z_2$,
we are accepting the classical idea that all a quantum measurement
does is to reveal a pre-existing property of a system, independent of
other properties we may wish to measure.

Now, the fact that a singlet is an
eigenvector with eigenvalue -1 of the three operators above means
there is a constraint on these hypothetical pre-existing values:

\begin{equation}
x_1 x_2 = y_1 y_2 = z_1 z_2 = -1. \label{constreigen}
\end{equation}

Let us consider the outcomes on some other possible measurements on this
state. Consider the observables $A=\sigma_{x1} \otimes \sigma_{y2}$ and
$B=\sigma_{y1} \otimes \sigma_{x2}$. Note that $[A,B]=0$, which means that
we can measure $A$ and $B$ simultaneously. This can be done in two
ways. One possibility is to measure the observable $AB$ directly. To
see what the measurement outcome will be, note that

\begin{equation}
AB=\sigma_{x1} \otimes \sigma_{y2} \sigma_{y1} \otimes \sigma_{x2} = \sigma_{z1}
\otimes \sigma_{z2} ,
\end{equation} 
which, as we have seen, always results in -1 for the singlet
state.

The second way to measure the product $AB$ is to measure $A$ and $B$
separately, and multiply the results. Quantum mechanics predicts that
a simultaneous measurement of $A$ and $B$ will yield results
consistent with the prediction above. The non-contextual assumption,
on the other hand, tells us that the result of
measuring $A=\sigma_{x1} \otimes \sigma_{y2}$ should be $x_1 y_2$, and of
measuring $B=\sigma_{y1} \otimes \sigma_{x2}$ should be $y_1
x_2$, where $x_1,x_2,y_1$ and $y_2$ are the same that appear in
eq. (\ref{constreigen}), in a different context. Multiplying these two
predicted measurement outcomes we obtain
\begin{equation}
x_1 y_2 y_1 x_2= (x_1 x_2)(y_1 y_2)=(-1)(-1)=+1, 
\end{equation}
[because of constraint (\ref{constreigen})], which is the exact opposite
of the outcome we experimentally measure (-1), as we have seen above. This contradiction, obtained by Peres in \cite{Peres90}, shows that we cannot assign definite results to
outcomes of non-commuting observables in a non-contextual way.

\subsubsection{Mermin's contextuality proof}

The above proof relies on some properties of the particular state given by
eq. (\ref{firstphiminus}). Building on Peres' simple proof, Mermin \cite{Mermin90b} was
able to obtain a contextuality proof which is state-independent. It
goes as follows.

Consider a two-qubit system in any quantum state. Now consider the
following array of operators:

\begin{center}
\begin{tabular}{|c|c|c|} \hline
\mbox{ $1\hspace{-1.0mm}{\bf l}$ } $\otimes \sigma_z$ & $\sigma_z
\otimes\mbox{$1 \hspace{-1.0mm}  {\bf l}$}$ & $\sigma_z\otimes\sigma_z$
\\ \hline
$\sigma_x \otimes \mbox{$1 \hspace{-1.0mm}  {\bf l}$}$ & \mbox{$1
\hspace{-1.0mm}  {\bf l}$} $\otimes \sigma_x$ & $\sigma_x \otimes \sigma_x$
\\ \hline
$\sigma_x \otimes \sigma_z$ & $\sigma_z \otimes \sigma_x$ & $\sigma_y
\otimes \sigma_y$ \\ \hline

\end{tabular}
\end{center}

Note that each of the nine operators have eigenvalues $\pm 1$, and
that the three operators in each row or column commute. It is also easy to
check that each operator is the product of the other two operators on
the same row or column, with a single exception: the third operator in
the third column equals \textit{minus} the product of the other two.

Suppose that there exist pre-defined values (1 or -1) for the outcomes
of these nine operators. Then we can substitute each of the nine
operators above by the pre-defined value of its outcome. This
leads us to an impossible situation: because of the minus
sign that appears in the third column, there is no consistent way of
assigning pre-defined values to the nine operators listed
above. These pre-defined measurement outcomes simply cannot exist, if
we do not make explicit reference to the complete set of commuting
observables being measured in the particular experiment under
consideration.

\subsection{Other results on contextuality}

The two proofs we have presented above make use of a 4-dimensional
Hilbert space. Proofs using only 3-dimensional systems are also
possible, and in fact appeared much earlier than the two proofs above.

Building on earlier work by Gleason \cite{Gleason57}, Kochen and
Specker \cite{KochenS67} provided the first proof of quantum
contextuality. They used a very complicated geometrical construct,
involving 117 operators on a 3-dimensional space. They were able to
show that this set of projectors could not be consistently assigned
values 0 and 1 in a non-contextual way. I chose to present Peres' and
Mermin's proofs because they are conceptually and algebraically much
simpler than the results of Kochen and Specker. For a clear discussion
of Kochen and Specker's theorem and its philosophical relevance, see
the article by Held \cite{Held00}.

As we will soon see, quantum non-locality involves a special type of
quantum contextuality. Because of the great number of
applications of quantum entanglement that have been found recently,
the discussion of non-locality has eclipsed that of quantum
contextuality. Nevertheless, much of the discussion on non-locality
can be relatively easily adapted to non-contextuality, in particular
many of the simple non-locality proofs are also contextuality proofs
\cite{GreenbergerHZ89, GreenbergerHSZ90, Mermin90b, Hardy93}.

More recently, some effort has been put into devising very simple contextuality tests, with a view to experimental
realization \cite{CabelloG-A98, MichlerWZ00, SimonZWZ00}. Generally,
it is easier experimentally to perform a contextuality test than a
non-locality one. We will further explore this issue in chapters
\ref{chap qcc} and \ref{chap feasible}.

\subsection{Making sense of contextuality}

The proofs above tell us something very unusual about Nature. In classical physics the measurement of one property
of a system does not interfere with measurements of other
properties. We can thus think of different properties of a classical
system as existing independently of our actual experimental setup.

In quantum mechanics the situation is different. We have just seen
that it is \textit{impossible} to consistently assign pre-existing
values to measurement outcomes of observables in a non-contextual
way. This quantum property is called contextuality
because the outcome of an observable $A$ in general must depend on the
experimental `context', i.e. on which other commuting observables are
measured together with $A$. In the array appearing in Mermin's proof,
for example, each operator appears in two different contexts,
representing two different possibilities of commuting operators to be
measured simultaneously with it. The fact that we used a single symbol
to represent the outcome of an observable in these two different
`contexts' leads to the inconsistency pointed out in the proof.

Quantum contextuality
forbids us, in general, to assign pre-defined elements of reality to individual
observables. This is in marked contrast to the classical
case, where the momentum of an object has an objective
existence, independently of its angular momentum, for
example. Contextuality tells us that if we still want to stick to the
idea that quantum systems have well-defined properties, the only
consistent way of assigning them would be with respect to the full
experimental set-up used to measure them. As remarked earlier, this is
not so unreasonable, given that an apparatus to measure just the
momentum of a particle is different from an apparatus that measures both
momentum and angular momentum, for example. Of course, if we measure a
complete set of commuting observables there is no contextuality
possible, because by definition a complete set of observables provides
the full experimental context. In Peres' book \cite[chapter
7]{Peres93}, he suggests that we accept the
ambiguity involved in defining the value of the measurement outcome of
a single operator (or incomplete set of them), and simply refrain from
saying a single observable reveals a well-defined
property of a system. Only complete sets of commuting observables
could be said to reveal objective properties of a system.

In quantum information processing,
it is sometimes convenient \textit{not} to measure a complete set of
operators, measuring only a few. This can be used to prepare a
state, which is then sent to a second party. The measurements that the
second party can perform will have contextual outcomes, i.e. dependent on
the measurement outcomes previously obtained by the first party. This
way, we can \textit{explore} quantum contextuality as a resource for
quantum information tasks. We will develop this idea in chapter \ref{chap
qcc}.

Understanding that quantum mechanics is contextual was an important
step towards realizing it is non-local as well. In the next section we
discuss quantum non-locality, which is an essential ingredient to a
number of quantum information applications we will investigate in
future chapters.

\section{Quantum non-locality \label{sec nonlocality}}

In all of classical physics, the description of a system's state
depends only on itself and its immediate surroundings; in other words, classical
physics is local. In 1935 Einstein, Podolsky and Rosen (EPR) published an
influential paper \cite{EinsteinPR35} in which they showed that
spatially extended quantum systems were apparently non-local. Fearing
that this non-locality would conflict with causality, they concluded
that quantum mechanics was necessarily incomplete, and would need to
be replaced in the future with some hidden-variable theory which still
preserved causality and locality.

At some point it became clear that the apparent non-locality of
quantum mechanics did not allow for faster-than-light communication. The next major step
was taken by Bell, who in 1964 showed that \textit{any}
hidden-variable theory compatible with quantum mechanics must
necessarily be non-local \cite{Bell64}. This provided a definitive negative answer to
EPR's suggestion that quantum mechanics might still be compatible with
a local theory yet to be developed.

Conceptually, quantum non-locality follows naturally once we accept
contextuality. In section \ref{sec contextuality} we have seen that
the measurement outcome of an observable $A$ depends not only on $A$ and the state of the system, but also on the
particular choice of commuting observable $B$ to be measured together with
$A$. This, in a nutshell, is the essence of what we call quantum
contextuality. It turns out that this contextuality is still present
for commuting $A$ and $B$ that are each measured on \textit{spatially
separated} particles. The crucial difference now is that the
outcome of observable $A$ will necessarily depend on a distant
(possibly even space-like separated) set-up to measure observable
$B$. Quantum mechanics exhibits non-locality when it remains contextual even in this spatially separated setting.

There have been many experimental tests of quantum non-locality, and
although some possible loopholes remain to be closed, the evidence has
been overwhelmingly in support of quantum mechanical
predictions. Besides its theoretical interest, many information
processing and communication applications have been found for
non-local effects, as we will see in the chapters ahead. In the rest
of this section we will review some quantum non-locality proofs, which
will be useful for the discussion in the rest of this thesis.

\subsection{The Clauser-Horne-Shimony-Holt inequality \label{sec chshineq}}

The first proof of quantum non-locality was provided by Bell in 1964
\cite{Bell64}. His proof involved devising an
inequality that simple measurement outcomes on two distant systems
needed to satisfy, if the outcomes at each location could not depend on
the setting at the second location (i.e. if locality holds). In this
section we will present a related inequality derived by Clauser, Horne, Shimony and Holt (CHSH) in 1969
\cite{ClauserHSH69}. In chapter \ref{chap qcc} we will see that quantum
non-locality associated with violation of the CHSH inequality can be
used in some quantum communication protocols.

\vspace{2 mm}
\textbf{The inequality.} Consider a situation in which we perform two-outcome ($\pm1$)
measurements on two spatially separated systems $A$ and $B$. It is
enough to consider just two different measurements at each
location. If the outcomes at $A$ are independent on the settings at
the distant $B$, then CHSH showed the following inequality
must be satisfied \cite{ClauserHSH69}:
\begin{equation}
\left|  \left\langle a_{1}b_{1}\right\rangle +\left\langle a_{2}
b_{1}\right\rangle +\left\langle a_{2}b_{2}\right\rangle -\left\langle
a_{1}b_{2}\right\rangle \right|  \leq2, \label{CHSH ineq}
\end{equation}
where $a_{i}$ and $b_{j}$ denote the outcomes of measurement $i$ on
system $A$ and $j$ on system $B$, and the $\left\langle
{}\right\rangle $ indicate an average over many runs of the
experiment. A simple, if tedious, way to check that the inequality is
true is to substitute all sixteen possible combinations of values for
$a_1, a_2, b_1, b_2$ in the left-hand side of the CHSH inequality
(\ref{CHSH ineq}).

Note that the inequality makes no reference to
quantum mechanics, it just puts a bound on a particular combination of
averages over dichotomic variables. By deriving an inequality which
must be obeyed by \textit{averages} of experimental runs, both Bell
and CHSH made it possible a direct comparison of any local theory with
quantum mechanics, which only predicts probabilities and averages
(expected values of observables). We will soon see that quantum
mechanical predictions conflict with the CHSH locality inequality
(\ref{CHSH ineq}).

A different way of expressing inequality (\ref{CHSH ineq}) results if we use $p^{e}(a_{i},b_{j})$ [$p^{d}(a_{i},b_{j})$] to denote the
probability of obtaining equal [different] outcomes for measurements
$a_{i}b_{j}$. Then it is easy to obtain the following inequality from
(\ref{CHSH ineq}):
\begin{equation}
p^{d}(a_{1},b_{1})+p^{d}(a_{2},b_{1})+p^{d}(a_{2},b_{2})+p^{e}(a_{1}
,b_{2})\leq3. \label{hardy ineq.}
\end{equation}
These two inequalities are equivalent, and must be obeyed by
two-outcome experiments predicted by \textit{any} local theory.

\vspace{2 mm}
\textbf{Quantum violation.} Now let us see how quantum mechanics violates the inequalities
above. For that, we will start with a maximally entangled state of two
qubits, one in each spatial region $A$ and $B$:
\begin{equation}
\left|\phi^{-}\right\rangle=\frac{1}{\sqrt{2}}\left(
\left|0_{A}1_{B}\right\rangle - \left|1_{A}0_{B}\right\rangle \right).
\end{equation}

At each location, one out of two possible projective measurements will
be chosen at random. Being projections, there are only two possible
outcomes for each measurement, which we will denote by $\pm 1$. Let us
represent the projector operators on each qubit by $\hat{P}(\theta,\phi)=\left|\psi(\theta,\phi)\right\rangle\left\langle\psi(\theta,\phi)\right|$, where
\begin{equation}
\left|\psi(\theta,\phi)\right\rangle=\cos(\theta/2)\left|0\right\rangle+\exp(i\phi)\sin(\theta/2)\left|1\right\rangle.\label{psithetaphi}
\end{equation}
This parametrization of pure states of a qubit assigns each point on a
sphere (the `Bloch sphere') to a unique pure quantum state. This
 qubit parametrization is very useful, and will be used many times
in this thesis.

Maximal violation of the CHSH inequality (\ref{CHSH ineq}) happens when
we consider the outcomes of projectors on qubits $A$ and $B$ that
alternate on the same plane (crossing the sphere at its centre), at equally spaced angles separated by
$\pi/4$ radians. For example, we can choose to measure the following
set of operators: $\hat{P}_{a_1}=\hat{P}(0,0),
\hat{P}_{a_2}=\hat{P}(\pi/2,0), \hat{P}_{b_1}=\hat{P}(\pi/4,0)$, and
$\hat{P}_{b_2}=\hat{P}(3\pi/4,0)$. For these observables we can easily
calculate the following expectation values:
\begin{equation}
\left\langle\Psi\right|\hat{P}_{a_1}\hat{P}_{b_1}\left|\Psi\right\rangle=
\left\langle\Psi\right|\hat{P}_{a_2}\hat{P}_{b_1}\left|\Psi\right\rangle=\left\langle\Psi\right|\hat{P}_{a_2}\hat{P}_{b_2}\left|\Psi\right\rangle=-\left\langle\Psi\right|\hat{P}_{a_1}\hat{P}_{b_2}\left|\Psi\right\rangle=\cos(\pi/4)
\end{equation}
If we now assume that each two-qubit operator above is just revealing the
pre-existing values $a_1=\left\langle\Psi\right|\hat{P}_{a_1}\left|\Psi\right\rangle$
and similarly for $a_2,b_1$, and $b_2$, we can substitute those
in the CHSH inequality (\ref{CHSH ineq}), obtaining:
\begin{eqnarray}
\left|
\left\langle\Psi\right|\hat{P}_{a_1}\hat{P}_{b_1}\left|\Psi\right\rangle+
\left\langle\Psi\right|\hat{P}_{a_2}\hat{P}_{b_1}\left|\Psi\right\rangle+\left\langle\Psi\right|\hat{P}_{a_2}\hat{P}_{b_2}\left|\Psi\right\rangle-\left\langle\Psi\right|\hat{P}_{a_1}\hat{P}_{b_2}\left|\Psi\right\rangle
\right|\\
=4\cos(\pi/4)=2\sqrt{2}>2,
\end{eqnarray}
which shows that the CHSH inequality is violated by these quantum
measurements.

Instead of using the expectation values in inequality (\ref{CHSH ineq}),
we can equivalently substitute the outcome probabilities in inequality
(\ref{hardy ineq.}):
\begin{eqnarray}
p_{q}^{d}(a_{1},b_{1})  &  =p_{q}^{d}(a_{2},b_{1})=p_{q}^{d}(a_{2}
,b_{2})=p_{q}^{e}(a_{1},b_{2})=\cos^{2}(\pi/8)\label{prob ineq}\\
&  \Rightarrow p_{q}^{d}(a_{1},b_{1})+p_{q}^{d}(a_{2},b_{1})+p_{q}^{d}
(a_{2},b_{2})+p_{q}^{e}(a_{1},b_{2})=4\cos^{2}(\pi/8)>3. \label{ineq viol}
\end{eqnarray}
The violation of these two equivalent inequalities shows that
measurements on a maximally entangled state violate locality.

\subsection{Non-locality without inequalities \label{sec nlwi}}

In the CHSH non-locality proof above we used either expectation values
of quantum experiments, or equivalently the probability of a given
outcome. To ascertain these quantities, many realizations of the
experiment are necessary until we reach a statistical violation of the
relevant inequality. In this section I review Hardy's simple proof of quantum non-locality \cite{Hardy92, Hardy93,
Hardy97}, which makes do without statistical considerations. Instead,
Hardy showed that we reach a logical contradiction if we assume:

a- that quantum mechanics is correct; and

b- that locality holds.

This type of reasoning offers a different way of uncovering
non-locality in quantum measurements, one which is perhaps more
satisfying intellectually. This is because most people recognise a
flagrant contradiction when they see one, but have more trouble
doing the same when the contradiction only appears at a statistical level. Before Hardy, a similar inequality-free contradiction
between quantum mechanics and locality had been found for four
\cite{GreenbergerHZ89} and three particles \cite{Mermin90}. Hardy simplified the reasoning and
showed that a similar proof was also possible for two
particles only. Curiously, his proof does not use maximally entangled
states, as references \cite{GreenbergerHZ89, Mermin90} do.

Before presenting the actual proof, I will dramatise the
situation by discussing a hypothetical real-life scenario (a duel) which is
paradoxical if we rule out non-locality. This is an extended,
adapted  version of a similar discussion by Hardy himself that
appeared in \cite{Hardy97}. When we come to the non-locality proof, we
will see that its structure mirrors that of the duel, and that quantum
non-locality offers a solution to the apparent paradox.

\subsubsection{A strange duel \label{sec duel}}

Suppose that we are running a business which provides duellists with rooms in which they can settle their
scores. Our duelling stand consists of a very long room, at whose ends
the two duellists will aim a pistol and shoot. After allowing them in, we
close the soundproof room and let the duel take place. After each
duel, Alice enters the room at end A and Bob at end B to attend to the
duellists.

\vspace{2 mm}
\textbf{Observation outcomes.} After each duel, it is always the case
that both gunmen are lying down, with their eyes closed. Since
sometimes these same gunmen will be found to be alive, we assume they
lie down even when not shot just to relax after the stressful
event. The observers Alice and Bob can choose to perform one of two
tests, but \textit{not} both at the same time. These can be: find out
if the man is alive (by listening to his heartbeat), or find out if
his gun has fired (by measuring the temperature of the gun).

Alice and Bob attend to many duels, and by comparing notes they find
out the following three puzzling facts:
\begin{description}
\item{\textbf{Observation 1}-} if both Alice and Bob checked their client's heartbeat, they
discover that in all duels at least one gunman was killed;
\item{\textbf{Observation 2}-} whenever Alice found A's gun unused and
Bob checked B's heartbeat, he always found him alive, and vice-versa; and
\item{\textbf{Observation 3}-} when both decided to examine the guns, sometimes it happened that
neither gun was used.

\end{description}

How can there be duels in which neither gun was used, but in which at
least one gunman always dies? Perhaps one of the duellists killed
using a concealed weapon, which could explain observations 1 and
3. This explanation, however, is inconsistent with observation 2.

A bit of thought shows that there is no way to have duels consistent
with observations \textbf{1},\textbf{2} and \textbf{3}. One way to see this is to make a table of
all possible combinations for experiment outcomes (there are sixteen
of them). From these sixteen, there are only five which comply with
observations \textbf{1} and \textbf{2}, and \textit{none} of these satisfy observation \textbf{3}, i.e. involve two cold guns.

\vspace{2 mm}
\textbf{Hidden variables.} But wait, there is perhaps one way to
explain these strange duel outcomes. The
normal duelling rules involve many assumptions: that they will try to
shoot each other, the only weapons they have are the pistols,
and so on. But perhaps the appearances here deceive us, and the
duellists are actually maniacs who are just trying to puzzle Alice and
Bob at any cost. If that is the case, then perhaps they have agreed on
a sophisticated, morbid pact just to fool the two innocent observers into thinking that an apparently impossible task is being performed.

One way of doing this would be for them both to lie down, while
secretly glancing at the other end of the stand through half-closed
eyes. When Alice and Bob come in, both gunmen have a chance to see
what sort of equipment each is bringing into the stand. The equipment
necessary for each measurement is different, being a stethoscope to
check the heartbeat or a thermometer to find the gun's
temperature. When both A and B realize what is going to be measured,
they can set things up so that observations \textbf{1},\textbf{2} and \textbf{3} are true, no
matter what the cost. One simple, if contrived, way of doing this is
for A and B never to shoot; and if both Alice and Bob come in with
stethoscopes, then either A or B must quickly kill him/herself with a
hidden cyanide tooth-cap.

\vspace{2 mm}
\textbf{Non-locality.} Note, however, that some communication was required between the two
duellists in order to conform to the three observations above. In the
case we have just presented, this communication was conveyed by visual
means. As we have noted above, however, any successful cheating
behaviour by the duellists must involve information about what is
happening at the other end of the stand. But we can assume that the
duelling stand is very long, and that Alice and Bob come in simultaneously to check on the duellists (i.e., the two
ends of the stand have a space-like separation). In such conditions
the two separated gunmen cannot communicate, and therefore there seems
to be no way for them to conform to the three experimental observations of Alice
and Bob.

For space-like separated measurements, there is only one way for the
gunmen to fool the observers. The gunmen must have access to some
device which reacts \textit{instantaneously} to what is being measured
at the other end of the stand. Short of a faster-than-light
signalling device, there is only one known way of doing this. This
involves exploring quantum non-locality, as
we will see in the next section.

Suppose that, after many duels on a stand with space-like separated
ends, Alice and Bob gather experimental data consistent with observations 1,2 and 3. Then, strange as it may seem, we will be forced
to conclude that whatever the mechanism behind the strange gunmen
behaviour, it must react instantaneously to distant events -- it must
be non-local.

In the next section I will review Hardy's non-locality proof which
uses non-local measurements mirroring exactly the duelling scenario
I have just presented. We will see that a simple two-qubit entangled
state can provide the non-locality necessary for space-like separated
gunmen to successfully account for the three observations above.

\subsubsection{Hardy's quantum non-locality proof}

Instead of finding out information about two duelling gunmen, let us
here consider measurements that Alice and Bob can make on a pair of
entangled qubits. In the duelling scenario the observers have two
mutually exclusive measurements options, one which
distinguishes between alive and dead (men), and one between hot or
cold (guns). In the quantum situation this corresponds to Alice and Bob
each choosing to measure one out of two dichotomic observables. To
keep the quantum discussion running parallel to the duelling paradox,
let us denote these observables $H$ (for health) and $T$ (temperature).

Observable $H$ has two eigenvectors, corresponding to the
two possible eigenvalues obtained in any measurement. Let us call
these eigenvectors $\left|a\right\rangle$ and $\left|d\right\rangle$,
for alive and dead, respectively. Similarly, observable $T$ has the
two eigenvectors $\left|h\right\rangle$ and $\left|c\right\rangle$
corresponding to hot and cold. The fact that Alice and Bob cannot
measure these two simultaneously shows up in the quantum scenario from
the
non-commutativity of observables $H$ and $T$ at each location (note,
however, that $[H_A,T_B]=[T_A,H_B]=0$ because they refer to different
particles). So we can express the $\left|h\right\rangle$ and
$\left|c\right\rangle$ eigenstates as linear superpositions of $\left|a\right\rangle$ and
$\left|d\right\rangle$:
\begin{eqnarray}
\left|h\right\rangle_i=\alpha_i \left|a\right\rangle_i + \beta_i
\left|d\right\rangle_i \label{eq had}\\
\left|c\right\rangle_i=\beta^{*}_i \left|a\right\rangle_i - \alpha^{*}_i
\left|d\right\rangle_i \label{eq cad}.
\end{eqnarray}
These two equations can be inverted:
\begin{eqnarray}
\left|a\right\rangle_i=\alpha^{*}_i \left|h\right\rangle_i + \beta_i
\left|c\right\rangle_i \label{eq ahc}\\
\left|d\right\rangle_i=\beta^{*}_i \left|h\right\rangle_i - \alpha_i
\left|c\right\rangle_i ,\label{eq dhc}
\end{eqnarray}
where the subindex $i$ indicates which particle we are referring to,
either A or B.

Using the two non-orthogonal states $\left|h\right\rangle$ and $\left|a\right\rangle$, we can define a particular
entangled state of two qubits, which we will use to demonstrate
non-locality:
\begin{equation}
\left|\psi\right\rangle=N\left( \left|h\right\rangle_A
\left|h\right\rangle_B - \alpha_A \alpha_B \left|a\right\rangle_A \left|a\right\rangle_B  \right),\label{psihardyproof}
\end{equation}
where $N(\alpha_A, \alpha_B, \beta_A, \beta_B)$ is a state-dependent
normalisation constant.

\vspace{2mm}
\textbf{Observation outcomes.} Now let us verify that the outcomes of observables $H$ and $T$ for
state $\left|\psi\right\rangle$ indeed conform to Alice and Bob's
paradoxical-looking observations. Let us start by observation \textbf{1}.

The first observation states that we never find both particles in the
`alive' state; this corresponds to saying that in eq. \ref{psihardyproof} the amplitude multiplying $\left|a\right\rangle_A \left|a\right\rangle_B$ must be zero. In spite of the appearances this is actually true, as we can check by
substituting eq. (\ref{eq had}) into eq. (\ref{psihardyproof}):
\begin{equation}
\left|\psi\right\rangle=N\left\{ \left( \alpha_A \left|a\right\rangle_A
+ \beta_A \left|d\right\rangle_A \right) \left( \alpha_B
\left|a\right\rangle_B +\beta_B \left|d\right\rangle_B
\right)-\alpha_A\alpha_B \left|a\right\rangle_A\left|a\right\rangle_B\right\}.
\end{equation}
We see that the $\left|a\right\rangle_A\left|a\right\rangle_B$ terms
cancel out in the above expression, which means
$\left|\psi\right\rangle$ always satisfies the first observation.

Let us now check whether a cold gun at A always means we find a live
person at B (observation two). In our quantum scenario this translates
as: is the amplitude multiplying the term
$\left|c\right\rangle_A\left|d\right\rangle_B$ zero?
The answer is yes, as we can check by
substituting eq. (\ref{eq ahc}) into eq. (\ref{psihardyproof}) for $\left|a\right\rangle_A$:
\begin{equation}
\left|\psi\right\rangle=N\left( \left|h\right\rangle_A
\left|h\right\rangle_B - \alpha_A \alpha_B \left(\alpha^{*}_A \left|h\right\rangle_A + \beta_A
\left|c\right\rangle_A \right)  \left|a\right\rangle_B \right).
\end{equation}
It is easy to perform a similar substitution and check
that a cold gun at B also implies that A is alive. These two perfect
correlations are what we need to satisfy observation \textbf{2} of Alice and
Bob.

Now all that is left to do is to check that observation \textbf{3} is true, and
we sometimes find both guns cold. For that we need to have a non-zero amplitude for the
$\left|c\right\rangle_A\left|c\right\rangle_B$ term. If we again substitute (\ref{eq ahc}) into (\ref{psihardyproof}) we
find:
\begin{equation}
\left|\psi\right\rangle=N\left\{ \left|h\right\rangle_A
\left|h\right\rangle_B - \alpha_A \alpha_B \left(\alpha^{*}_A \left|h\right\rangle_A + \beta_A
\left|c\right\rangle_A \right) \left( \alpha^{*}_B \left|h\right\rangle_B + \beta_B
\left|c\right\rangle_B\right) \right\}.
\end{equation}
Note that the amplitude multiplying
$\left|c\right\rangle_A\left|c\right\rangle_B$ is
$N\alpha_A\alpha_B\beta_A\beta_B$. This means that both guns will
be found to be cold with probability
$\left|N\alpha_A\alpha_B\beta_A\beta_B\right|^{2}$. This probability will
be non-zero whenever the two observables for each party fail to
commute. If we denote the golden mean by
$\tau=\frac{1}{2}\left(1+\sqrt{5}\right)$, the maximum value for this
probability can be shown to be equal to $1/\tau^{5}\simeq 9\%$
\cite{Hardy93,Hardy92}, a value which is obtained for a less than maximally entangled state.

\vspace{2 mm}
\textbf{Quantum non-locality.} As we have noted in the previous
section, both the duelling scenario and Hardy's non-locality proof
contradict the locality assumption. In
the duelling scenario, if the two observations are
time-like separated, it was possible for the two gunmen to communicate
and set things up in line with the three observations of Alice and
Bob. In the quantum case this is also true, and these strange
protocols go by the name of local hidden-variable theories.

For space-like separated measurements, however, there can be
no communication between the two space-time regions $A$ and $B$. This
restricts any local hidden-variable theory to define the measurement
outcomes without having access to information about the distant
measuring apparatus. The
only way of making sense of these quantum measurement results is to
accept that quantum measurement outcomes in one place can be
influenced instantaneously by distant experimental setups,
i.e. quantum theory is non-local.

\vspace{ 2mm}
\textbf{Using quantum non-locality.} Now maybe we should go back to the duelling scenario to convince those who
are still sceptical about quantum non-locality. There, because we make
reference to more day-to-day properties than projections on qubits, the apparent paradox
seems to be more vivid and strange. A sceptical may be unwilling to accept
that people and guns can exist in superposition states and behave
non-locally in the same way as particles, and thus may think that the
seemingly impossible observations of Alice and Bob cannot possibly be
obtained for space-like separated measurements.

It is important to stress, however,
that the duellists \textit{can} trick Alice and Bob by using quantum non-locality. For that, before each duel they need
to share instructions on how to proceed (local hidden variables)
\textit{plus} two qubits in the entangled state
$\left|\psi\right\rangle$, maybe kept in futuristic matchbox-sized
ion traps in their pockets. When the
observer comes in, each gunman quickly presses a button, which
implements the measurement of observable $T$ or $H$, according to what
equipment the observer is bringing in the stand. The ion-trap
experiment outcome is dependent on what is being (was,
will be) measured at the other end of the stand, and thus can be used by the gunmen to mimic the quantum mechanical
correlations and trick the two observers.

This last point illustrates a general idea that brings quantum
non-locality closer to everyday life. The idea is to make use of
non-locality to make some seemingly impossible communication tasks
possible. This turns non-locality from a curious theoretical trait of
quantum theory,
into a useful resource that can be explored in quantum computing and
communication. What is more, it becomes harder to doubt the existence
of non-local phenomena once useful applications of non-locality can be
successfully tested in the lab. In chapters \ref{chap qcc} and
\ref{chap feasible} I will explore some applications of quantum
non-locality.

\section{Hardy's axioms for quantum theory \label{sec hardyaxioms}}

One way to try to understand quantum mechanics is to find
simple axioms from which one can derive the whole theory. An
axiomatisation of a theory is a way to compress its information
content to a small core of axioms, whose status and
meaning can then hopefully be used to shed light on the essential
characteristics, rather than the complex structure arising from them.

Recently, Hardy proposed a set of five simple axioms from which he
deduced the mathematical structure of quantum theory for discrete
systems \cite{Hardy01, Hardy01b}. Since his axioms will prove
important in our discussion of quantum communication in chapter
\ref{chap qubit}, I am now going to reproduce them, adapting slightly
from \cite{Hardy01b}. 

Before we proceed, let us recall what pure states are. We say a physical system is in a pure state when its measurement outcomes cannot be reproduced by any statistical mixture of other states. In this sense, pure states are extremal in the set of all allowed states. Operationally speaking, pure states embody the idea of states for which some property is known for certain, without statistical uncertainty. This is because for every pure state there exists an experiment which always results in a predictable outcome. 

Now we need a few definitions:

\begin{description}
\item[Dimension]: The dimension of a classical or quantum system
(denoted by $N$ below) is
the maximum number of distinguishable states in which we can prepare the system.
\item[Degrees of freedom]: The number of degrees of freedom of a
system (denoted by $K$ below) is the number of probability
measurements necessary to completely describe a state.
\end{description}

More generally, the number of degrees of freedom is the number of
independent real parameters needed to completely characterise the
state of a system. With these two simple definitions, we can state the axioms:

\begin{description}
\item[Axiom 1] {\it Probabilities}.  Relative frequencies (measured by
taking the proportion of times a particular outcome is observed)
tend to the same value (which we call the probability) for any case
where a given measurement is performed on a ensemble of $n$ systems
prepared by some given preparation in the limit  $n \to \infty$.
\item[Axiom 2] {\it Subspaces}. There exist systems for which
$N=1,2,\cdots$, and, furthermore, all systems of dimension $N$, or
systems of higher
dimension but where the state is constrained to an $N$ dimensional
subspace, have the same properties.
\item[Axiom 3]  {\it Composite systems}. A composite system consisting of
subsystems $A$ and $B$ satisfies $N=N_AN_B$ and $K=K_AK_B$.
\item[Axiom 4] {\it Continuity}. There exists a continuous reversible
transformation on a system between any two pure states of that
system for systems of any dimension $N$.
\item[Axiom 5] {\it Simplicity}. For each given $N$, $K$ takes the
minimum value consistent with the other axioms.
\end{description}

Rigorously speaking, the fifth axiom is not an axiom, as it makes
reference to the other axioms. Nevertheless, the five propositions stated above are
simple and can be motivated by our experience with classical
probability and physical reality, with the exception of Axiom 4. In
finite-dimensional classical systems, we can only jump discontinuously
between one pure state and another, for example different digital
values of a computer memory, or different sides of a die. However, if the word
``continuous'' is dropped from Axiom 4, it is possible to show (see
\cite{Hardy01,Hardy01b}) that the five axioms give rise to classical probability theory.

By rejecting these discontinuous jumps between pure states of a finite-dimensional system, and re-introducing the word ``continuous''
in Axiom 4, Hardy showed that we can derive the
mathematical structure of quantum theory instead. This includes the linearity of
quantum mechanics, the trace formula for expected values of
observables, and the general form of allowed super-operators.

In this approach, the crucial difference between quantum and classical
physics is the continuous nature of the set of quantum pure states, as
opposed to the discrete nature of the set of classical pure states. As
we will see in chapter \ref{chap qubit}, it is exactly this which
allows for an unbounded separation in communication power between a
qubit and any finite-dimensional classical system.

\section{The invariant information of Brukner and Zeilinger \label{sec invinfo}}

Another approach to the foundations of quantum mechanics was taken by Brukner and Zeilinger in a series of papers
\cite{BruknerZ99, Zeilinger99, BruknerZ99b, BruknerZ02,
BruknerZ01}. It is based on taking as a foundational principle the
following statement: `an
elementary system represents the truth value of one proposition'. This
amounts to using what is basically Holevo's theorem as a foundational
principle for quantum mechanics.

Brukner and Zeilinger have been investigating how much of quantum
theory can be derived, or better understood, using this simple
principle and little else. As an example of their approach,
let us have a look at how they analyse the simple case of two-qubit states \cite{Zeilinger99, BruknerZZ01}.

According to the principle, two qubits can encode a maximum of
only two bits. However, there is a continuous set of possible encoding
strategies.  In one extreme, the two bits could be encoded only in the
correlations of outcomes for Alice and Bob. One way to do this would
be to prepare the two qubits in one of the four Bell states, which can
be discriminated through local operations and classical communication
between the two parties. The two bits of information can only be
gathered when the parties communicate to compare their measurements,
obtaining the tell-tale correlations. Because all the information is
contained in the correlations, however, there must not be any way for
them to gather more information through their local measurements. This
would then be the reason why the local measurement outcomes are
completely random (corresponding to a completely mixed reduced density
matrix $\rho=\mbox{$1 \hspace{-1.0mm}  {\bf l}$}/2$).

To continue with this simple example of two qubits, we could instead
prepare a two-qubit state so that the two encoded bits are available
locally. This would entail preparing definite states for each qubit
separately. According to the principle, if the two bits are available
locally, there must not be any further information to be gathered in
the correlations. No entanglement is allowed, and we are dealing now
with two qubits in a product state.

Instead of dwelling further on the
consequences of their principle, let us now turn to an essential
ingredient of their approach, which is a mathematical expression
quantifying the amount of information in a general state. This
`invariant information' $I_{total}$ will be useful when
we investigate the so-called quantum random access codes in chapter
\ref{chap qcc}. Before defining the invariant information $I_{total}$, let us
review the Shannon and von Neumann entropies, and propose a quantum
measurement scenario for which they are \textit{not} the  most adequate measure
of information.

\subsection{Shannon and von Neumann entropies}

Let us start by considering a completely classical setting, before
moving on to the quantum case. When we perform measurements on a classical system with $d$ distinguishable states, we obtain probabilistic outcomes whose
information content can be characterised by the following simple
expression involving the Shannon entropy $H(p_1,p_2,...,p_d)$: 
\begin{equation}
I_{S}=\log(d)-H(p_1,p_2,...,p_d)=\log(d)+\sum_{i=1}^d p_i \log(p_i). \label{shannoninfo}
\end{equation}
A completely predictable outcome corresponds to $H=0$, and therefore
to maximal information $I_{S}=\log(d)$ bits. Intuitively, this is
quantifying the number of bits of information that can be transmitted
between two parties by preparing this system in one of its $d$
distinguishable states. This would correspond to
optimal use of the $d$ distinguishable states for information
transmission.

There exist other less efficient ways of using the same classical system for
information transmission, for example by preparing it in a statistical
mixture of its distinguishable states. In this case $H>0$ and
$I_S<\log(d)$; the less than optimal $I_S$ reflects the non-optimality
of the preparation procedure.

In classical physics all measurements commute and the information
present in the state has a pre-defined, objective existence prior to
and independently of the observation. Quantum systems are different in
this respect. As we have seen in the previous sections, contextuality
and non-locality in general prevent quantum systems from having
experimental outcomes which are pre-defined independently on the
choice of compatible observables being measured, even in space-like
separated regions. This subtlety behind quantum measurements requires us to be careful
when trying to quantify information encoded in quantum systems.

When thinking about entropy or information for quantum systems, the
first thing one could do would be to try to generalise the Shannon
entropy to the quantum case. This is achieved by the von Neumann
entropy $S$, which is simply the Shannon entropy of the eigenvalues of
the $d\times d$ density matrix $\rho$. $S(\rho)$ is given by
\begin{equation}
S(\rho)=-Tr(\rho)\log(\rho),
\end{equation}
an expression which is invariant under change of basis. For an extensive
discussion of the mathematical properties of $S$ see the review article by Wehrl \cite{Wehrl78}.

The von Neumann entropy $S$ is central in
many results in quantum information theory: channel capacities,
entanglement dilution and concentration, teleportation, and so on. It
appears in the simplest possible quantum information application:
how to send information from Alice to Bob by using mixed states
$\rho_i$ with respective probabilities $p_i$. The von Neumann entropy
of this `alphabet' of states $S(\rho=\sum_i p_i \rho_i)$ is an upper
bound on the amount of information per message that Alice can send
Bob. A tight bound was obtained by Holevo (see \cite{Holevo73} and
\cite[chapter 12]{NielsenC00}).

Despite its importance and interest, we cannot expect the von Neumann
or Shannon entropies to be the most adequate measures of information
in all quantum information applications. Let us imagine a situation
where a single quantum system can be measured with either of two projectors $\hat{P}_1$ or
$\hat{P}_2$, which do not commute. We can think of preparing quantum states
such that $\hat{P}_1$'s measurement outcomes have maximum entropy, and
$\hat{P}_2$'s minimum, or some balance of the two. In this context, the
invariant information of Brukner and Zeilinger offers an interesting
way to put bounds on the entropies of \textit{alternative}, and
mutually exclusive, measurements performed on a single system, as we
will see next.

\subsection{Defining invariant quantum information \label{sec definvinfo}}

Brukner and Zeilinger's approach to quantifying quantum information
started from an analysis of the axioms used by Shannon for deriving
what became known as Shannon entropy. They then proceeded to a
criticism of the applicability of these axioms in the context of
quantum measurements \cite{BruknerZ01}, which motivated their proposed
measure of quantum information.

Here I shall motivate Brukner and Zeilinger's invariant information in
a somewhat different, and simpler, way than their original
proposal. Instead of considering quantum information in its generality
and criticising the use of Shannon's entropy, I will concentrate on a
simple quantum measurement scenario and look for an expression which
properly describes the information gathered in it.

Suppose that we are merely trying to quantify the
amount of information that can be gathered in non-commuting
measurements of a single quantum system. Because of the non-commutativity, we
know that the measurement of a single system necessarily `projects' it
into one eigenvector of the first observable, ruining our possibility
of obtaining further information through measurement of other
observables. Nevertheless, there should be a balance between the
information that can be gathered using alternative, exclusive choices
of  projective measurements. Our aim then is to describe
quantitatively this balance between alternative measurement choices.

To simplify things, let us keep to the most symmetric possible sets
of observables, which are projections on
\textit{mutually unbiased bases} (MUB's). These are bases such that
preparation of a quantum state along any basis vector of basis $i$
yields equally probable outcomes for a projection on basis $j \ne i$. For
dimensions $d$ which are powers of primes it has been shown
\cite{Ivanovic81,WoottersF89, WoottersF89b} that it is possible to
define $d+1$ MUB's. In the qubit case, the three MUB's can consist,
for example, of the eigenvectors of projectors $\hat{P}_x$, $\hat{P}_y$ and $\hat{P}_z$
along the three Cartesian axes in the Bloch sphere.

To quantify the amount of information, or the randomness of the
resulting outcomes, one immediately thinks of using the  Shannon
entropy $H$ of the measurement outcomes. For the example of one qubit,
we may define the total entropy of these outcomes by the
sum of the three Shannon entropy-based $I_S$ [eq. (\ref{shannoninfo})]
associated with the outcomes of the three projectors:
\begin{equation}
I_S^{total}=3-H(\hat{P}_x)-H(\hat{P}_y)-H(\hat{P}_z).\label{Itotalsha}
\end{equation}
For a qubit prepared in an eigenstate of $\hat{P}_z$, for
example, $I_S^{total}=1$. We expect that a useful measure of information
in this simple case should be invariant under unitaries. It is easy to
see, however, that $I_S^{total}$ does \textit{not} satisfy this simple criterion:
it depends on the pure state being used, and it takes different values
depending on the choices of MUB's. This suggests that defining
information for such alternative settings by the sum of Shannon
entropies of the outcomes of alternative measurements may not be a
good idea after all. How can we understand this unsuitability of
Shannon entropy in this case?

The sum of Shannon entropies is adequate for describing the entropies
of independent events, which exactly for being independent have
\textit{additive} entropies. The non-commutativity of the measurements
we are considering imply they are \textit{not} independent, and for
this reason the sum of the Shannon entropies is not adequate. Using
Shannon entropies in this case amounts to assuming that projections on
the three MUB's have pre-determined, independent outcomes, as would
happen with three hidden dice in three different boxes, only one of
which can be opened.

As we have seen earlier in this chapter, such previously existing
values cannot even in principle be assigned to outcomes of
non-commuting quantum measurements. Thus, in order to go ahead with
the idea of quantifying the amount of information available through
measurements on MUB's, we need to find a more suitable entropy
function.

Shannon's $\sum_{i=1}^n p_i \log(p_i)$ entropy formula has the virtue
of being additive for independent systems. As argued
above, this is not an appropriate assumption in the case of
alternative measurements on a single
quantum system. Instead, we would like to have a \textit{sub-additive}
entropy, to reflect the fact that such projections over MUB's are not
independent: a large information (low entropy) of outcomes of
one projector necessarily implies a reduced information
(larger entropy) associated with other non-commuting projectors.

While the general functional form of the Shannon and von Neumann
entropies is the only one that is additive, other entropic functions
have been proposed without this property. These include
\begin{eqnarray}
M_{\alpha}&=&\left( \sum_{i=1}^{n} p_i^{\alpha} \right)^{\alpha-1}, \label{na entropy1}\\
N_{\alpha}&=&\frac{1}{1-\alpha}\sum_{i=1}^{n}(p_i^{\alpha}-1), \label{na
entropy2}\\
O_{\alpha}&=&\frac{1}{1-\alpha}\sum_{i=1}^{n}(p_i^{\alpha}),\label{na entropy3}
\end{eqnarray}
which can be defined for a wide range of the parameter $\alpha$. These
were proposed respectively by Hardy, Littlewood, and P\'{o}lya
\cite{HardyLP52}; Tsallis \cite{Tsallis88}; and R\'{e}nyi
\cite{Renyi62}.

The invariant information introduced by Brukner and Zeilinger in
\cite{BruknerZ99, BruknerZ01} uses the sub-additive entropy
\begin{equation}
U(p_1,p_2,...,p_n)=\frac{n-1}{n}-\sum_{i=1}^{n}\left(p_i-\frac{1}{n}    \right)^2,
\end{equation}
which is quadratic in the probabilities and similar to the entropies
(\ref{na entropy1})-(\ref{na entropy3}) above. The factor $1/n$ is
introduced as this is the probability associated with a uniformly random
variable with $n$ outcomes. Intuitively, this entropy function
measures a sort of `distance' from a probability distribution to the
maximally random one. I introduced the additive term $\frac{n-1}{n}$ here for convenience: this way the entropy is zero when one of the
probabilities $p_i=1$, with all the other $p_j$'s being equal to zero.

We will see now that this non-additive entropy can be used to define
an information measure with the desired property, i.e. invariance under
unitaries. Let us define the information $I$ of an $n$-outcome measurement
as:
\begin{equation}
I(p_1,p_2,...,p_n)=\frac{n-1}{n}-U(p_1,p_2,...,p_n)=\sum_{i=1}^{n}\left(p_i-\frac{1}{n}    \right)^2.
\end{equation}
As defined above, the information takes minimum value
zero when all outcomes have equal probabilities $p_i=1/n$, and a
maximum value of $(n-1)/n$ when one of the probabilities is one, and
all the others zero.

We can use this definition of information to
characterise the total information content of outcomes of projections
on a set of $d+1$ alternative MUB's that exist for a $d$-dimensional
system. For that, we just add the information obtained for $d$-outcome
measurements on $d+1$ different MUB's. This constitutes the \textit{quantum
invariant information} $I_{total}$ of Brukner and Zeilinger:
\begin{equation}
I_{total}=\sum_{j=1}^{d+1}I(p_1^j,p_2^j,...,p_d^j) =
\sum_{j=1}^{d+1}\sum_{i=1}^{d} \left(p_i^j -\frac{1}{d} \right)^2. \label{Itotalddim}
\end{equation}
Using the properties of MUB's it is possible to evaluate
eq. (\ref{Itotalddim}) explicitly:
\begin{equation}
I_{total}=Tr(\rho^2)-1/d
\end{equation}
(see \cite[chapter 2]{Brukner99} for details). Now we see that, unlike
the information measure using the Shannon entropy
[eq. (\ref{Itotalsha})], $I_{total}$ is invariant under unitaries,
depending only on the purity of the system, as measured by
$Tr(\rho^2)$. Note that $I_{total}$ assumes a minimum value of zero for
maximally mixed states $\rho=\mbox{$1 \hspace{-1.0mm}  {\bf l}$}/d$,
and a maximal value of $\frac{d-1}{d}$ for a pure $d$-dimensional
state.

$I_{total}$ quantifies in a neat way the
balance between information that can be gathered through different
measurements that do not commute. It also provides an inequality-free
version of Heisenberg's principle (see \cite{Durr01,Hall01}). More
generally, there have been many uses of non-additive entropic
functions, for example to analyse decoherence
\cite{VidiellaBarrancoMC01} and to quantify entanglement
\cite{TsallisLB01}.

These results suggest that in at least some applications, it may be
more convenient to consider non-additive forms of entropy for
characterisation of quantum information. In chapter \ref{chap qcc} we
will see an example of this, where $I_{total}$ naturally provides us
with means to quantify the advantage of using quantum systems instead
of classical systems in a simple information transmission task.


\chapter{Quantifying entanglement \label{chap ent}}

When we use non-locality in a number of quantum information
applications, it is natural to see it as a resource to be consumed and
manipulated in different ways. This view led to many attempts
to better understand and mathematically characterise entanglement in all its forms and types.

Today quantum entanglement represents a wide research subject which has been flourishing vigorously in the
last few years. In this chapter I will make no attempt to describe
all the main results in the area. Instead, I will just give a bird's
eye view of the subject of quantifying entanglement, mostly referring
to work which will be useful later on. This will be particularly
handy in chapter \ref{chap tri}, where I describe some of my
work on tripartite pure-state entanglement.

We start in section \ref{sec entvssep} by distinguishing separable and
entangled states, and reviewing some of the known results on
separability criteria. In section \ref{sec 2approaches} I introduce
the two main approaches towards quantifying entanglement developed in the
literature. These involve either discussing entanglement in single
states, or considering the asymptotic limit of manipulations on large
number of copies. Some of the main results of these two approaches are
presented in sections \ref{sec singlecopy} and \ref{sec
asymplim}. Finally, in section \ref{sec qreintro} I introduce the
quantum relative entropy, which can be used to formulate a measure of entanglement that will be useful
when we discuss tripartite entanglement in chapter \ref{chap tri}.

\section{Entangled versus separable states \label{sec entvssep}}

From a mathematical point of view, an $N$-partite pure quantum state is
defined as being separable when it can be written as a product of
states of the individual parts:
\begin{equation}
\left|\psi\right\rangle_{sep}=\left|\phi\right\rangle_A
\left|\phi\right\rangle_B \cdots \left|\phi\right\rangle_N,
\end{equation}  
where $\left|\phi\right\rangle_j$ represents a pure state belonging
to party $j$. Similarly, a $N$-partite mixed state $\rho$ is separable
when it can be represented as a convex combination of tensor products
of (possibly mixed)
states of its subsystems:
\begin{equation}
\rho_{sep}=\sum_i p_i \left(\rho_A \otimes \rho_B \otimes \cdots
\otimes \rho_N \right), \hspace{3 mm}\sum_i p_i=1. \label{mixaltsep}
\end{equation} 
Note that any mixed separable state representable as in
eq. (\ref{mixaltsep}) can also be put in the alternative form
\begin{equation}
\rho_{sep}=\sum_i p_i \left|\phi_i\right\rangle
\left\langle\phi_i\right|_A \left|\phi_i\right\rangle \left\langle\phi_i\right|_B
\cdots \left|\phi_i\right\rangle 
\left\langle\phi_i\right|_N, \hspace{3 mm}
\sum_i p_i=1. \label{gensepstate}
\end{equation} 
Whenever the state is not separable, we say it is
\textit{entangled}.

The definition above has a simple operational characterisation behind
it: the subsystems of a separable
state can be produced separately by the different parties, whereas
this is not possible for entangled states. To produce the separable
state (\ref{gensepstate}), each party $j$ prepares state
$\left|\phi_i\right\rangle_{j}$ in a classically correlated
fashion with the others. The states must be prepared according to the probability
distribution $p_i$ characterising the separable state $\rho_{sep}$,
which can be done if the parties previously share instructions on the
appropriate sequence of states to prepare. This procedure is a
perfectly valid way of creating the separable state $\rho_{sep}$
above. Entangled states, on the other hand, exhibit correlations among
measurement outcomes which simply cannot be emulated by any such
locally prepared (i.e. separable) state.

As we have seen in section \ref{sec
nonlocality}, measurements on entangled states can reveal
non-locality. More precisely, for any bipartite pure state there exist
measurements whose outcomes are incompatible with any local
hidden-variable theory (a result known as Gisin's theorem
\cite{Gisin91, GisinP92}). In the case of multipartite states the
situation is more complicated, but Popescu and Rohrlich have shown
\cite{PopescuR92} that provided the parties can collaborate and use
post-selection, then any multipartite pure entangled state violates
some Bell-type inequality. The question of which inequality to use in
order to reveal the non-locality of a given entangled state is hard;
see \cite{ZukowskiBLW02,WernerW01,CollinsGPRS02,Acin02} for some
recent results.

For mixed states, the relation between entanglement and non-locality is
less well understood, and there are many open problems. In
\cite{Werner89} Werner showed that there exist mixed states of two
parties which are not separable, and yet admit a local hidden-variable
model. If a state is distillable, then with some probability one can
distill pure-state entanglement from it, and violate some Bell
inequality by the results of
\cite{Gisin91,GisinP92,PopescuR92}. However, as we will see soon,
there exist \textit{bound entangled states} which cannot be distilled
by local operations and classical communication into standard pure
entangled states. Some multipartite bound entangled states have been shown
to violate Bell-type inequalities \cite{Dur01,SenDeSZ02}, but it is not
known whether this is the case for all bound-entangled states \cite{Terhal00}.

Unfortunately, the definition (\ref{gensepstate}) of separable states gives no clue as to how to test a
given density matrix for separability. A great deal of research has
been done on the important problem of finding separability
criteria. Let us briefly review one of the simplest yet strongest
tests, the Peres-Horodecki criterion.

\subsection{The Peres-Horodecki separability criterion \label{sec phsep}}

A remarkable early result
was the Peres criterion, which is a powerful
necessary criterion for bipartite separability \cite{Peres96}. Soon
after its publication, the Horodecki family proved it was also
sufficient for systems of dimension $2 \times 2$ and $2 \times 3$
\cite{HorodeckiHH96}.

To motivate the criterion, let us consider the non-unitary operation
of transposition. Despite being non-unitary, the transpose of a
density matrix is also a density matrix. The transposition is an
example of a mathematical operation that cannot be physically
implemented, but which takes a density matrix $\sigma$ into another
density matrix $\sigma^T$ representing a physically possible state.

Now let us consider what happens when we transpose the states
corresponding only to subsystem $A$ of a bipartite composite quantum
system $\sigma_{AB}$. This partially transposed matrix $\rho$ of a bipartite density matrix
$\sigma$ is defined as
\begin{equation}
\rho_{m \mu,n \nu}=\sigma_{n \mu, m \nu}, \label{parttransp}
\end{equation}
where the Latin indices representing subsystem $A$ where transposed,
while the Greek indices of subsystem $B$ were not. In the case of
separable states, this operation always results in a physical
state. To see this, we can think of a simple way of preparing any
separable state of two parties:
\begin{equation}
\rho_{sep}=\sum_i p_i \left|\phi_i\right\rangle
\left\langle\phi_i\right|_A \left|\phi_i\right\rangle \left\langle\phi_i\right|_B,
\sum_i p_i=1. \label{sep2party}
\end{equation} 
One way to prepare this state is to issue instructions to Alice and
Bob, so that they release the classically correlated states
$\left|\phi_i\right\rangle_A$ and $\left|\phi_i\right\rangle_B$ with
the right probability distribution $p_i$. This can always be done for
separable states, because they are exactly those states admitting a
description in terms of definite subsystem states.

The partial transpose of separable state (\ref{sep2party}) is
another separable state, in which each of the states describing Alice's subsystems are transposed. We thus see that partial
transposition takes separable states into separable states, even
though again it is not a physical operation that can be performed on
Alice's system. Rather, it is a mathematical operation resulting in
another separable density matrix which can be physically prepared
using classical correlations only, as described above.

Peres noted, however, that when applied to entangled states, the
partial transposition need not result in a physical state. This is
related to the fact that entangled states do not admit a description
in which each subsystem's state is specified without reference to
other subsystems, as is the case with separable states. The partial
transposition of entangled states can result in negative operators,
i.e. operators with negative eigenvalues. This leads to Peres'
necessary criterion for separability:
\begin{description}
\item{\bf Peres' separability criterion \cite{Peres96}}: Positivity under partial
transposition (PPT for short) is a necessary condition for
separability.
\end{description}
This separability test has the great advantage that it is
computationally simple to perform. All that is needed is to swap some elements of the
density matrix [corresponding to the partial transposition operation
(\ref{parttransp})], followed by a simple diagonalisation to find
whether the resulting matrix has negative eigenvalues. If it does, the
state is entangled.

This raises the question of whether the condition is also sufficient to guarantee separability. The Horodeckis answered this question in the negative
\cite{HorodeckiHH96} by proving that it is sufficient only for systems
of dimensionality $2 \times 2$ and $2 \times 3$. Interestingly, they
showed that there exist \textit{bound entangled} states, which despite being
PPT (positive under partial transposition), are not separable. In
\cite{Horodecki97} it was shown that the entanglement of these states
cannot be distilled into a useful form for protocols such as
teleportation, and it was this property which motivated the term `bound entanglement'.

Since then, many other results on separability were obtained;
for a review see Lewenstein \textit{et al.}'s article \cite{LewensteinBCKKSST00}. Some important criteria
have been derived recently, and were not reviewed in
\cite{LewensteinBCKKSST00}. These include the results reported in
\cite{Rudolph00, Rudolph02} and \cite{DohertyPS02}, for example.

\section{Different approaches to quantifying entanglement\label{sec 2approaches}}

When trying to quantify entanglement, it is important to specify which
type of system and approach we are taking. Some useful distinctions
among quantum systems are: bipartite versus multipartite states;
general mixed states or pure states only; and systems described either
by discrete or continuous variables, examples of which are a
spin-$1/2$ particle and the electric field, respectively. In this
chapter I will mention some results on all these areas, with the
exception of continuous variables.

An important question regards how to manipulate entanglement and
convert it from one form to another, using only local operations and classical communication (LOCC). If we allow operations more
general than these (for example by allowing quantum communication), it is obvious that we can create any amount of
entanglement. Since it is exactly the non-locality present in entangled
states that we want to quantify, we need to restrict ourselves to
finding out what transformations are possible under LOCC
only. Unfortunately, LOCC are a set of operations which is very hard
to characterise mathematically, unlike simple unitaries for
example. Nevertheless, many important results have been obtained, and they
pertain to two main approaches which we need to specify.

The first approach consists of investigating how to interconvert
the entanglement present in single copies of entangled states. The second approach is inspired in classical
information theory, and asks what is possible to do when the parties
have access to an ensemble of states, i.e. an arbitrarily large number
$N$ of identical copies. Let us now briefly review some important
results obtained in recent years along these two lines of research.

\section{Single-copy approach \label{sec singlecopy}}

This approach investigates which transformations are possible with
LOCC on single states. If necessary and sufficient conditions for this
type of manipulation on mixed states can be found, this would be
directly relevant to experiments involving small number of qubits. On
the other hand, there are clues that the overall picture of
entanglement simplifies only in the asymptotic limit of large
number of parties, much in the same way as it happens in the
asymptotic limit of classical information theory. This is
suggested by some results I will describe in this and the next
section. We will see that, in the single copy approach, the number of entanglement
classes increases rapidly with the number of parties. Moreover, in the bipartite pure-state case a
single number can be shown to encapsulate the entanglement properties
of any given state, but only in the asymptotic limit of many copies.

A key result in the single-copy regime is Nielsen's theorem
\cite{Nielsen99}, which establishes a simple necessary and sufficient
condition for interconvertibility of pure bipartite states under LOCC. In order
to state it, first we need to introduce the
concept of majorization. To do that, let us use the notation
$\vec{v}^{\downarrow}$ to indicate a re-ordered vector $\vec{v}$, in
such a way that the largest element comes first, the next-largest
second, and so on. Then we can say that $\vec{x}=(x_1,x_2,\ldots,x_d)$
is \textit{majorized} by $\vec{y}=(y_1,y_2,\ldots,y_d)$ (written
$\vec{x}\prec \vec{y}$) if $\sum_{j=1}^{k}x_j^{\downarrow} \le
\sum_{j=1}^{k}y_j^{\downarrow}$ for $k=1,2,\ldots,d$, with equality
when $k=d$.

Now, let us define $\vec{\lambda_{\rho}}$ to be the vector of
eigenvalues of density matrix $\rho$. We are now in position to state
Nielsen's theorem:

\begin{description}
\item[Theorem (Nielsen \cite{Nielsen99})] A bipartite pure state
$\left|\psi\right\rangle$ can be tranformed by LOCC to state
$\left|\phi\right\rangle$ if and only if $\vec{\lambda_{\psi}} \prec \vec{\lambda_{\phi}}$. 
\end{description}
This theorem basically closes the interconvertibility problem for
exact transformations of bipartite pure states. Soon after it was
published, Hardy found an alternative geometrical proof of the
theorem, and obtained the optimal strategy for distilling maximally
entangled states from any bipartite pure state
\cite{Hardy99}. Jonathan and Plenio derived the same result
independently in \cite{JonathanP99, JonathanP00} using other methods,
and also worked out other applications of Nielsen's theorem.

The theorem above only deals with exact transformations. Where an
exact transformation is not possible, we can always try to perform an
optimal, approximate transformation instead. This will be necessary in
most cases, as interconvertibility is not guaranteed in general. For
example, $\left|\phi\right\rangle$ cannot be converted into
$\left|\psi\right\rangle$ whenever the latter has a larger Schmidt
number than the former. 

It is thus important to address the question of approximate
transformations. More precisely, starting with state
$\left|\psi\right\rangle$, we may want to find the closest
approximation $\left|\phi_{approx}\right\rangle$ of target state
$\left|\phi\right\rangle$ that can be reached. In \cite{VidalJN00}
Vidal \textit{et al.} obtained this optimal approximate conversion
strategy for pure states, when we quantify the distance between
$\left|\phi\right\rangle$ and $\left|\phi_{approx}\right\rangle$ by
their mutual fidelity.

When we leave the pure states and consider mixed states, or even pure
states of more parties, the problem of interconvertibility becomes harder. Even for mixed states of two
parties, no necessary and sufficient criterion was found to
date. Some partial results are known: for example, Vidal introduced
entanglement monotones, which are functions that cannot decrease under
LOCC, for pure or mixed states of two or more parties
\cite{Vidal00b}. These monotones can be used to formulate necessary
criteria for a LOCC transformation to be possible.

\subsection{Multipartite entanglement}

An interesting result concerns the number of different types of pure
multi-party entanglement. It has been shown in \cite{Vidal00,
BennettPRST01} that two pure multipartite states
$\left|\psi\right\rangle$ and  $\left|\phi\right\rangle$ can be
obtained deterministically from each other by LOCC if and only if they are
related by local unitaries. Since even in the bipartite case this is
not usually the case \cite{Vidal00,LindenP98,CarteretLPS99}, if we define `different types of entanglement'
by mutual deterministic interconvertibility we will be stuck with
disjoint families of different types of entanglement parametrised by a
continuous parameter.

This motivated D\"{u}r, Vidal
and Cirac to relax somewhat this interconvertibility criterion,
grouping entangled states in the same class if they can be
interconverted by LOCC with any non-zero probability, instead of
deterministically. This makes sense, as states in the same class
can be used for the same applications, albeit with different
probabilities of success.

Using their newly defined classifying criterion for multipartite
entanglement, D\"{u}r, Vidal and Cirac were able to show that there
are two genuinely different kinds of tripartite entanglement for
three-qubit states \cite{DurVC00}. The generic family consists of
states which can be converted stochastically into GHZ states
\[
\left|\mbox{GHZ}\right\rangle=\frac{1}{\sqrt{2}}\left(
\left|000\right\rangle +  \left|111\right\rangle\right).
\]
Remarkably, there is another family of states with tripartite
entanglement which cannot be converted into GHZ states, and therefore
belong to a different class. A representative of this class is the
so-called \textit{W state}\footnote{Named after Wolfgang D\"{u}r, one
of the authors of \cite{DurVC00}.}:
\begin{equation}
\left|\mbox{W}\right\rangle=\frac{1}{\sqrt{3}}\left(
\left|001\right\rangle +  \left|010\right\rangle +
\left|100\right\rangle \right). \label{Wstatedef}
\end{equation}
This state has an interesting entanglement structure which we will
explore further in chapter \ref{chap tri}.

Following this approach, in \cite{VerstraeteDdMV02} the nine different
classes for a four-qubit system were worked out, the generic class
being again the GHZ state of four qubits. More generally, in
\cite{DurVC00} it was shown that the number of different entanglement
classes increases rapidly with the number of parties.

\section{Asymptotic limit approach \label{sec asymplim}}

A second approach to quantification of entanglement takes its
inspiration from classical information theory, and considers the
transformations under LOCC on an arbitrarily large number $n$ of identical copies of the same state. It turns out
that we can use a single number to quantify the amount of entanglement
of any pure bipartite state in this asymptotic limit.

In order to discuss this problem and introduce multipartite
entanglement, first we need to review concepts such as asymptotic
equivalence and Minimal Reversible Entanglement Generating Sets
(MREGS).

\subsection{Asymptotic equivalence \label{sec asympequ}}

Let us now present the idea of asymptotic reducibility between two
multipartite pure states, discussed by Bennett \textit{et al.} in \cite{BennettPRST01}. Let us denote by $F$ the fidelity between
two states $\left|\psi\right\rangle$ and $\left|\phi\right\rangle$:
\begin{equation}
F=|\langle\psi|\phi\rangle|^2.
\end{equation}
Also, let us use $\cal{L}$ to denote any locally implementable
super-operator, i.e. any manipulation of the quantum state which can be
performed by using LOCC only. Then state $\left|\psi\right\rangle$ is said to be
\textit{asymptotically reducible} to state $\left|\phi\right\rangle$
if and only if
\begin{eqnarray}
\forall {\delta>0,\epsilon>0} \hspace {3 mm}\exists {n,n',{\cal L}}\\
\left|(n/n')-1 \right|<\delta, \hspace{3 mm} F({\cal L}(\psi^{\otimes
n'}),\phi^{\otimes n}) \ge 1-\epsilon,
 \label{defasred}
\end{eqnarray}
In words: $\left|\psi\right\rangle$ is said to be
asymptotically reducible to state $\left|\phi\right\rangle$ if there
exists a LOCC operation that takes $n'$ copies of
$\left|\psi\right\rangle$ into an arbitrarily good approximation of
$n$ copies of $\left|\phi\right\rangle$, for large enough $n',n$.

With the concept of asymptotic reducibility, we can define asymptotic
equivalence. States $\left|\psi\right\rangle$ and
$\left|\phi\right\rangle$ are \textit{asymptotically equivalent} if
and only if $\left|\psi\right\rangle$ is asymptotically reducible to
$\left|\phi\right\rangle$, and vice-versa.

To illustrate the usefulness of these concepts, let us re-state the
results of \cite{BennettBPS96} using them. In \cite{BennettBPS96}, Bennett \textit{et al.} showed that all bipartite pure states are asymptotically equivalent
to a pair of
qubits in a Bell state, for example
\begin{equation}
\left|\psi^{-}\right\rangle=\frac{1}{\sqrt{2}}\left(
\left|0_{A}0_{B}\right\rangle - \left|1_{A}1_{B}\right\rangle
\right). \label{psiminushere}
\end{equation}
Moreover, it was shown that a simple function quantifies this
reducibility relation. To present this result, let us first note that asymptotically reversible transformations can have non-integer yields. More precisely, we say $\left|\psi\right\rangle^{\otimes x}$ is asymptotically equivalent to $\left|\phi\right\rangle^{\otimes y}$ (with $x,y$ real numbers) if and only if

\begin{eqnarray}
\forall {\delta>0,\epsilon>0} \hspace {3 mm}\exists {n,n',{\cal L}}\\
\left|(n/n')-x/y \right|<\delta, \hspace{3 mm} F({\cal L}(\psi^{\otimes
n'}),\phi^{\otimes n}) \ge 1-\epsilon.
 \label{defasred}
\end{eqnarray}

Now we can consider asymptotic transformations of copies of a less-than-maximally entangled state $\left|\eta\right\rangle$ and maximally entangled states $\left|\psi^{-}\right\rangle$:
\begin{equation}
\left|\eta\right\rangle^{\otimes x} \stackrel{LOCC}{\rightleftharpoons }
\left|\psi^{-}\right\rangle^{\otimes y}.
\end{equation}
Bennett \textit{et al.}
showed that this transformation is possible in the asymptotic limit, with the asymptotic yield $x/y$ approaching $S(\rho_A)=S(\rho_B)$, the von Neumann entropy of the reduced density matrices of state $\left|\eta\right\rangle$.

Transformations from EPR pairs to
less entangled states are referred to as \textit{entanglement dilution}, and the reverse is called
\textit{entanglement concentration}. As we discussed above, the same
problem can be discussed in the finite-copy approach, but in that case
the transformations are in general irreversible.

Based on this result, we can say that in the asymptotic limit there is
only one type of bipartite pure-state entanglement, and this is
conveniently represented by an EPR pair of qubits. This entanglement
can be manipulated reversibly in the asymptotic limit, which is a very
nice result and gives support to the idea that entanglement is a
resource to be quantified and manipulated for different quantum information processing protocols.

\subsubsection{Mixed state entanglement}

When we consider entanglement manipulations with mixed states, the
situation becomes intriguing even for bipartite states. In the
single-copy case, the Horodeckis showed
\cite{Horodecki97,HorodeckiHH98} that there existed entangled
states from which no entanglement could be distilled. In bipartite
systems, these peculiar states exist only in dimensions $3 \times 3$
and higher. They were named \textit{bound entangled} states, because
the entanglement in them seems not to be in a `free form', as it cannot be
distilled into a form useful for a task such as
teleportation. Surprisingly, in \cite{MuraoV01} a quantum information
task was shown to have a better performance when using a tripartite
bound-entangled state, than would be possible with classically
correlated states. The properties and possible uses of bound entangled
states are yet not clear; a good reference discussing this problem and
separability is the review by Lewenstein \textit{et al.} \cite{LewensteinBCKKSST00}.

Because they are not separable, some entanglement needs to be invested to create bound entangled states, even though none
can be distilled in the single-copy regime. Although it seemed
unlikely, the possibility remained that in the asymptotic limit the
required entanglement to create these states would tend to
zero, and the distillation/concentration processes would recover the
reversibility of the pure-state case. It was only through the recent
work of Vidal and Cirac \cite{VidalC01,VidalC02} that the question was
definitely settled in the asymptotic regime. They showed that also in
the asymptotic regime a non-vanishing amount of entanglement is
required to create some entangled states, while no entanglement can be
distilled from them.

\subsection{Minimal Reversible Entanglement Generating Sets (MREGS)}

In reference \cite{BennettPRST01} the concept of Minimal Reversible Entanglement Generating Sets
(MREGS for short) was introduced as a natural way of tackling the
pure-state entanglement problem in a multipartite setting. This
concept sets a framework which I use in chapter \ref{chap tri} to
explore tripartite pure-state entanglement in the asymptotic limit.

\begin{description}
\item[Definition: ]A Reversible Entanglement Generating Set
(REGS) for $N$-partite pure states is a set $G$ with $j$ pure $N$-partite
states $G=\{\left|\phi_1\right\rangle,
\left|\phi_2\right\rangle,\dots,\left|\phi_j\right\rangle\}$  having the property that
any $N$-partite pure entangled state is asymptotically equivalent to
tensor products of states in $G$ (the definition of asymptotic
equivalence can be found in section \ref{sec asympequ}).

\item[Definition: ]A Minimal Reversible Entanglement Generating Set
(MREGS) is a REGS with smallest possible cardinality.
\end{description}

In section \ref{sec asympequ} above we have reviewed the results of
\cite{BennettBPS96}, which can be conveniently summarised using the
concept of an MREGS for bipartite states. Their fundamental result was
to prove that any single bipartite pure entangled state is an MREGS
for two-party entanglement. The most natural such MREGS for bipartite
pure states consists of a single EPR pair, for example in state (\ref{psiminushere}).

The idea of an MREGS is not useful only for bipartite states; it
provides a solid starting point for addressing important questions
regarding multipartite entanglement. Among them, we can ask: is there
a finite MREGS for multipartite entanglement manipulations? Which
states must be present in such an MREGS? In chapter \ref{chap tri} I
describe my work on this problem for tripartite states.

A useful measure of entanglement in a bipartite or multipartite
setting is the relative entropy of entanglement, which we will use in
chapter \ref{chap tri}. In the next section we motivate its
definition and review some of its properties.

\section{Quantum relative entropy \label{sec qreintro}}

In this section we will briefly review some properties of a very
useful measure of entanglement called \textit{relative entropy of
entanglement} \cite{VedralPRK97, VedralP98}. It can be applied either
in the finite-copy case, or in the asymptotic case. What is more interesting, however, is that it is
based on a well-founded idea, that of quantifying distinguishability
between entangled and separable states. Before we discuss the quantum
version, let us see how relative entropy arises in classical
information theory as a measure of distinguishability of probability
distributions.

For an accessible introduction to relative entropy in classical
information theory see the book by Cover and Thomas
\cite{CoverT91}. As for quantum relative entropy, its many uses in
quantum information theory have been reviewed recently by Vedral in
\cite{Vedral02}.

\subsection{Classical relative entropy}

In many situations we need to compare two probability distributions
$p_i$ and $q_i$, and quantify how different they are. The relative
entropy $D(p||q)$ was first defined by Kullback and Leibler \cite{KullbackL51},
and quantifies this `distance' between two probability distributions:
\begin{equation}
D(p||q)=\sum_i p_i \log(\frac{p_i}{q_i}),
\end{equation}
with the convention that $0\log\frac{0}{q}=0$ and $p \log
\frac{p}{0}=\infty$.

$D(p||q)$ has many properties: it is always
non-negative; it is zero only when $p=q$; and it is \textit{not} symmetric. This
last point may be surprising, as we have mentioned it can be
interpreted as a `distance', and a properly defined mathematical
distance must have the symmetry property. In order to perceive
intuitively the necessity for this asymmetry, it is enough to consider
the problem of distinguishing between tosses of a fair coin and those
of a coin so loaded it always comes out Tails. If we do not
know which coin we have, the assumption that the coin is loaded can be
dismissed completely after just a few tosses, whereas the assumption
that it is fair (when in fact it is not) always has a finite
probability $2^{-N}$ of being upheld after $N$ tosses. This simple
example shows that $D(p||q)$ must be asymmetric, as it depends on
which probability distribution we expect ($q$), and which is actually
obtained ($p$).

The reason why the relative entropy $D(p||q)$ is unique is because it
quantifies how hard it is to distinguish two probability
distributions, in a very operational sense. It follows from Sanov's
theorem \cite{Sanov57} that the probability of mistaking events
happening according to probabilities $p$ for events that would arise
from probability distribution $q$ decreases exponentially with the
number of trials $N$, times the relative entropy $D(p||q)$:
\begin{equation}
p(\mbox{taking $p$ for $q$})=2^{-N D(p||q)}, \label{pmistpq}
\end{equation}
which is valid for large $N$. In our simple example
we toss a loaded coin ($p_1=0,p_2=1$) and ask the question: is it
possibly a fair one ($q_1=q_2=1/2$)? After $N$ tosses we can be sure it is loaded with
probability roughly $1-p(\mbox{taking $p$ for $q$})=1-2^{-N
D(p||q)}=1-2^{-N}$. The relative entropy $D(p||q)=1$ appears in the
exponent of this small depart from certainty, to quantify the probability of error in distinguishing $p$ and $q$.

If we still keep the coin example, let us see what happens when we reverse the
situation, tossing a fair coin and asking whether it may be a loaded
coin always yielding Tails. In this case, a single Heads result gives us complete certainty
that the coin is not loaded, and thus the exponential approach to
certainty (given by eq. \ref{pmistpq}) is not valid anymore. This is
reflected by the fact that $D(q||p)=\infty$, as can be easily checked.

\subsection{Relative entropy of entanglement \label{sec relentent}}

The relative entropy has a straightforward quantum generalisation. We
define the von Neumann relative entropy between two density matrices
$\sigma$ and $\rho$ by:
\begin{equation}
D(\sigma||\rho)=Tr \sigma(\log\sigma - \log\rho).
\end{equation}
Hiai and Petz \cite{HiaiP91} have shown that $D(\sigma||\rho)$ has the
same statistical interpretation as its classical analogue, indicating
how hard it is to distinguish measurement outcomes on state $\sigma$
from our expectations of outcomes from state $\rho$.

More recently, Vedral and Plenio \cite{VedralPRK97,VedralP98} have suggested the use of
the quantum relative entropy to quantify entanglement. They defined
the relative entropy of entanglement $E(\sigma)$ to be:
\begin{equation}
E(\sigma)=\min_{\rho\in S}D(\sigma||\rho),
\end{equation}
where $S$ is the set of separable states. The motivation for this
definition is, as argued above, its interpretation as a measure of
distinguishability between quantum states. By taking the minimum over the set of
separable states, we are finding the separable state
$\rho^{sep}_{min}$ whose measurement statistics most closely resemble that
of an entangled state $\sigma$, and quantifying the distinguishability
between $\sigma$ and this closest separable state $\rho^{sep}_{min}$.

The relative entropy of entanglement was also proved to satisfy four basic requirements for entanglement measures: it is zero for
separable states; it is invariant under local unitaries; it cannot
be increased by local operations and classical communication; and it reduces to the von Neumann entropy
$S(\sigma)$ for pure states $\sigma$ \cite{VedralP98,Vedral02}.

One property which may be desirable from measures of entanglement is
\textit{additivity}. In order to define additivity of the relative
entropy of entanglement $E$, we need to discuss how it can be applied
to $N$ copies of a given state, or even in the asymptotic limit $N\to\infty$. To discuss this limit, we can define a \textit{regularised}
relative entropy of entanglement $E^{reg}$ by
\begin{equation}
    E^{reg}(\sigma_{AB})=\mbox{lim}_{n \rightarrow\infty}(1/n)
    \min_{\rho_{AB}\in S}D\left(  \sigma_{AB}^{\otimes
    n}||\rho_{AB}\right) \; .
\end{equation}

If $E^{reg}(\rho)$ = $E(\rho)$ we say $E$ is \textit{asymptotically
additive} for state $\rho$; if
$E(\rho_{1}\otimes\rho_{2})=E(\rho_{1})+E(\rho_{2})$
$\forall\rho_{1},\rho_{2}$ then we would say that $E$ is a
\textit{fully additive} measure.

Until very recently the relative entropy of entanglement $E$ was
conjectured to be fully additive, as numerical investigations had
failed to identify sub-additivity. However, Vollbrecht and Werner
found a counter-example to the full additivity conjecture
\cite{VollbrechtW01}  for two subsystems of dimension $d \ge 3$
each. This was a surprising result, bearing on the work I report in chapter \ref{chap tri}.


\chapter{Quantum information applications \label{chap applic}}

In the previous chapters we have reviewed some general properties of
quantum systems, such as contextuality and non-locality. We have also
sketched some different approaches to the problem of entanglement
characterisation. In this chapter we will
review three quantum information applications which draw on these
quantum characteristics for their better-than-classical efficiency.

A great number of quantum information applications have been
developed in the last few years. I will
make no attempt to review the very many ideas about how to use quantum
phenomena for information processing. Instead, I will concentrate on
describing the main features of just three applications: quantum communication
complexity protocols in section \ref{sec ccintro}; quantum random
access codes in section \ref{sec intro qracs}; and quantum cloning in
section \ref{sec qclointro}.

In later chapters I will investigate many
characteristics of these three applications, with an emphasis on
finding out why they are better than their classical counterparts, and
what they can tell us about quantum theory. I will also discuss the
feasibility of experimentally implementing some instances of the first two applications.

Let us start by discussing communication complexity problems, which
illustrate in a clear way the communication requirements for a type of
distributed computation.

\section{Communication complexity \label{sec ccintro}}

The notion of communication complexity was introduced by Yao
\cite{Yao79}, who investigated the following problem involving two
separated parties (Alice and Bob). Alice receives a $n$-bit string $x$
and Bob another $n$-bit string $y$. They are then allowed to perform
local computations and to exchange some communication, in such a way
that one of them (say Bob) can announce the value of a function
$f\left(x,y\right)$. The resource we will try to minimise here is the
amount of \textit{communication} between them, hence the name
`communication complexity'. Note that here we are
not concerned about the number of computational steps, or the size of
the computer memory used. Communication complexity tries to quantify
the amount of communication required for such distributed
computations.

Of course they can always succeed by having Alice send her whole
$n$-bit string to Bob, who then computes the function, but the idea
here is to find clever ways of calculating $f$ with less than $n$ bits
of communication.

This abstract problem is relevant in many contexts: in Very Large
Scale Integrated (VLSI) circuit design, for example, one wants to
minimise energy use by decreasing the amount of electric signals
required between the different components during a distributed
computation. The problem is also relevant for the study of data
structures, and in the optimisation of computer networks. For a survey
of the field, see the book by Kushilevitz and Nisan
\cite{KushilevitzN97}.

This simple two-party scenario can be generalised in many different
ways. We can think of more parties involved in the distributed
computation; consider deterministic or probabilistic protocols; or
allow for quantum phenomena to be used. This last possibility
generalises communication complexity to the quantum realm, where we
can use quantum communication or entanglement.

\subsection{Quantum communication complexity \label{sec qccintro}}

Quantum communication can be more efficient than classical
communication for some tasks. Dense coding \cite{BennettW92}, for
example, is a process in which one
previously shared pair of maximally entangled states allows one party
to communicate two bits to a second one by sending only a single qubit. This `two bits per qubit' figure is, however,
misleading, as an entangled qubit had to be shared between the parties
beforehand. More generally, Holevo has shown \cite{Holevo73} that the
mutual information between two unentangled parties cannot increase by more than
one bit per qubit of quantum communication between them. This is still
true when the two parties can exchange qubits both ways
\cite{Nielsen98, ClevevDNT99}.

So even if we allow for previously shared entanglement, quantum
communication cannot be more than twice more efficient than classical
communication at sending information between two parties. It may come
as a surprise, then, that for distributed computations quantum
communication and entanglement can result in even exponential gains
over optimal classical protocols. Quantum communication complexity
tries to quantify the communication reduction possible by using
quantum effects during a distributed computation.

At least three quantum generalisations of communication complexity have been
proposed; for a survey see Brassard's article \cite{Brassard01}. The first one is the
\textit{qubit-communication model} introduced by Yao \cite{Yao93} and
Kremer \cite{Kremer95}. In this model the communication is performed
by exchanging qubits between the parties, instead of classical
bits. In chapters \ref{chap qcc} and \ref{chap feasible} I will
analyse some simple communication complexity tasks which can be better performed with
quantum communication than with classical communication. In chapter
\ref{chap feasible} we will see that the quantum advantage can be so
pronounced that there exist feasible schemes to experimentally
demonstrate it even with very low quantum detection efficiencies.

In the second model the communication is still performed with
classical bits, but the parties are allowed to manipulate an unlimited
supply of quantum entangled states as part of their protocols. This is
the \textit{entanglement-based model} introduced by Cleve and Buhrman
\cite{CleveB97}, and further investigated by various authors
\cite{BuhrmanCvD97,BuhrmanvDHT99,Raz99,HardyvD99}. In chapter
\ref{chap qcc} and \ref{chap feasible} I will analyse some simple entanglement-based quantum
protocols, and show how they relate to quantum non-locality and
contextuality tests. We will also see that such protocols naturally
give us some bounds on the detection efficiencies and background rates
necessary for experimentally establishing these quantum
characteristics.

The third model involves access to previously shared entanglement in
addition to qubit communication, and is the least explored of the
three quantum models.

Quantum communication complexity protocols feature prominently in this
thesis, appearing in chapters \ref{chap qcc} and \ref{chap
feasible}. They also served as an inspiration for the work I report in
chapter \ref{chap qubit}.

\section{Quantum random access codes (QRAC's)\label{sec intro qracs}}

There are many ways in which quantum
communication can be superior to classical communication. Imagine a
situation in which Alice encodes $m$ classical bits into $n$ bits
($m>n$), which she sends to Bob, who will need to know that value of a \textit{single} bit (out of the $m$ possible ones) with a
probability of at least $p$. We may represent such an
encoding/decoding scheme by the notation:
$m\overset{p}{\rightarrow}n$.

Prior to sending the $n$-bit message,
however, Alice does not know which of the $m$ bits Bob will need to
read out. To maximise the least probability of success $p$, Alice and
Bob need to agree on the use of a particular, efficient $m\overset{p}{\rightarrow}n$ encoding.

We can consider the
quantum generalisation of this situation, in which Alice can send Bob
$n$ qubits of communication, instead of $n$ bits. The idea behind
these so-called \textit{quantum random access codes} is very old by
quantum information standards; it already appeared in a paper written
circa 1970 and published in 1983 by Stephen Wiesner
\cite{Wiesner83}.

These codes were re-discovered in
\cite{AmbainisNT-SV99}, where the explicit comparison with classical
codes was made. One way to think about the possibility of a quantum
advantage is to think of the (generally) non-commuting measurements we
can use in the decoding process. Because of the non-commutativity,
each particular measurement disturbs the
system, destroying some of the information that could have been
revealed by another choice of measurement. Thus, we see that the
constraints on what can be achieved by a single measurement on a
quantum system of $n$ qubits are less severe than in the classical
case, possibly allowing for a higher quantum encoding efficiency.

In chapter \ref{chap qcc} we will investigate the role played by
quantum contextuality and non-locality in these quantum random access
codes. We will also see how the invariant information of Brukner and
Zeilinger helps us to bounds the performance of more general QRAC's.

\subsection{A simple $2 \to 1$ QRAC \label{sec 21qrac}}

Let us illustrate the concept with the simplest such scheme, which
encodes Alice's two-bit string $b_0b_1$ into a single qubit's states
(a $2{\rightarrow}1$ quantum random access code). Before we
discuss the quantum code, let us find out what the optimal
probability of success is for \textit{classical} $2\overset{p}{\rightarrow}1$ codes.

\subsubsection{Optimal classical protocol}

Alice and Bob need to decide on a protocol defining which bit-valued
message is to be sent by Alice, for each of the four possible values of
her two-bit string $x$. Each of these $2^{4}=16$ different
deterministic protocols $prot_i$ has a probability of success
$p_c(prot_i)$ which can be evaluated in a straightforward way. The
optimal deterministic classical protocols can then be shown to have a probability of success $p_c=3/4$.

Now we need to address the possibility of probabilistic protocols. Any
probabilistic protocol can be represented as a convex combination of
the 16 deterministic protocols: $\sum_{i=1}^{16}(q_i prot_i)$, with
weights $q_i$ satisfying $\sum_{i=1}^{16} q_i=1$. The corresponding
probability of success $p_c$ for any such probabilistic protocol will
be given by the weighted sum of the probabilities of success of the
individual deterministic protocols: $p_c=\sum_{i=1}^{16}\left[q_i
p_c(prot_i)\right]$. This makes it clear that the optimal
probabilistic protocols can at best be as efficient as the optimal
deterministic protocol, which we have already shown can have a
probability of success of at most $p_c=3/4$.

This is a much simpler proof than the one in the original
paper by Ambainis \textit{et al.} \cite{AmbainisNT-SV99}. Note that there, the definition of
the probability of success $p$ is slightly different: they define $p$
as the least probability of success, whereas I define $p$ as the
average probability of success for a uniformly random distribution of Alice's bit-strings.

\subsubsection{Quantum protocol}

\begin{figure}
\begin{center}
{\includegraphics[scale=0.45]{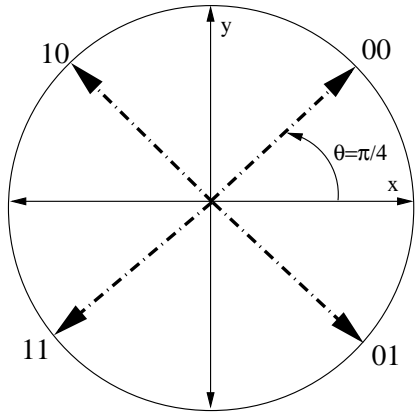}}
\caption[Qubit states used in a $2 \to 1$ QRAC.]{\label{fig applic1}Qubit states used in the $2 \to 1$ QRAC. Alice prepares the
qubit in one of the four states pictured above, which lie on the
equator of the Bloch sphere. Bob's decoding procedure will consist of
projecting either on the $x$-axis or the $y$-axis, depending on which
of Alice's bits he wants to read. }
\end{center}
\end{figure}

Ambainis \textit{et al.} \cite{AmbainisNT-SV99} also showed that there is a quantum
$2\overset{p}{\rightarrow}1$ encoding with
$p=\cos^{2}(\pi/8)\simeq0.85$. It works as follows: depending on her
two-bit string $b_0b_1$, Alice prepares one of the four states
$\left|  \phi_{b_0 b_1}\right\rangle $. These states are chosen to
be on the equator of the Bloch sphere, separated by equal angles of
$\pi/2$ radians (see figure \ref{fig applic1}).

Using the usual Bloch sphere parametrisation
\[
\left|\psi(\theta,\phi)\right\rangle=\cos(\tfrac{\theta}{2})\left|0\right\rangle+\exp(i\phi)\sin(\tfrac{\theta}{2})\left|1\right\rangle
,
\]
we can represent the encoding states depicted in figure \ref{fig applic1} as:
\begin{eqnarray}
\left|  \phi_{00}\right\rangle &=&
\left|\psi(\pi/2,\pi/4)\right\rangle,\\
\left|  \phi_{01}\right\rangle &=&
\left|\psi(\pi/2,7\pi/4)\right\rangle,\\
\left|  \phi_{10}\right\rangle &=&
\left|\psi(\pi/2,3\pi/4)\right\rangle,\\
\left|  \phi_{11}\right\rangle &=&
\left|\psi(\pi/2,5\pi/4)\right\rangle.
\end{eqnarray}

Bob's decoding
measurements will depend on which bit he wants to obtain. To decode
$b_0$, he projects the qubit along the $x$-axis, and to read $b_1$ he
projects it along the $y$-axis. The then associates a result `0' if
the measurement outcome was along the positive direction of the axis
measured, and `1' otherwise. It is easy to check that the
probability of successfully retrieving the correct bit value is the same in all cases:
\[
p_q=\cos^{2}(\pi/8)\simeq0.85,
\]
which is higher than the optimal probability of success $p_c=0.75$ of
the classical random access code using a bit of communication.

In chapter \ref{chap qcc} I will analyse this simple quantum random
access code and show that its higher-than-classical efficiency arises
from quantum contextuality and non-locality. Using Brukner and
Zeilinger's invariant information (see section \ref{sec invinfo}), in
section \ref{sec invinfoqrac} I will derive some bounds on the
probability of success of more general codes.

We will also see that the simplest quantum random access codes involve
the same quantum operations as other quantum information applications;
this does not seem to have been noticed before, and helps in
understanding the fundamental basis for the quantum advantage in a
number of similar applications.

\section{Quantum cloning \label{sec qclointro}}

One of the first results stressing the peculiarities of quantum information was the no-cloning theorem obtained in 1982 by Wootters
and Zurek \cite{WoottersZ82} and independently by Dieks
\cite{Dieks82}. In simple terms, they showed that unlike classical
information, an arbitrary quantum state cannot in general be perfectly
copied. In section \ref{sec noclone} we will review a simple version of this
important theorem, and briefly discuss its relevance.

From a computational point of view, the no-cloning theorem exemplifies
one of the fundamental differences between quantum and classical
information processing. A classical memory register can be read out
and operated upon at any point during a computation; in particular, it is always
possible, and indeed necessary in many cases, to perform the COPY
operation on a subset of the registers. A quantum register, however,
can encode a large amount of classical information (see chapter
\ref{chap qubit}), which
cannot in general be read out and copied during the unitary evolution
that characterises quantum algorithms.

It may be impossible to make perfect copies of quantum states, but
\textit{approximate} copies are not ruled out by the no-cloning
theorem. This idea led Bu\v{z}ek and Hillery to develop
the so-called \textit{universal quantum cloning machines} (UQCM's for
short) \cite{BuzekH96}, which are
unitary quantum operations resulting in good-quality (but not perfect)
copies of a given quantum state. Since then many authors have further
investigated the limitations that quantum mechanics imposes on unitary
distribution of quantum information. I will describe their main
findings in section \ref{sec UQCM}.

Another strategy for quantum cloning is to use a procedure combining
unitary evolution and measurement to obtain perfect clones with some probability $p<1$. This approach was developed by Duan and Guo
\cite{DuanG98, DuanG99} in 1998. In section \ref{sec probclo} I briefly describe
these \textit{probabilistic cloning machines}.

These two cloning approaches have something in common: they are
necessarily imperfect. The limitations of UQCM's show up in the noise
present in the imperfect clones created. In the case of probabilistic
cloning, the problem is that the process only works with probability $p<1$.

In chapter \ref{chap cloning} I will present my work on the use of
cloning for quantum computation. We will see that despite their
imperfections, during quantum computations cloning may offer an
advantage over approaches that do not resort to quantum information
distribution. For example, we will see that in some cases the noise
introduced by UQCM's represents a lesser problem than the irreversible
loss of quantum information represented by alternatives involving
extraction of classical information through measurement.

Despite its importance from a theoretical point of view, until
recently very few, if any, applications of cloning procedures had been
found. Distributing information during quantum computations is an
application that stresses the fundamental differences between quantum
and classical information.

\subsection{The no-cloning theorem \label{sec noclone}}

To get the basic idea behind the no-cloning theorem
\cite{WoottersZ82,Dieks82}, suppose we are given a qubit in an unknown
pure quantum state $\left|
\psi\right\rangle$ and we are asked to make a perfect copy of it. We
start with our original $\left| \psi\right\rangle$, a qubit in a known
pure state $\left| \mbox{blank}\right\rangle$, and an arbitrary
ancillary system (the Cloning Machine) also in a known state, $\left|
\mbox{M}_0\right\rangle$. Our objective is then to
find a unitary transformation $U$ such that
\begin{equation}
U(\left|\psi\right\rangle \otimes \left|
\mbox{blank}\right\rangle \otimes \left|
\mbox{M}_0\right\rangle) = \left|\psi\right\rangle \otimes \left|
\psi\right\rangle \otimes \left| \mbox{M}_{\psi}\right\rangle.
\end{equation}
The role of the ancilla is to give us
more freedom in the unitaries which we can perform; we allow it to end
up in a state $\left| \mbox{M}_{\psi}\right\rangle$ which may depend
on the original state $\left|\psi\right\rangle$. If we want this unitary to work for any state, in particular it should
work for two particular states $\left| \psi\right\rangle$ and $\left|
\phi\right\rangle$. In other words, we require that
\begin{eqnarray}
U(\left|\psi\right\rangle \otimes \left|
\mbox{blank}\right\rangle \otimes \left|
\mbox{M}_0\right\rangle)& =& \left|\psi\right\rangle \otimes \left|
\psi\right\rangle\otimes \left|
\mbox{M}_{\psi}\right\rangle, \\
U(\left|\phi\right\rangle \otimes \left|
\mbox{blank}\right\rangle\otimes \left|
\mbox{M}_0\right\rangle)& =& \left|\phi\right\rangle \otimes \left|
\phi\right\rangle\otimes \left|
\mbox{M}_{\phi}\right\rangle.
\end{eqnarray}

Unitary transformations preserve inner products; if we take the
inner product of these states before and after the unitary $U$ we find
the necessary condition:
\begin{equation}
\left|\left\langle\psi|\phi\right\rangle \right|=
\left| \left\langle\psi|\phi\right\rangle^{2}\left\langle\mbox{M}_{\psi}|\mbox{M}_{\phi}\right\rangle\right|
\le \left|\left\langle\psi|\phi\right\rangle^{2}\right|.
\end{equation}
This can be true only when $\left|\psi\right\rangle$ and
$\left|\phi\right\rangle$ are either equal or orthogonal to each
other. This shows the impossibility of a universal, unitary cloning
process that is perfect. The simple proof above also highlights the
fact that orthogonal states can be cloned -- which is not surprising,
given our experience of the classical world and classical computation.

There are other ways of showing that perfect cloning is
impossible. One particularly nice proof consists of showing that a
perfect cloning machine would enable faster-than-light
signalling. To see why, let us start by providing two very distant
partners, Alice and Bob, with a pair of qubits in a maximally
entangled state:

\begin{equation}
\left|\psi^{+}\right\rangle=\frac{1}{\sqrt{2}}(\left|\mbox{00}\right\rangle
+\left|\mbox{11}\right\rangle).
\end{equation}

Alice would like to send a superluminal message to Bob. If she wants
to send a zero, she projects her qubit along a certain direction
$\vec{m_{0}}$ in the Bloch sphere, otherwise she projects along a
second direction $\vec{m_{1}}$. Alice's measurement effectively projects
Bob's qubit along one of two possible directions, $\vec{m_{0}}$ or
$\vec{m_{1}}$. If Bob has a perfect cloning machine, he can make a
large number of copies of his state and do measurements on them to
distinguish these two possibilities with high accuracy. This would
enable him to learn which measurement Alice performed, resulting in a
faster-than-light communication of one bit from Alice to Bob.

If the cloning procedure is just a little imperfect, still it is
intuitively reasonable to expect that for high enough `copy
quality', one would still be able to signal faster than light. This
simple idea led Gisin \cite{Gisin98} to calculate the maximum cloning
fidelity which still avoids superluminal
communication. Interestingly, he found that the limits imposed by the
no-signalling condition are the same as those imposed by the
mathematical structure of quantum mechanics, which we will investigate
in the next section.

Next I will present the main features of the known approximate quantum cloning procedures.

\subsection{Quantum cloning machines \label{sec qcloning}}

We have just seen that perfect copying of an unknown quantum state is
impossible. In this section we will have a look at some of the known
cloning procedures, developed to allow optimal distribution of quantum
information.

There are two main types of cloning procedures. The first approach
consists of using ancillary quantum systems and a global unitary operation to obtain multiple imperfect clones of a
given, unknown quantum state. These \textit{universal quantum cloning
machines} (UQCM's) were first proposed by Bu\v{z}ek and
Hillery \cite{BuzekH96}, and are discussed in section \ref{sec UQCM}.

A second approach combines a unitary evolution step with a measurement
on a subsystem, which may sometimes collapse the rest of the system into
perfect copies of the original state. This does not violate the
no-cloning theorem, as it only works with
probability $p<1$. Moreover, we will see that for it to work we need
to be sure that the original qubit is in one out of a known set of linearly
independent states; thus, it is not universal, unlike the UQCM's. These \textit{probabilistic cloning machines} were first analysed by Duan and
Guo \cite{DuanG98}, and are described in section \ref{sec probclo}.


\subsection{Universal quantum cloning machines \label{sec UQCM}}

In this section we will have a look at the universal quantum cloning
machines (UQCM's), introduced by Bu\v{z}ek and Hillery in 1996
\cite{BuzekH96} and developed by other authors \cite{GisinM97,
Werner98, KeylW99, BrussM99, BrussEM98, BuzekHB98, Cerf00, BuzekBHB97}.

For simplicity, let us start by discussing the original UQCM that
makes two clones of a single qubit state
$\left|\psi_{in}\right\rangle$ \cite{BuzekH96}. The UQCM consists of a
two-step procedure. Firstly, a $4$-dimensional
ancilla quantum system in a known state $\left|\mbox{blank}\right\rangle$ is
coupled to the state $\left|\psi_{in}\right\rangle$ to be cloned. Then, a
carefully chosen unitary $U$ is performed over the combined
$8$-dimensional system, resulting in a final output state
$\left|\psi_{out}\right\rangle$ containing the two clones:
\begin{equation}
U(\left|\psi_{in}\right\rangle \otimes \left|\mbox{blank}\right\rangle) =\left|\psi_{out}\right\rangle. 
\end{equation}

The final state $\left|\psi_{out}\right\rangle$ `contains' the clones
in the following sense: each of the first 2 qubit's reduced density
matrices is in the state
\begin{equation}
\rho_{out}=\eta\left|\psi_{in}\right\rangle\left\langle\psi_{in}\right|
 +(1-\eta)\frac{1}{2}\mbox{$1 \hspace{-1.0mm}  {\bf l}$},\label{rhoout}
\end{equation}
where $\eta$ is a yet undefined parameter which indicates the
`quality' of the clones, i.e. how closely they resemble the original
state. The third qubit is
not a clone; it is present to make possible the unitary that creates
the two clones.

If it were possible to find
$U$ such that $\eta=1$ in eq. (\ref{rhoout}), then we would have a perfect cloning
machine. An optimal UQCM can be found by obtaining a unitary $U$
that maximises $\eta$ and the fidelity of the clones with respect to the original state:
\begin{equation}
F=\left\langle\psi_{in}|\rho_{out}|\psi_{in}\right\rangle = \eta+(1-\eta)\frac{1}{2}=\frac{1}{2}+\frac{\eta}{2}.
\end{equation}
The fidelity, as defined above, gives the probability that the clones
will pass a test designed to check whether they are equal to
$\left|\psi_{in}\right\rangle$.

The development of the theory of cloning machines started with
Bu\v{z}ek and Hillery \cite{BuzekH96}, who showed that there exists a UQCM that makes two clones of a single qubit with
$F=\frac{5}{6}$. Gisin and Massar \cite{GisinM97} and Bru{\ss}  \textit{et al.} \cite{BrussDEFMS98} proved that this fidelity could
not be surpassed, and therefore that this cloning machine was optimal. In addition, Gisin and Massar
obtained fidelities for a family of $N\to M$ UQCM's for a
qubit, that is, cloning machines that have as an input $N$ copies of
the original qubit, and which output $M$ clones. Bru\ss, Macchiavello and
Ekert \cite{BrussEM98} proved that these UQCM's were actually
optimal. The optimal clone fidelity for $N\rightarrow M$ UQCM's for qubits
was found to be:

\begin{equation}
F=\frac{NM+N+M}{M(N+2)} \label{qubitclonef}.
\end{equation}

Cloning machines for $d$-dimensional systems ($d>2$) were studied by Werner
and Keyl
\cite{Werner98, KeylW99}. They showed that the optimal clone fidelity
for a $N \to M$ machine for $d$-dimensional systems is:

\begin{equation}
F^{N\rightarrow M}_{clone}=\frac{M-N+N(M+d)}{M(N+d)}.\label{fcloopt}
\end{equation}

This last equation summarises and extends the previous results on
optimal clone fidelities.

Other work on cloning includes various analyses of quantum networks
for implementing cloning machines \cite{CheflesB99, BuzekBHB97, BuzekH99} and some experimental
realizations of cloning \cite{SimonWZ00, DeMartiniMB00, HuangLLZJG01,CumminsJFSMPJ02,LamasLinaresSHB02}.

Gisin and Massar \cite{GisinM97} make an interesting connection
between cloning machines and state estimation, which will be important
for the application to quantum computing we will develop
later.

State estimation is a very important problem in quantum information
theory, and concerns the strategies that must be used to obtain
maximum information about given properties of quantum states, through
measurements on them. The simplest quantum state estimation scenario
involves just a single qubit in an unknown state
$\left|\psi_{in}\right\rangle$. Our task is then to make some
measurement on the qubit, in order to obtain classical information
about its state. This classical information must enable us to prepare
a guess state $\left|\psi_{guess}\right\rangle$, which is to be `as
close as possible' to the original state
$\left|\psi_{in}\right\rangle$. The distance measure we will use is
again the fidelity of the guess with respect to the original:

\begin{equation}
F=|\left\langle\psi_{in}|\psi_{guess}\right\rangle|^{2}.
\end{equation}

Massar and Popescu \cite{MassarP95} found the optimal fidelity
obtainable when we do state estimation in the above sense, using $N$
qubits prepared identically in the same state
$\left|\psi_{guess}\right\rangle$. They obtained

\begin{equation}
F=\frac{N+1}{N+2}.\label{fMP95}
\end{equation}

They noticed that eq. (\ref{fMP95}) is the limit, as $M\rightarrow
\infty$, of the expression for the fidelity of the optimal
$N\rightarrow M$ cloning machine, eq. (\ref{qubitclonef}). As they observed (and was further discussed in
\cite{BrussEM98}), the two processes are equivalent. To see this, it is
enough to note that an infinite
number of clones with fidelity $F$ provides us with means to
accurately describe the state of each clone (by characterising the
states through measurements on
the infinite ensemble of clones). This, in itself, provides a
recipe to create a mixed-state approximation with fidelity $F$ for the
state estimation problem.

It is important to note that the state estimation fidelity for a qubit
[eq. (\ref{fMP95})] is \textit{always} smaller than the cloning fidelity
(\ref{qubitclonef}). This reflects a fundamental characteristic of
quantum information: the measurement process only reveals part of the
information content of a single quantum state. By cloning a state, instead of
measuring it, we choose to preserve as much as possible the quantum
information encoded in it, which can be useful in subsequent quantum
information processing tasks. This is the main idea behind the
use of cloning in quantum computation tasks, an issue I will discuss
in chapter \ref{chap cloning}. 

\subsection{A curious fidelity balance result \label{sec curfbr}}

Before we resume our review of quantum cloning processes, I
would like to take a short detour and discuss a curious fidelity
balance result I found for general, universal quantum cloning
machines. Let us start by having a closer look at the simplest UQCM, the $1 \to 2$ machine
for a single qubit. The original qubit state can be said to contain
maximal quantum information, and by cloning it we will be distributing
this quantum information unitarily among the two copies. Is there a
way to characterise quantitatively this quantum information distribution?

First, we need to distinguish between the merely classical and the genuinely quantum information content present in the qubit
state. There is no unique way of doing this; here I will proceed in a
intuitively reasonable way, which reveals a curious fidelity balance
result for optimal UQCM's.

There is only a limited amount of classical information that can be gathered
from a single qubit state through measurement. One way to quantify
this classical information content is by using the fidelity of the
state an optimal measurement procedure enables us to create, with respect to the original state. As
we have seen in the previous section, this fidelity is simply the
optimal estimation fidelity, and for a single qubit is equal to $2/3$
[from eq. (\ref{fMP95})]. Let us use this optimal estimation fidelity
as our measure of classical information present in a given pure state,
before cloning occurs:

\[
I^{before}_{c}=2/3.
\]

The original state, however, also contains quantum information which
is not available through the measurement process. A measure of this
quantum information content is given simply by the original fidelity
($=1$) minus this classical estimation fidelity ($=2/3$). We thus
define the quantum information content of the original state as the
difference between these two fidelities:
\[
I^{before}_{q}=1-2/3=1/3.
\]

After the cloning process, we obtain two imperfect clones, each with
fidelity $5/6$ [see eq.(\ref{qubitclonef})]. The classical information
content is still the same; one way to see that is to note that the
cloning process is unitary, and therefore can be reversed, resulting
in the original state which we can then optimally estimate. Therefore:
\[
I^{after}_{c}=I^{before}_{c}=2/3.
\]

The equality of this classical information content is just reflecting
the fact that a unitary on the original state (even with ancillas)
preserves the original classical information, as available by optimal state
estimation.

What is the quantum information content now? Each clone still has a
higher fidelity than is possible to attain by classical information
extraction through measurement from the original state. The difference
between this fidelity and the optimal state estimation fidelity will
be the intrinsically quantum information content of each clone. Since
we have two clones, we obtain:
\[
I^{after}_{q}=2(5/6-2/3)=1/3.
\]
We see that the `quantum information content' (as I defined above) is
conserved in this simple $1 \to 2$ qubit UQCM.

This might be seen as a numerical coincidence, as I have not presented
an argument for why this should hold in general. It can be checked,
however, that this conservation between classical and quantum
fidelities also holds in the \textit{general case} of
symmetric $N \to M$ cloning machines in any dimension, as can be verified using eq. (\ref{fcloopt}). Intuitively, what we see is that the original quantum
information content (which cannot be obtained through measurement) is
conserved in the unitary evolution characterising the cloning process,
being equally distributed among the clones.

This is a curious result. The way I defined the quantum information
content was somewhat arbitrary, but still it is intriguing that it should
be conserved during cloning, and decreased if we do any non-unitary
(i.e. measurement) operation on the system. Unfortunately, the $I_{q}$
quantity I defined above is \textit{not} conserved in
asymmetric cloning procedures. Still, this type of approach may prove
fruitful in quantifying the genuinely quantum information content of
states undergoing different types of operations.

\subsection{Probabilistic cloning \label{sec probclo}}

In section \ref{sec noclone} we saw that there exists no unitary operation which
deterministically obtains two perfect copies of an unknown state (the
no-cloning theorem). This, however, does not preclude the existence of
a probabilistic non-unitary process which succeeds only part of the
time (that is, with probability $p<1$). These \textit{probabilistic
cloning machines} were discovered by Duan and Guo in 1998
\cite{DuanG98}. In this section I describe their main results.

The basic idea can be understood with the following simple
example. Suppose we are given qubit $A$, which is
known to be in one out of two
non-orthogonal states: either $\left|\psi_{0}\right\rangle$ or
$\left|\psi_{1}\right\rangle$. As with the simplest UQCM, we start by
adding an ancilla system, consisting of two qubits (which we denote
by $B$ and $C$). Qubit
$B$ starts in a blank state; if the cloning is successful, at the end it will
be in the same state as the undisturbed qubit $A$. Qubit $C$ will be used as a probe
to be measured at the end. Its role is two-fold: first, it enables us
to collapse the state of the other two qubits into two perfect copies
of the original state; besides, the measurement outcome will inform us
whether the cloning process was successful or not.

For this scheme to work, we must find a unitary $U$ over the three
qubits, which has the following action:

\begin{eqnarray}
U ( \left| \psi_0\right\rangle_A \left| \mbox{blank} \right\rangle_B
\left|p_0\right\rangle_C ) =a_{00}\left| \psi_0\right\rangle_A
\left| \psi_0\right\rangle_B \left| p_0\right\rangle_C +a_{01}\left|
\phi_0\right\rangle_{AB} \left| p_1\right\rangle_C  , \\
U ( \left| \psi_1\right\rangle_A\left| \mbox{blank}
\right\rangle_{B} \left|p_0\right\rangle_C ) =a_{10}\left|
\psi_1\right\rangle_A \left| \psi_1\right\rangle_B \left|
p_0\right\rangle_C +a_{11}\left| \phi_1\right\rangle_{AB} \left| p_1\right\rangle_C .
\end{eqnarray}

If we can perform the above unitary and measure qubit $C$, whenever we
obtain the outcome $\left| p_0\right\rangle_C$, we will have succeeded
in collapsing qubits $A$ and $B$ into two copies of the original state
of qubit $A$. The question is, then, to build the unitary $U$ and to
obtain the optimal probability of success, which will depend on the
coefficients $a_{00}$ and $a_{10}$.

In ref. \cite{DuanG98} Duan and Guo showed that it is possible to build such
a unitary to clone a set of non-orthogonal states if and only if the
states are linearly independent. Their proof was constructive, and
enables us to find the optimal cloning probability. The proofs are
lengthy and can be found in \cite{DuanG98}; here I will just reproduce
the results which will be useful for the discussion of the uses of
cloning in quantum computing.

\begin{description}
\item[Theorem 4.1 (Duan and Guo \cite{DuanG98})] A set $S$ of pure quantum states can be
probabilistically cloned if and only if the states in $S$ are linearly
independent.
\end{description}

This theorem shows that, unlike the UQCM's of section \ref{sec UQCM}, the
probabilistic cloning machines are state-dependent and only work for
states from a previously defined set of linearly-independent states.

Let the above-mentioned set $S$ consist of the $n$ linearly
independent pure states $S =
\left\{ \psi_1 , \psi_2, \cdots , \psi_n \right\}$. Let us define the 
$n\times n$ matrices
\begin{eqnarray}
 X^{\left( 1\right) }_{i,j}= \left\langle \psi _i|\psi _j\right\rangle
 ,\\
X^{\left( 2\right) }_{i,j}= \left\langle \psi _i|\psi _j\right\rangle
^2.
\end{eqnarray}

Now we also define a diagonal efficiency matrix $\Gamma $ as 
\begin{equation}
\Gamma=diag\left( \gamma_1,\gamma_2,... ,\gamma_n\right).
\end{equation}

The coefficients $\gamma_1, \gamma_2,..., \gamma_n$ appearing in
the definition above are the efficiencies (i.e. probability of
success) associated with cloning each of the states
$\psi_1,\psi_2,...,\psi_n$. From the definition for the matrix $\Gamma$
it follows that $\sqrt{%
\Gamma }=\sqrt{\Gamma }^{+}=diag\left( \sqrt{\gamma_1},\sqrt{\gamma_2}%
,\cdots ,\sqrt{\gamma_n}\right) $.

All these definitions are necessary to state the theorem which gives a
necessary and sufficient condition for probabilistically cloning a set
$S$ of linearly independent pure states:

\begin{description}
\item[Theorem 4.2 (Duan and Guo \cite{DuanG98})] The linearly-independent states in $S$ can be probabilistically cloned with
a diagonal efficiency matrix $\Gamma$ if and only if the matrix
$T=X^{\left( 1\right) }-\sqrt{\Gamma }X_P^{\left( 2\right) }
\sqrt{\Gamma }^{+}$ is positive semidefinite.
\end{description}

The semi-positivity condition on matrix $T$ results in a series
of inequalities involving the efficiencies $\gamma_i$. Solving these
inequalities gives us the achievable cloning probabilities.

Suppose we take one of the linearly-independent states in $S$ and
manage to probabilistically clone it into an arbitrarily large number
of copies. By performing optimal state estimation on them, we can
perfectly identify the original state. This connection between cloning
and state estimation enabled Duan and Guo to establish the following
theorem:

\begin{description}
\item[Theorem 4.3 (Duan and Guo \cite{DuanG98})] The
linearly-independent states in $S$ can be unambiguously identified with
probabilities given by the matrix $\Gamma$ if and only if the matrix
$X^{\left( 1\right) }-\Gamma$ is positive semidefinite.
\end{description}

Again, the semi-positivity of matrix $X^{\left( 1\right) }-\Gamma$
results in inequalities expressing the achievable state identification probabilities.


\chapter{Tripartite entanglement and relative entropy \label{chap
tri}}

In chapter \ref{chap ent} we have reviewed some approaches to the
problem of 
quantifying pure-state entanglement. We have
also seen that the problem of quantification of entanglement for three
or more parties is much harder than the bipartite case, in part because many of the theorems
that apply to the bipartite case (such as the existence of a Schmidt
decomposition \cite[section 2.5]{NielsenC00}) do not carry over to many
parties.

In this chapter we will investigate pure state tripartite
entanglement in the asymptotic limit of manipulations of a large
number of copies of states. More specifically, we will try to characterise a Minimal
Entanglement Generating Set (MREGS) for tripartite entanglement. Basically,
we would like to find a set of tripartite states that play the role of
maximally entangled EPR pairs in the bipartite case,
i.e. from which any pure tripartite state can be obtained through
reversible local operations with classical communication (LOCC) in the asymptotic regime.

We start in section \ref{sec mregsm} with an introduction to this
problem, followed in section
\ref{sec trirel} by a review of some previous results by other authors. We will see that there exist relations between
entanglement measures that need to be obeyed by tripartite states
which can be reversibly obtainable from EPR pairs and GHZ
states. These conditions establish relations between the problem of
additivity for the relative entropy of entanglement, and the problem
of finding an MREGS for tripartite entanglement. The first problem is
important because it relates to an unsuspected mathematical property
of one of the few entanglement measures that can be readily
generalised for multi-particle states. The second, apparently unrelated
problem will be shown to lead to the identification of tripartite
states with interesting entanglement structures, which may be useful
in new quantum information processing protocols.

In section \ref{sec calent} I present the results of numerical
investigations I did, identifying families of states which either
violate additivity of relative entropy, or which fail to be asymptotically equivalent to collections of GHZ and EPR states. When this work was finished,
there were no examples of states with either of these interesting
properties. Subsequently, other authors managed to prove sub-additivity
of the relative entropy for some states \cite{VollbrechtW01}, and also
that GHZ and EPR pairs do not form an MREGS for tripartite states
\cite{AcinVC02}. This last paper in particular succeeded by further
investigating a family of states we pointed out in the published
version of this chapter \cite{GalvaoPV00}, which I review in section \ref{sec susp}. This shows my
approach here seems to have been fruitful, as it drew attention to
the relation between these two problems and spurned further research
into the entanglement structure of some families of tripartite states.

By numerically calculating the relative entropy, in section \ref{sec
wvsghz} I show that there is a tripartite state which is more
entangled than the three-qubit state usually referred to as `maximally entangled',
the GHZ state. In section \ref{sec conctri} I close
the chapter with a summary of the main results.

\section{MREGS for multipartite entanglement \label{sec mregsm}}

In this section we address the problem of asymptotically
reversible entanglement transformations for multipartite entanglement. In section \ref{sec asymplim} we reviewed the theoretical
framework proposed in \cite{BennettBPS96}, defining the notion of a
Minimal Reversible Entanglement Generating Set (MREGS) for creating multipartite pure entangled states. In the
bipartite case we have seen that any pure entangled state can be
reversibly created with LOCC only on copies of a maximally entangled pair of
qubits, in the asymptotic limit of manipulations of many copies. This
means a single EPR pair of qubits qualifies as an MREGS for bipartite
pure-state entanglement.

The question we will investigate here is whether it is possible to
define an MREGS for multipartite entanglement generation. In
particular, we will address the problem of \textit{tripartite} entanglement.

Before I describe my work on this subject, it is convenient to review
some results by other researchers. A key motivation for my work was
the proof that generalised GHZ states alone do not constitute an MREGS
for four-particle entanglement. This was obtained by Wu and Zhang
\cite{WuZ01}, and can be more precisely stated as follows.

Suppose that, in analogy with the bipartite case, we can create any
four-partite entangled state using reversible LOCC on a large number
of states of only a few types, each of which is represented in an
MREGS for 4-partite states, which we will call $G_4$. From our
experience with the bipartite case, it is natural to conjecture that
$G_4$ may consist of:
\begin{eqnarray}
G_4=\{&&\left|\psi^{+}\right\rangle_{AB} , \left|\psi^{+}\right\rangle_{AC} ,
\left|\psi^{+}\right\rangle_{AD} , \left|\psi^{+}\right\rangle_{BC},
\left|\psi^{+}\right\rangle_{BD}, \left|\psi^{+}\right\rangle_{CD},\nonumber\\
&&\left|\mbox{GHZ}\right\rangle_{ABC},
\left|\mbox{GHZ}\right\rangle_{ABD},\left|\mbox{GHZ}\right\rangle_{ACD},\left|\mbox{GHZ}\right\rangle_{BCD},\nonumber\\
&&\left|\mbox{GHZ}\right\rangle_{ABCD} \quad \}.
\end{eqnarray}
In the set above we included all maximally entangled EPR pairs between
all pairs of parties, plus GHZ states
\[
\left|\mbox{GHZ}\right\rangle_{ijk}=1/\sqrt{2}\left(\left|0_i0_j0_k\right\rangle
+\left|1_i1_j1_k\right\rangle\right)
\]
of all possible combinations of three parties $i,j$ and $k$, plus the  GHZ state of all four parties
\begin{equation}
\left|\mbox{GHZ}\right\rangle_{ABCD}=1/\sqrt{2}\left(\left|0_A0_B0_C0_D\right\rangle
+\left|1_A1_B1_C1_D\right\rangle\right).
\end{equation}
The conjecture here is that
from the states in $G_4$, it is possible to generate \textit{any}
pure entangled state of four parties, reversibly in the asymptotic
limit.

In \cite{WuZ01} Wu and Zhang showed that this conjecture is
\textit{false}, i.e. set $G_4$ is not an MREGS for pure states of four
parties. Their argument relied on calculating some combinations of sums of von Neumann
entropies of arbitrary tensor products of states in $G_4$ for all possible cuts into two
parties, and showing these were inconsistent with any possible
combination of states in $G_4$ for the particular state
\begin{equation}
\left|\eta\right\rangle=\frac{1}{2}\left( \left|0000 \right\rangle +
\left|0110 \right\rangle+\left|1001 \right\rangle-\left|1111
\right\rangle \right). \label{wuzhangstate}
\end{equation}

Besides the fundamental importance of the result, this approach is
also promising because it identifies states with interesting
entanglement structures, which can be useful for new quantum
information applications. For example, state $\left|\eta\right\rangle$ above has
appeared in the work of Briegel and Raussendorf \cite{BriegelR01,
RaussendorfB01}, who proposed a quantum computation
model based solely on a sequence of single-qubit measurements on
carefully designed, highly entangled \textit{cluster states} of many qubits. State $\left|\eta\right\rangle$ above
was pointed out in \cite{BriegelR01} as a simple example of such a
cluster state of four qubits. The same state reappeared in
\cite{VerstraeteDdMV02}, this time as a representative of the generic
class of four-qubit entangled states under single-copy LOCC manipulations. 

The result of Wu and Zhang prompts the question about the structure
for the MREGS for tripartite states. Linden \textit{et al.} addressed
the tripartite case in \cite{LindenPSW99}, which actually appeared before the results of Wu and Zhang. They showed that the three-party
GHZ state and collections of two-party EPR states between the three
pairs of parties are indeed asymptotically inequivalent. This means an
MREGS for tripartite entanglement cannot consist only of EPR states
between all pairs of parties.

In the next section we review the results of Linden \textit{et al.}'s
paper \cite{LindenPSW99}, so that I can state my main results on this
problem. 

\section{Tripartite entanglement and relative entropy \label{sec trirel}}

Given the asymptotic inequivalence of 3-party GHZ states and pairs of
EPR states, we can state a conjecture:
\begin{eqnarray}
\mbox{\textbf{Conjecture :}}&&\mbox{Set }G_3=\{ \left|\psi^{+}\right\rangle_{AB} , \left|\psi^{+}\right\rangle_{AC} ,
\left|\psi^{+}\right\rangle_{BC} ,
\left|\mbox{GHZ}\right\rangle_{ABC}\} \nonumber \\
&&\mbox{ is an MREGS for tripartite pure states.}\label{conjg3} 
\end{eqnarray}
Some evidence supporting this conjecture was provided in
\cite{VidalDC00}, which describes a class of $G_3$-equivalent states
(for the definitions of asymptotic equivalence see section \ref{sec asymplim}).

Regarding this conjecture, Linden \textit{et al.} showed in \cite{LindenPSW99} that any
$G_3$-equivalent state must obey certain relations between two
different entanglement measures: the von Neumann entropy $S$ of each
party's subsystem, and the regularised relative entropy of
entanglement $E^{reg}$ of two-party reduced density matrices (see
section \ref{sec relentent} for the relevant definitions). Since those
relations will be important for the work I present in this chapter,
let us see how we can establish them.

Let us start by remembering that a tripartite pure state $\left| \phi \right\rangle _{ABC}$ is
said to be $G_3$-equivalent if there exist LOCC operations that
convert $\left| \phi \right\rangle_{ABC}$ into arbitrarily good
approximations of tensor products of the states in $G_3$, reversibly
in the asymptotic limit of many copies. These asymptotically
reversible transformations can be represented as:
\begin{equation}
\left| \phi \right\rangle _{ABC}^{\otimes N}\stackrel{LOCC}{\rightleftharpoons }\left|\psi^{+}\right\rangle _{AB}^{\otimes Ns_{AB}}\otimes \left|\psi^{+}\right\rangle
_{AC}^{\otimes Ns_{AC}} \otimes \left|\psi^{+}\right\rangle
_{BC}^{\otimes Ns_{BC}}\otimes \left|\mbox{GHZ}\right\rangle _{ABC}^{\otimes Ng} \label{g3es}
\end{equation}
where $Ns_{AB},Ns_{AB},Ns_{AB}$ and $Ng$ represent the number of
copies of each state in $G_3$ necessary for the reversible creation of
$N$ copies of $\left| \phi \right\rangle _{ABC}$, as $N\to\infty$.

In \cite{LindenPSW99} Linden \textit{et al.} showed that the
regularised version of the relative entropy of entanglement $E^{reg}$
for any two-party cut of a multipartite state must be invariant under
asymptotically reversible LOCC. It is also known that the von Neumann
entropy must also remain the same during asymptotically reversible
LOCC transformations. In the case of $G_3$-equivalent states
(\ref{g3es}), this means that the two-party $E^{reg}$ and von Neumann
entropies must be the same on the two sides of equation (\ref{g3es}).

To see more explicitly the consequences of this result, let us first
calculate the reduced density
matrix $\sigma_{AB}$ of the state on the right-hand side of
eq. (\ref{g3es}). It can be readily shown to be
\begin{eqnarray}
\sigma_{AB}=&&\left|\psi^{+}\right\rangle\left\langle\psi^{+}\right|^{\otimes
s_{AB}}\otimes \left(\tfrac{1}{2} |0\rangle\langle 0|_A + \tfrac{1}{2} |1\rangle\langle 1|_A
\right)^{\otimes s_{AC}} \nonumber\\
&&\otimes \left(
\tfrac{1}{2}|0\rangle\langle 0|_B + \tfrac{1}{2}|1\rangle\langle 1|_B
\right)^{\otimes s_{BC}}\otimes \left(
\tfrac{1}{2}|00\rangle\langle 00|_{AB} + \tfrac{1}{2}|11\rangle\langle
11|_{AB} \right)^{\otimes g}. \label{rhoabg}
\end{eqnarray}
Since the von Neumann entropy is additive for tensor products, it is
easy to calculate it for $\sigma_{AB}$ above, and also for the other
similar two-party reduced density matrices. We obtain
\begin{eqnarray}
S(\sigma_{AB})=S(\sigma_C)&=&s_{AC}+s_{BC}+g,\label{6r1}\\
S(\sigma_{AC})=S(\sigma_B)&=&s_{AB}+s_{BC}+g,\\
S(\sigma_{BC})=S(\sigma_A)&=&s_{AB}+s_{AC}+g.
\end{eqnarray}

Now let us calculate $E^{reg}$ for the same two-party density matrices
of the right-hand side of eq. (\ref{g3es}). For that we need two
results: for pure states $E$ is equal to the von Neumann entropy, and
therefore is additive in this case \cite{VedralP98}; and $E(\sigma_{AB}\otimes
\rho_{sep})=E(\sigma_{AB})$ for separable $\rho_{sep}$ and arbitrary
$\sigma_{AB}$ \cite{Rains99,Rains01}. Using these two facts we can
calculate $E^{reg}$ for the two-party density matrices $\sigma_{AB}, \sigma_{AC}$ and $\sigma_{BC}$:
\begin{eqnarray}
E^{reg}(\sigma_{AB})&=&E(\sigma_{AB})=s_{AB},\\
E^{reg}(\sigma_{AC})&=&E(\sigma_{AC})=s_{AC},\\
E^{reg}(\sigma_{BC})&=&E(\sigma_{BC})=s_{BC}.\label{6r6}
\end{eqnarray}

Because of the invariance of $S$ and $E^{reg}$ under asymptotically
reversible LOCC, equations (\ref{6r1})-(\ref{6r6}) must hold also for
the left-hand side of eq. (\ref{g3es}), i.e. for any $G_3$-equivalent
state. These equations can be succinctly summarised as:
\begin{align}
    E^{reg}(\sigma_{ij})  &  =s_{ij}\label{corcondr}\\
    S(\sigma_{A})  &  =g+s_{AB}+s_{AC}\label{corcond2r}\\
    S(\sigma_{B})  &  =g+s_{AB}+s_{BC}\\
    S(\sigma_{C})  &  =g+s_{AC}+s_{BC}, \label{corcond3r}
\end{align}
where $E^{reg}(\sigma_{ij})$ is the regularised relative entropy of
entanglement of the reduced density matrix $\sigma_{ij}$ of parties $i,j$.

Let us now see how we can use the relations established above to
make a connection between the problem of $G_3$-equivalence and the
question of additivity of the relative entropy of entanglement.

\subsection{Sub-additivity of $E$ and tripartite MREGS}

Suppose we find a state incompatible with relations
(\ref{corcondr})-(\ref{corcond3r}). Because of the results reviewed above,
this state would be asymptotically irreducible to
the set $G_3$, and therefore would represent asymptotic tripartite
entanglement of a type genuinely different from that of a GHZ
state. Unfortunately, only very recently were methods to compute
$E^{reg}$ developed \cite{AudenaertEJPVdM01}, and they only work for a
restricted class of states, and for a slight modification of the
relative entropy we will discuss later.

Despite this, I
managed to make some progress in establishing relations between
additivity questions and the structure of a possible MREGS for
tripartite pure states. My approach was to search for states that fail
to obey the following relations:
\begin{align}
    E(\sigma_{ij})  &  =s_{ij}\label{corcond}\\
    S(\sigma_{A})  &  =g+s_{AB}+s_{AC}\label{corcond2}\\
    S(\sigma_{B})  &  =g+s_{AB}+s_{BC}\\
    S(\sigma_{C})  &  =g+s_{AC}+s_{BC}, \label{corcond3}
\end{align}
which are identical to the relations (\ref{corcondr})-(\ref{corcond3r}),
except that here we have the single-copy relative entropy of
entanglement $E$, instead of its regularised version. At the time this
work was completed, it was conjectured that $E$ was fully additive. If
that were the case, any state violating relations
(\ref{corcond})-(\ref{corcond3}) would automatically violate
(\ref{corcondr})-(\ref{corcond3r}) as well, as in that case
$E=E^{reg}$ (see footnote \footnote{The original paper
\cite{LindenPSW99} stated relations (\ref{corcond})-(\ref{corcond3})
instead of the correct (\ref{corcondr})-(\ref{corcond3r}). This relied
on the implicit assumption that $E(\sigma_{ij})$ is asymptotically
additive, which was subsequently shown to be false in \cite{VollbrechtW01}.}).

Through numerical calculations and analytical work I identified
families of states which fail to comply with relations
(\ref{corcond})-(\ref{corcond3}), and are likely candidates for
$G_3$-irreducibility. These will be presented in section \ref{sec
susp}.

So what does it mean to show that there are states which are
incompatible with relations (\ref{corcond})-(\ref{corcond3})? Basically,
it means that from the two statements below, \textit{at least one}
must be false:
\begin{description}
\item{\bf{1}}- all tripartite pure states of three qubits (an in particular the
states I discuss) are asymptotically reducible to set $G_3$,
i.e. $G_3$ is an MREGS for asymptotic tripartite entanglement;

\item{\bf{2}}- the relative entropy of entanglement is generally additive, or at
least it is for the states I present in section \ref{sec susp}.
\end{description}

When most of the work reported in this chapter was published
\cite{GalvaoPV00}, it drew attention to the relation between these two problems, which were then open. In relation to statement \textbf{2}, a definitive proof
that the relative entropy of entanglement is sub-additive for some $3
\times 3$-dimensional states was subsequently found by Vollbrecht and
Werner \cite{VollbrechtW01}. Despite extensive numerical work
\cite{HorodeckiSTT99,VedralP98}, the additivity question is still open
for two-qubit states, and in particular for all such states studied
numerically in section \ref{sec susp} below.

This suggests statement \textbf{1} above is likely to be false. Indeed,
following our suggestion presented in section \ref{sec purbe} (and
published in \cite{GalvaoPV00}), Ac\'{\i}n, Vidal and
Cirac showed that purifications of a certain bound-entangled state
violate relations (\ref{corcondr})-(\ref{corcond3r}) and are not $G_3$-equivalent \cite{AcinVC02}. This
resulted in a definitive refutation of conjecture (\ref{conjg3}),
without the assumption of additivity that makes $E^{reg}=E$ and
relations (\ref{corcond})-(\ref{corcond3}) equivalent to relations (\ref{corcondr})-(\ref{corcond3r}).

\section{Calculating the relative entropy of entanglement \label{sec calent}}

In this section I will describe how we can compute the relative
entropy $E$ numerically for
general states. With $E$ of the two-party reduced
density matrices $\sigma_{ij}$ and the von Neumann
entropies $S(\sigma_j)$ we can check whether relations
(\ref{corcond})-(\ref{corcond3}) hold or not. Using this method, in
the next section I will explicitly list families of states which do not
obey those relations. As discussed above, this has implications for
the question of additivity of $E$ and also for the possible structure
of an MREGS for tripartite pure states.

But how do we go about numerically calculating $E$? This problem was
discussed in \cite{VedralP98}, and here we describe some of the
properties of $E$ which allows us to do this for
small-dimensional systems. In section \ref{sec entvssep} we have seen
that a general separable density matrix can be written as
\begin{equation}
\rho_{sep}=\sum_{i=1}^K p_i \left|\phi_i\right\rangle
\left\langle\phi_i\right|_A \left|\phi_i\right\rangle\left\langle\phi_i\right|_B
\cdots \left|\phi_i\right\rangle 
\left\langle\phi_i\right|_N,\hspace{4 mm}
\sum_{i=1}^K p_i=1,
\end{equation}
where $\left|\phi_i\right\rangle_j$ represents a pure state belonging
to party $j$. The first question we can ask is how many terms $K$ this
expression needs to have. If $K$ cannot be upper-bounded or is too
high, no numerical procedure can be efficiently implemented to compute
$E$. Fortunately, for two-qubit systems it is known
\cite{Wootters98,SanperaRV98} that $K \le 4$. More generally, it has
been shown \cite{Horodecki97} that for
systems of dimensionality $d_A \times d_B$ we need at most $K \le (d_A
d_B)^2$ terms.

Let us now consider how to determine the relative entropy
$E(\sigma_{AB})$, where $\sigma_{AB}$ is a general two-qubit density
matrix. The results cited above show that any two-qubit separable state
$\rho^{sep}_{AB}$ can be parametrised by a convex combination of at
most four separable states:
\begin{equation}
\rho^{sep}_{AB}=\sum_{i=1}^4 p_i \left|\phi_i\right\rangle
\left\langle\phi_i\right|_A \left|\phi_i\right\rangle 
\left\langle\phi_i\right|_B,\hspace{4 mm}
\sum_{i=1}^4 p_i=1. \label{2qubitsep}
\end{equation}
Each two-qubit product state in the equation above can be parametrised
by $2 \times 2=4$ real parameters, and because $\sum_i p_i=1$, the
total number of real parameters in the parametrization for
$\rho_{AB}^{sep}$ is $3+4 \times 4=19$.

To calculate the relative entropy of entanglement we need to do a
search over this $19$-parameter space to find the separable state
$\rho_{AB}$ which minimises the following function:
\begin{eqnarray}
E(\sigma_{AB})&=&\min_{\rho_{AB}^{sep}}D(\sigma_{AB}||\rho_{AB})\\
 &=&Tr \left(\sigma_{AB}\log\sigma_{AB}\right) -
 \min_{\rho_{AB}^{sep}}Tr \left(
 \sigma_{AB}\log\rho_{AB}\right)\\
&=& -S(\sigma_{AB})-\min_{\rho_{AB}^{sep}}Tr
 \left(\sigma_{AB}\log\rho_{AB} \right)
\end{eqnarray}
For a given $\rho_{AB}^{sep}$, the evaluation of the function above involves the diagonalisation of
$\rho_{AB}^{sep}$, taking the logarithms of its eigenvalues, and
calculating $Tr \left(\sigma_{AB}\log\rho_{AB}^{sep} \right)$ in the
basis which diagonalises $\rho_{AB}^{sep}$.
 
Finding the global minimum of a function defined in such a large
parameter space can be very hard. Fortunately, the relative entropy has the useful
property of \textit{convexity} \cite{Donald86,Donald87}. This,
together with the convexity of the set of separable states, guarantees
that any local minimum of the relative entropy is also a global
minimum. This makes a simple gradient search algorithm efficient to
find the minimum for the function above, and calculate the relative
entropy of entanglement. We start with a given separable state and
calculate the approximate gradient of the relative entropy, for small
changes in all directions in the parameter space. With this
approximation of the gradient we can move towards a new separable
state which is strictly closer to the absolute minimum, and start the
process again. After a few thousand iterations we reach the local
minimum, which will also be the global minimum because of convexity,
as we have just mentioned.

For the gradient search to work it is important to choose a
parametrization that does not have discontinuities for 
close separable states. In my program I used the suitable
parametrization presented in \cite{VedralP98}.

\section{Suspicious states \label{sec susp}}

In this section I present tripartite states I found to be
incompatible with relations (\ref{corcond})-(\ref{corcond3}) using the
numerical methods described in the last section. These
include the class of states known as $W$-type, and also a family which
I called $\Lambda$ states.

I  also provide an analytical argument identifying a
large class of states as likely candidates to violate relations
(\ref{corcond})-(\ref{corcond3}) and also
(\ref{corcondr})-(\ref{corcond3r}), and thus contradict conjecture
(\ref{conjg3}). These are the purifications of bound-entangled states,
and they were in fact proven to be $G_3$-inequivalent in
\cite{AcinVC02}, following publication of the argument presented in
section \ref{sec purbe} in \cite{GalvaoPV00}.

\subsection{$\Lambda$ states}

These are states of the form:
\begin{equation}
|\Lambda(a,b)\rangle = a|000\rangle + b|100\rangle + b|101\rangle +
 b|110\rangle + b|111\rangle. \label{lambdas}
\end{equation}

An interesting property of states
$|\Lambda(a,b)\rangle$ is that $\sigma_{BC}$ is separable for all
$a,b$. This follows from the fact that $\sigma_{BC}$ is positive
under partial transposition (see section \ref{sec phsep}).

Numerical tests show this family violates relations
(\ref{corcond})-(\ref{corcond3}). For example, when $a=b=1/\sqrt{5}$
we have
\begin{eqnarray}
\sigma_{AB}=\sigma_{AC}= \left( \begin{array}{c c c c}
1/5 & 0 & 1/5 & 1/5\\
0 & 0 & 0 & 0      \\
1/5 & 0 & 2/5 & 2/5\\
1/5 & 0 & 2/5 & 2/5\\
\end{array}\right)\\
\sigma_{BC}= \left( \begin{array}{c c c c}
2/5 & 1/5 & 1/5 & 1/5\\
1/5 & 1/5 & 1/5 & 1/5      \\
1/5 & 1/5 & 1/5 & 1/5\\
1/5 & 1/5 & 1/5 & 1/5\\
\end{array}\right)
\end{eqnarray}

If state $|\Lambda(1/\sqrt{5},1/\sqrt{5})\rangle$ obeys equations
(\ref{corcond})-(\ref{corcond3}), we would need to have
$E(\sigma_{AC})$ equal to $S(\sigma_A)-S(\sigma_B) \simeq 0.1541$. My numerical calculation found $E(\sigma_{AC}) \simeq 0.1971$, which is over
$25\%$ higher than the required $0.1541$ for agreement with equations
(\ref{corcond})-(\ref{corcond3}).

These states violate relations (\ref{corcond})-(\ref{corcond3}). Now let us assume
that $E$ is asymptotically additive for all two-qubit states. From our
previous discussion it follows that they would also violate
(\ref{corcondr})-(\ref{corcond3r}) and
would be $G_3$-inequivalent. This  would bring interesting
consequences for the minimal cardinality for a tripartite entanglement
MREGS.

An MREGS for tripartite entanglement must contain at least the states
in $G_3$. The GHZ state is necessary to asymptotically generate states
which are GHZ-equivalent: all Schmidt-decomposable states \cite{BennettPRST01}. The three EPR pairs are needed to
asymptotically generate all states with bipartite entanglement
only. If the $|\Lambda(a,b)\rangle$ states (\ref{lambdas}) are indeed
$G_3$-inequivalent, then at least one of them would have to be added
to set $G_3$ if we are trying to form an MREGS. Let us add one such
state to set $G_3$, and see whether the new set can be an MREGS. Our new
candidate for an MREGS is now:
\begin{equation}
G_{\Lambda}=G_3 \cup \{|\Lambda(1/\sqrt{5},1/\sqrt{5})\rangle\}.
\end{equation}

Consider now the states obtainable from
$|\Lambda(1/\sqrt{5},1/\sqrt{5})\rangle$ by swapping parties $A$ and
$B$. By the same argument given in this section, this new state
would also be $G_3$-inequivalent. Note that now the separable
two-party reduced density matrix is $\sigma_{AC}$,
i.e. $E^{reg}(\sigma_{AC})=0$.  Because it has separable
$\sigma_{AC}$, it is also $G_{\Lambda}$-inequivalent, as relations
(\ref{corcondr})-(\ref{corcond3r}) forbid it to be equivalent to any
tensor product of states with non-separable $\sigma_{AC}$.

This argument shows that if $|\Lambda(a,b)\rangle$ is indeed
$G_3$-inequivalent, then we would need to expand the candidate for an
MREGS $G_3$ by adding three more states, corresponding to one member
of the family $|\Lambda(a,b)\rangle$ and the two other states
obtainable from it by swapping either $A$ and $B$, or $A$ and $C$.  This
would mean that the cardinality of a tripartite MREGS would
have to be at least seven -- it would have to contain the GHZ, three
EPR pairs, one $|\Lambda(a,b)\rangle$ state and the two
$|\Lambda(a,b)\rangle$-type states obtained from
$|\Lambda(a,b)\rangle$ through swapping two parties.

\subsection{W-type states}

Another class of states violating relations
(\ref{corcond})-(\ref{corcond3}) are the so-called \textit{W states}, which have
already appeared when we discussed single-copy entanglement
manipulations in section \ref{sec singlecopy}.

The class of states we will consider are of the form:
\begin{equation}
|W(e,f)\rangle=e|000\rangle+f|101\rangle+f|110\rangle, \label{Wclass}
\end{equation}
where $e$ and $f$ are real parameters. The two-party reduced density
matrices are given by:
\begin{align}
    \sigma_{AB}(e^{2},f^{2})= \sigma_{AC}(e^{2},f^{2}) = \left(
    \begin{array}[c]{cccc}%
    e^{2} & 0 & 0 & ef\\
    0 & 0 & 0 & 0\\
    0 & 0 & f^{2} & 0\\
    ef & 0 & 0 & f^{2}
\end{array}
\right) \label{rhoab}\\[0.1cm]
\sigma_{BC}(e^{2},f^{2})= \left(
\begin{array}[c]{cccc}
    e^{2} & 0 & 0 & 0\\
    0 & f^{2} & f^{2} & 0\\
    0 & f^{2} & f^{2} & 0\\
    0 & 0 & 0 & 0
\end{array}
\right)
\end{align}
Let us first check the predictions for $E$ given by equations
(\ref{corcond})-(\ref{corcond3}). For those relations to hold we would
need
\begin{equation}
E(\sigma_{AB})+S(\sigma_{AB})=E(\sigma_{BC})+S(\sigma_{BC}) \label{neccond}.
\end{equation}
The relative entropy of state $\sigma_{BC}$ has been computed in
example 1 of \cite{VedralP98}:
\begin{equation}
E(\sigma_{BC})=2(f^{2}-1)\log_{2}(1-f^{2})+(1-2f^{2})\log_{2}(1-2f^{2})
\end{equation}
and by direct computation we find that
\begin{align}
S(\sigma_{BC})  &  =-(1-2f^{2})\log_{2}(1-2f^{2})-2f^{2}\log_{2}(2f^{2}),\\
S(\sigma_{AB})  &  =-f^{2}\log_{2}f^{2}-(1-f^{2})\log_{2}(1-f^{2})\;.
\end{align}
Combining these equations with eq. (\ref{neccond}), we see that
if $E$ is asymptotically additive for $\sigma_{AB}$, then $G_{3}$ can
only be an MREGS for states of the form (\ref{Wclass}) if
\begin{eqnarray}
E(\sigma_{AB})&=& \nonumber\\
&&\hspace*{-8ex}(f^{2}-1)\log_{2}(1-f^{2})-2f^{2}\log_{2}(2f^{2})+f^{2}%
\log_{2}(f^{2}) \label{necnew}
\end{eqnarray}
With $e=\sqrt{2/3}$ and $f=\sqrt{1/6}$ eq. (\ref{necnew}) predicts
$E \simeq 0.3167$. A numerical calculation yields $E \simeq 0.3548$,
in disagreement with the prediction obtained from relations
(\ref{corcond})-(\ref{corcond3}). Moreover, in \cite{GalvaoPV00} an
analytical argument was found to corroborate this numerical result, in
joint work with Martin Plenio and Shashank Virmani \footnote{I decided
to omit this proof here since my collaborators' share in it was bigger
than mine, and it involved more sophisticated mathematical arguments
which would require extensive background material to be added to
this thesis.}.

\subsection{Purifications of bound-entangled states \label{sec purbe}}

There is a very wide class of states which are good candidates for not
being $G_3$-equivalent. These are simply all tripartite purifications
of bipartite bound-entangled states (see section \ref{sec phsep}). Let
me now present an argument for why this is the case.

When we defined the relative entropy of entanglement $E_S(\sigma_{AB})$ and its regularised version
$E^{reg}_S(\sigma_{AB})$ in section \ref{sec qreintro}, it was with
respect to the
set $S$ of separable states. The definition makes sense because we
wanted to quantify distinguishability of entangled states with respect
to any
state in that set, which is invariant under local unitary operations. We can,
however, define the relative entropy of entanglement with respect to other local-unitary invariant sets, such as the set of PPT states,
i.e. states which remain positive after partial transposition. This
represents a wider class of states, which includes separable states as
a subset, but also the non-separable PPT states known as bound-entangled
states. Because the set of PPT states is easier to characterise
mathematically than the set of separable states, some results have
been obtained recently for the relative entropy with respect to PPT
states (which we shall denote here by $E_{PPT}$), for example
calculations of its regularised version $E_{PPT}^{reg}$ \cite{AudenaertEJPVdM01,AudenaertdMVW02}.

As we have seen in the previous sections, any $G_3$-equivalent state
$\left| \phi \right\rangle _{ABC}$ can be obtained from $G_3$ through
reversible LOCC in a process like:
\begin{equation}
\left| \phi \right\rangle _{ABC}\stackrel{LOCC}{\rightleftharpoons }\left|\psi^{+}\right\rangle _{AB}^{\otimes s_{AB}}\otimes \left|\psi^{+}\right\rangle
_{AC}^{\otimes s_{AC}} \\
\otimes \left|\psi^{+}\right\rangle _{BC}^{\otimes s_{BC}}\otimes
\left|\mbox{GHZ}\right\rangle _{ABC}^{\otimes g}. \label{magiceq}
\end{equation}
The relations (\ref{corcondr})-(\ref{corcond3r}) were derived in \cite{LindenPSW99} by
proving that the relative entropy $E_S^{reg}$ of two-party subsystems
in such asymptotically reversible transformations must remain
constant. Their proof relies on general properties of the relative
entropy, and also goes through for the alternative definition
$E^{reg}_{PPT}$ with respect to PPT states.

The two-party reduced density matrices $\sigma_{ij}$ of the
right-hand-side of eq. (\ref{magiceq}) above are tensor products of pure states with
separable states [see eq. (\ref{rhoabg})]. For such states it has been shown that $E^{reg}_S(\sigma_{ij})=E^{reg}_{PPT}(\sigma_{ij})$, being equal to
the von Neumann entropy $S(\sigma_{ij})$ \cite{VedralP98, Rains99,
Vedral99}. Because $E^{reg}(\sigma_{ij})$ is conserved in reversible
transformations, this must be true also of the two-party reduced
density matrices of the left-hand side of
eq. (\ref{magiceq}), i.e. of any $G_3$-equivalent state.

Now consider a tripartite state $\left| \phi \right\rangle _{ABC}$
whose reduced density matrix $\sigma_{AB}$ is bound entangled, i.e. $E_{PPT}^{reg}(\sigma_{AB})=0$ but $E_S(\sigma_{AB})\ne
0$ (because it is non-separable). Despite the fact that
$E_S(\sigma_{AB}) \ne 0$, to be $G_3$-equivalent $\left| \phi
\right\rangle _{ABC}$ needs to have $E^{reg}_S(\sigma_{AB})=0$, so that
$E^{reg}_S(\sigma_{AB})=E^{reg}_{PPT}(\sigma_{AB})=0$ as required. This is a very peculiar
requirement: we need the non-zero $E_S(\sigma_{AB})$ to vanish in
the asymptotic limit of many copies. Note here that for $G_3$ to be an
MREGS, this requirement would need to hold for arbitrary tripartite
purifications of arbitrary bound-entangled states.

This shows that purifications of arbitrary bipartite bound-entangled
states are candidates for non-$G_3$ reducibility. Subsequently to
the published version of this argument \cite{GalvaoPV00}, Ac\'{\i}n,
Vidal and Cirac proved that indeed $G_3$ is not an MREGS for
tripartite states by showing that $E^{reg}_S \ne 0$ for a particular
bipartite bound-entangled state \cite{AcinVC02}.

\section{Entanglement of W versus GHZ \label{sec wvsghz}}

It is easy to modify the program for calculating the relative entropy
of entanglement $E$ for two-qubit
states and make it work for arbitrary pure-state systems. I changed
the program slightly to investigate $E$ for three-qubit states. In
this section I will show that in a very operational sense, the $W$ state
\begin{equation}
|W\rangle=\frac{1}{\sqrt{3}} \left(|000\rangle+|101\rangle+|110\rangle
 \right) \label{Wstateagain}
\end{equation}
can be shown to be more entangled than the three-qubit GHZ state,
contrary to the usual claims that the GHZ state is the `maximally
entangled state' of three qubits.

The $W$ state (\ref{Wstateagain}) has many interesting properties. In
\cite{KoashiBI00,DurVC00} it was shown that it is the three-qubit state that is
maximally robust under disposal of one of the three qubits. More
precisely, the two-party reduced density matrices of the $W$ state
maximise several measures of bipartite mixed-state entanglement. For
example, the concurrence was defined by Wootters \cite{Wootters98} as a measure of
bipartite entanglement and can be calculated exactly
for any two-qubit mixed state. It turns out that the $W$ state
maximises the average concurrence of the three two-party reduced
density matrices \cite{KoashiBI00,DurVC00}. This is in great contrast with the
$|\mbox{GHZ}\rangle$ state, which has separable reduced density
matrices.

So we see that the bipartite entanglement present in the reduced
density matrices is maximal for the $W$, and zero for the GHZ
state of three qubits. This, however, may be an unfair way to measure
the overall entanglement for tripartite states. We know the GHZ state
is highly entangled, allowing proofs of non-locality without
inequalities \cite{GreenbergerHZ89,GreenbergerHSZ90,Mermin90}, and improving the efficiency of quantum
information applications, an example of which we will see in chapter \ref{chap
qcc}. What we can safely say is that the types of entanglement present
in the $W$ and the GHZ states are different (see section \ref{sec
singlecopy}), and accordingly can be used for different tasks.

This having been said, there is a very suggestive way of defining
what a maximally entangled state should consist of. It should be the
state which, upon measurements, is easiest to distinguish from any separable state. In other words, we can call a state ``maximally
entangled'' when it is the state most different from any separable
state, i.e. the state which displays entanglement in the most manifest
way. As we have seen in section \ref{sec qreintro}, Sanov's theorem
can be used to show that the relative entropy of entanglement is a
good measure of distinguishability between a given entangled state and
\textit{any} separable state. It is therefore reasonable to call a
state maximally entangled when it maximises the relative entropy of
entanglement, with respect to the set of separable states.

In order to compute $E$ for three-qubit states, we need to parametrise
all tri-separable states, i.e. states which can be expressed as
\begin{equation}
\rho_{sep}=\sum_{i=1}^K p_i \left|\phi_i\right\rangle_A
\left\langle\phi_i\right|_A \left|\phi_i\right\rangle_B
\left\langle\phi_i\right|_B \left|\phi_i\right\rangle_C 
 \left\langle\phi_i\right|_C, \hspace{4 mm} \sum_{i=1}^{K} p_i=1. \label{trisep}
\end{equation}
By the arguments of \cite{VedralP98} it can be shown that an upper
bound for the number of terms $K$ is $(d_A \times d_B \times
d_C)^2=8^2=64$. Including the probabilities for each term, the total
number of real parameters describing a three-qubit separable state is
given by $63+6 \times 64=447$. A gradient search over this large
parameter space takes a long time but it is still feasible.

While for the GHZ and simple EPR states $E=1$ exactly, for the
three-qubit $W$ state (\ref{Wstateagain}) I numerically found
$E(W)=2\log(3)-2\simeq 1.170$, which is definitely higher than for the
GHZ state. We thus see that when we use the relative entropy as a measure of entanglement, it is fair to claim that
the $W$ state is more entangled than the GHZ state.

It would be interesting to do some more work on this problem, for
example by explicitly describing the tests necessary to show this
higher degree of entanglement. One interesting approach
to this problem was put forward by Peres \cite{Peres00}, who suggested that
a Bayesian analysis of experiment outcomes can be used to quantify the
`strength' of locality violation. Basically, he uses a Bayesian
approach to calculate how many experimental runs are necessary to
convince a skeptical observer that locality cannot hold, up to a
certain level of statistical confidence. It would be interesting to
perform such an analysis for non-locality tests using $W$ states (for
example \cite{Cabello02}), and compare the results with experiments
done with GHZ states.

\section{Chapter summary \label{sec conctri}}

In this chapter we have identified a number of tripartite states for
which either the relative entropy is sub-additive, or which represent
a type of genuine tripartite entanglement different from the GHZ
type, in the asymptotic limit. Further work by other authors has
proved that indeed some of those states are not $G_3$-equivalent.

These states were singled out because of their interesting
entanglement structure, which may be useful in new quantum information
applications, as has already happened with states (\ref{Wstateagain})
and (\ref{wuzhangstate}) above. The fact that purifications of
bound-entangled states are not $G_3$-equivalent suggests that
multipartite entanglement generation may be genuinely irreversible,
i.e. it may be impossible to find MREGS with finite cardinality for
entanglement of more than two parties.

In order to identify candidates for $G_3$-inequivalence I resorted to a numerical
calculation of the relative entropy of entanglement for two-qubit
systems. By applying the same method for three-qubit states, I showed
that the $W$ state (\ref{Wstateagain}) is more entangled than the GHZ
state, in the sense that its measurement statistics are more easily
distinguishable from those of any separable state.

\chapter{Cloning and quantum computation \label{chap cloning}}

\begin{figure}
\begin{center}
{\includegraphics[scale=0.4]{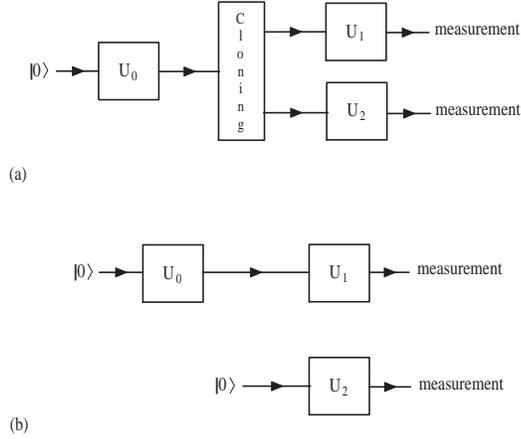}}
\caption[Cloning and no-cloning approaches.]{\label{fig clonoclo}(a) This circuit represents the use of cloning to obtain
information about computations $U_{1}U_{0}$ and $U_{2}U_{0}$ using
$U_{0}$ only once. (b)This is one of the no-cloning strategies
available for the same task. There are other possibilities: for
example, we could embed each of the $U_{j}$ in a different quantum circuit, designed specifically for obtaining
information about $U_{j}$. }
\end{center}
\end{figure}

The COPY operation in classical computing is very useful, as it allows
one to make multiple copies of the output of some computation, that
can be fed as the input to further multiple processes. As we have seen
in section \ref{sec qcloning}, the quantum copying (i.e. cloning) is
necessarily imperfect, introducing some noise in the second round of
computation. This situation is pictured in fig. \ref{fig clonoclo}(a).

The imperfect nature of the quantum cloning procedure results in lower
chances of getting the correct computational outputs at the
end. Nonetheless, in this chapter I will show that in some cases cloning improves our chances of correctly computing both branches,
if there are constraints on the number of times we can run the first
part $U_{0}$. I will do this by presenting two examples of
computations for which cloning offers an advantage which cannot be
matched by any approach that does not resort to quantum information
distribution [as in fig. \ref{fig clonoclo}(b)].

A reasonable question to ask is in which situations we can expect
cloning to be helpful during a quantum computation. One might think
that cloning could be helpful in state estimation. As we have seen in
section \ref{sec UQCM}, state estimation is equivalent to cloning,
when the number of copies $M\rightarrow\infty$ \cite{BrussM99,
BrussEM98}. In relation to this point, D'Ariano \textit{et
al.} \cite{DArianoMS01} have shown that cloning can be useful for obtaining
a good estimation of non-commuting observables of a single qubit,
by making a few clones and directly measuring  different observables
in each of them. However, optimal state estimation procedures for
the same task will
in general be different, and more efficient, than simply measuring the
clones of a Universal Quantum Cloning Machine (UQCM).

It seems that cloning for quantum computation is most efficient when
used in the middle of a coherent quantum computation. This is the case
because the full specification of an intermediary state of the
quantum register typically involves a very large amount of classical
information, which in general is not accessible through measurement. A
compelling demonstration of this will be presented in chapter
\ref{chap qubit}. By preserving the quantum information in a way
classical measurement cannot do, cloning ends up offering an advantage
over procedures which do not distribute quantum information.

This use of quantum cloning represents and idea more in the spirit of
fig. \ref{fig clonoclo}(a). In this chapter I will give two examples of
quantum computation tasks which can be better performed if we use
quantum cloning in this way. The task we
analyse in section \ref{sec clou} uses
the UQCM's presented in section \ref{sec UQCM}, whereas in section
\ref{sec compprobclo} the task relies on the probabilistic cloning procedure we discussed in section \ref{sec probclo}.

\section{Example with universal cloning \label{sec clou}}

The first example I present is based on the scenario introduced in
fig. \ref{fig clonoclo}(a). It models the general situation in which we want to perform
$M$ different quantum computations, all of them with some first
computational steps $U_{0}$ in common.

Suppose that we are constrained
to run $U_{0}$ only once. This may happen if $U_{0}$ is a complex,
lengthy computation. In this case, we will be forced to find a scheme
that obtains the $M$ computation results with as large a probability as possible,
using $U_{0}$ only once. Possible schemes may or may not resort to
cloning to distribute quantum information; the example below is one in
which cloning enables us to improve our performance, in relation to
any scheme in which there is no information distribution using
cloning.

In order to specify our task, suppose that we are given $(M+1)$
quantum black-boxes. What black-box $j$ does is to accept one $d$-level
quantum system as an input and apply a unitary operator $U_{j}$ to it,
producing the evolved state as the output. We may think of the
black-boxes as quantum oracles, or quantum sub-computations. The
$U_{j}$ are chosen randomly from all possible $U(d)$ unitaries, using
the unique uniform distribution invariant under action of $U(d)$ (see
\cite{Lubkin78}).

Our task will be to build quantum circuits that
use each $U_{j}$ at most once to create $M$ mixed quantum states
$\rho_{j}$, each as close as possible to $\left|\phi_{j}\right\rangle
=U_{j}U_{0}\left|  0\right\rangle ,$ $(j=1,2,...M)$, where
$\left|0\right\rangle $ is an arbitrary reference state. Our score
will be given by the average fidelity of these imperfect
approximations of the ideal $\left|  \phi_{j}\right\rangle$'s:
\begin{equation}
\overline{F}=\frac{1}{M}\sum_{j=1}^{M}\left\langle \phi_{j}\right|  \rho_{j}\left|  \phi_{j}\right\rangle .
\end{equation}

This choice of scoring function evaluates the `quality' of the
approximations we can make to each computation output, giving the same
weight for each of the $M$ computational branches.

\subsection{Score with cloning step}

Now let us see how cloning allows us to obtain a high score
$\overline{F}$. We accomplish this by using a quantum circuit that
first applies $U_{0}$ to the initial state $\left|  0\right\rangle $,
followed by an optimal universal cloning machine to obtain $M$
imperfect copies $\rho_{c}$ of state $U_{0}\left|  0\right\rangle
$. We then apply each $U_{j}$ $(j=1,2,...M)$ to a clone, obtaining reduced density matrices

\begin{equation}
\rho_{j}=\eta\left|\phi_{j}\right\rangle\left\langle\phi_{j}\right|
 +(1-\eta)\frac{1}{d}\mbox{$1 \hspace{-1.0mm}  {\bf l}$},
\end{equation}
where $\eta$ is such as to give us the optimal cloning fidelity given by eq. (\ref{fcloopt}) with $N=1$. Using the resulting $\rho_{j}
$'s as our guesses for states $\left|  \phi_{j}\right\rangle $
$(j=1,2,...M)$, we obtain an overall score

\begin{equation}
\bar{F}_{cloning}=\frac{2M+d-1}{M(d+1)}, \label{fcloning}
\end{equation}
which is just the fidelity for the optimal $1 \to M$ cloning
machine [see eq. (\ref{fcloopt})].

In the next section we will evaluate the score obtainable without
cloning, and we will see that it is always smaller than this. In fact,
eq. (\ref{fcloning}) represents the optimal score obtainable for this
task, at least in the case $M=2$. In order to see this, we first note
that asymmetric cloning (arising when the factors $\eta$ are in
general different for each copy) is of no help in raising the
score. This can be deduced from \cite{BuzekHB98} and \cite{Cerf00},
where the authors consider asymmetric cloning with $M=2$ and show that
the sum of the fidelities of the copies is maximised by symmetric
cloning. Furthermore, the optimality of the universal cloning
procedure we have used entails optimality for the fidelity of each of
the $\rho_{j}$, and therefore a maximal value of the score
$\bar{F}$. This shows that this task is optimally performed [with
optimal score given by eq. (\ref{fcloning})] if and only if we are
allowed to use cloning. It is straightforward to generalise the result
to the case where we are allowed to run $U_{0}$ $N$ times $(N<M)$,
instead of just once, and quantum cloning still offers an advantage.

\subsection{Score without cloning step}

Now let us evaluate the attainable scores if we do not resort to
cloning. If we are not allowed to clone the state, there are two
simple strategies possible.

The first strategy relies on
obtaining maximal \textit{classical} information about state
$U_{0}\left|0\right\rangle$, and using it to run all $M$ branches of
the computation as well as possible. It  starts by
running $U_{0}$, followed by measurements that accomplish an optimal
estimation of the resulting state $U_{0}\left|  0\right\rangle $. After this, we can use the
information gathered to build the $M$ imperfect copies necessary to
proceed to the second part of the computation with the $U_{j}$
$(j=1,2,...M)$. As we have seen in section \ref{sec UQCM}, optimal
state estimation yields `classical clones' with fidelity given by eq. (\ref{fcloopt}) with $N=1,M\rightarrow\infty$:

\begin{equation}
\bar{F}_{1}=\frac{2}{(d+1)}. \label{f1}
\end{equation}

We obtain our approximations of states $\left|  \phi_{j}\right\rangle $
$(j=1,2,...,M)$ by applying each $U_{j}$ $(j=1,2,...M)$ to one of
these classical clones,
resulting in a score also given by eq. (\ref{f1}). As we have seen in
section \ref{sec UQCM}, this fidelity is always inferior to the
symmetric cloning fidelity presented in the previous section. In other
words, it is best to preserve the quantum information present in the
intermediary state $U_{0}\left|  0\right\rangle$, distributing it
coherently through the cloning process.

Another strategy which avoids using cloning is to obtain as
\textit{little} classical information as possible about the intermediary state. This can be done
by obtaining $\left|\phi_{1}\right\rangle
=U_{1}U_{0}\left|0\right\rangle $ with fidelity one. Now that we have
used $U_{0}$ and $U_{1}$ once already, we must make guesses about the
other $(M-1)$ states $\left|\phi_{j}\right\rangle $ by using only the
remaining $(M-1)$ black-boxes. As they were drawn from an uniform random distribution, the best we can do is to make random guesses
(each on average with $F=1/d$), obtaining, on average, a score
\begin{equation}
\bar{F}_{2}=\frac{1}{M}\left(  1+\frac{(M-1)}{d}\right)  .
\end{equation}
This score is also always inferior to that of the cloning strategy.

The two strategies presented above are extremes of a continuous set of
strategies,  which rely on extracting \textit{some} classical
information from  state $U_{0}\left|0\right\rangle $, using it to run
$M-1$ of the computational branches, while preserving part of the
quantum information present in that same state, using it to run one of
the branches. Intuitively, one would expect that extraction of classical
information can only degrade the quality of the resulting clones, and
the overall score. This is indeed true, as was confirmed by work on this
fidelity-tradeoff balance problem, studied by Banaszek and Devetak
\cite{Banaszek01,BanaszekD01}. Their work shows that in order to
maximise the sum of the fidelities of the clones, it is necessary to
avoid doing any measurement that extracts classical information from
the state. As we have seen above, the symmetric cloning strategy is
optimal for the task, beating any such combined strategy.

With this we conclude the analysis of possible no-cloning strategies,
and reach the conclusion that the optimal strategy is the one using symmetric
cloning, which I presented in the previous section.

\subsection{Remarks}

This first example models the situation in which we have a
series of quantum computations with some computational steps $U_{0}$
in common. We must note that we have assumed complete lack of
knowledge about the intermediate state $U_{0}\left|  0\right\rangle $
and about the final target states $U_{j}U_{0}\left|  0\right\rangle $
$(j=1,2,...M)$.

In the general case this will not be a good
assumption, as many quantum computations will output states known to belong to a limited set of states. In some cases, this can be taken into account with
state-dependent quantum cloning and a different choice of scoring
functions. In the next section we give an example of this.

\section{Example with probabilistic cloning \label{sec compprobclo}}

In our second example we take the black-boxes of the previous example
to consist of arbitrary quantum circuits that query a given function
only once. The query of function $f_{i}$ is the unitary that performs

\[
\left|  x\right\rangle \left|  y\right\rangle
\rightarrow\left|x\right\rangle \left|  y\oplus f_{i}(x)\right\rangle
,
\]
where we have used the symbol $\oplus$ to represent the bitwise $XOR$
operation. For ease of analysis, we restrict ourselves to the case
$M=2$ and also restrict the set of possible functions $f_{0}$, $f_{1}$
and $f_{2}$. Our task will involve determining two functionals, one
which depends only on $f_{0}$ and $f_{1}$, and the other on $f_{0}$
and $f_{2}$. As in the previous example, we will compare the
performances of cloning and no-cloning strategies.

In order to precisely state our task, let us start by considering all
functions $h_{i}$ which take two bits to one bit. We may represent
each such function with four bits $a,b,c$ and $d$, writing
$h_{a,b,c,d}$ to represent the function $h$ such that
$h(00)=a,h(01)=b,h(10)=c,$ and $h(11)=d$. Let us now define some sets
of functions that will be helpful in stating our task:
\begin{align}
&  S_{f0}=\{h_{0010},h_{0101},h_{1001}\},\nonumber\\
&  S_{1}=\{h_{0001},h_{0010},h_{0100},h_{1000}\},S_{2}=\{h_{0000}%
,h_{0011},h_{0101},h_{1001}\}\nonumber\\
&  S_{f12}=S_{1}\cup S_{2},\nonumber\\
&  S_{0000}=\{h_{0000},h_{1111}\},S_{0011}=\{h_{0011},h_{1100}\},
\label{2sets}\\
&  S_{0101}=\{h_{0101},h_{1010}\},S_{1001}=\{h_{1001},h_{0110}%
\},\label{2setsmore}\\
&  S_{f}=S_{0000}\cup S_{0011}\cup S_{0101}\cup S_{1001}.\nonumber
\end{align}

Now we randomly pick a function $f_{0}$ $\in S_{f0}$, after which two
other functions $f_{1}$ and $f_{2}$ are picked from the set $S_{f12}$,
also in a random fashion but obeying the constraints:

\begin{equation}
f_{0}\oplus f_{1},\hspace{1 mm} f_{0}\oplus f_{2}\in
S_{f}\hspace{1 mm}.
\label{constraints}
\end{equation}
Here we use the symbol $\oplus$ to represent addition modulo 2, which
is equivalent to the bitwise $XOR$ operation. Constraints (\ref{constraints}) ensure that both functions $f_{0}\oplus f_{1}$ and $f_{0}\oplus f_{2}$ are either balanced or constant. The idea of imposing this requirement comes from the Deutsch-Josza algorithm \cite{DeutschJ92}, which works only for functions which are either balanced or constant.

Our task will be to find
in which of the four sets $S_{0000},S_{0011},S_{0101}$ and $S_{1001}$
lie each of the functions $f_{0}\oplus f_{1}$ and $f_{0}\oplus f_{2}$,
using quantum circuits that query $f_{0},f_{1}$ and $f_{2}$ at most
once each. Our score will be given by the average probability of successfully guessing both correctly.

\subsection{Score without cloning}

The best no-cloning strategy I have found goes as follows. Firstly,
note that if $f_{0}=h_{0010}$ then both $f_{1}$ and $f_{2}$ must be in
set $S_{1}$, because of the constraints given by
eq. (\ref{constraints}); similarly, if $f_{0}$ is either $h_{0101}$ or
$h_{1001}$, then $f_{1}$ and $f_{2}$ must be in set $S_{2}$. Since we
have drawn the function $f_{0}$ with a uniform random distribution, we will have both functions
$f_{1}$ and $f_{2}$ in set $S_{2}$ with probability $p=2/3$. We will
assume that this is the case; then we can discriminate between the two
possibilities for $f_{0}$ with a single, classical function
call. Furthermore, by using the quantum circuit in fig. \ref{fig
discri} twice (once with each of $f_{1}$ and $f_{2}$) we can
distinguish the four possibilities for functions $f_{1}$ and $f_{2}$.

\begin{figure}
\begin{center}
{\includegraphics[scale=0.7]{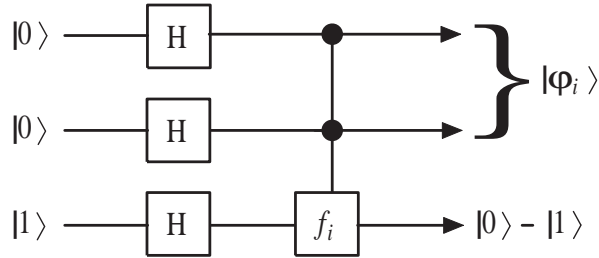}}
\caption[Quantum circuit to distinguish functions $f_i$.]{\label{fig discri}If function $f_{i}$ is guaranteed to be
either in set $S_{1}$ or in $S_{2}$, then this quantum circuit can be used to distinguish between
the four possibilities in each set. This is done by measuring the
final state $\left| \phi_{i}\right\rangle
=\sum_{x=00}^{11}(-1)^{f_{i}(x)}\left|  x\right\rangle $ in one of two
orthogonal bases, depending on which set contains $f_{i}$. The $H$
operations are Hadamard gates.}
\end{center}
\end{figure}

This happens because depending on which function in set $S_{2}$ was
queried, this quantum circuit results in one of the four orthogonal states
\begin{equation}
\left|  \phi_{i}\right\rangle
=\sum_{x=00}^{11}(-1)^{f_{i}(x)}\left|  x\right\rangle .
\end{equation}
This allows us to determine
functions $f_{0},f_{1}$ and $f_{2}$ correctly with probability
$p=2/3$, in which case we can determine which sets contain
$f_{0}\oplus f_{1}$ and $f_{0}\oplus f_{2}$ and accomplish our
task. Even in the case where our initial assumption about $f_{0}$ was
wrong, we may still have guessed the right sets by chance; a simple
analysis shows that our chances of getting both right this way are
only $1/16$. On average, then, by using this no-cloning strategy we
obtain a score:

\[
p_{1}=\frac{2}{3}+\frac{1}{3}\cdot\frac{1}{16}=0.6875.
\]
This is the best no-cloning score I could find for this task, and I
conjecture that it is optimal among no-cloning approaches.

\subsection{Score using cloning}

We can do better than that with quantum cloning. The idea now is to
devise a quantum circuit that queries function $f_{0}$ only once,
makes two clones of the resulting state and then queries functions
$f_{1}$ and $f_{2}$, one in each branch of the computation. Since we
have some information about the state produced by one query of
$f_{0}$, the best cloning strategy will no longer be a UQCM, but the
probabilistic cloning process described in section \ref{sec probclo}.

\begin{figure}

\begin{center}

{\includegraphics[scale=0.6]{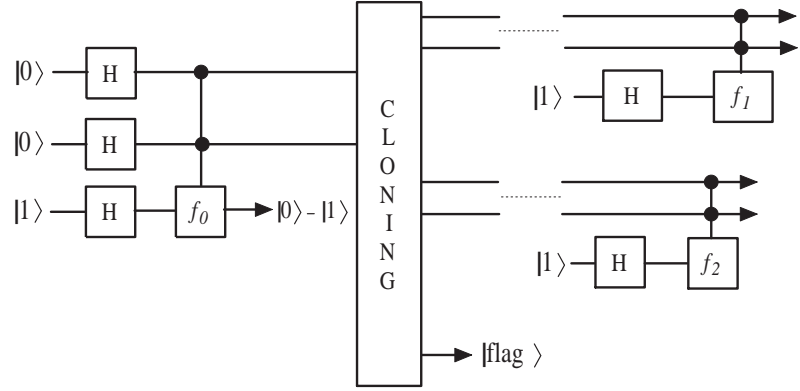}}

\caption[Probabilistic cloning in quantum computation.]{\label{fig probclo}The cloning procedure in this circuit is probabilistic; a
measurement on the state $\left|  Flag\right\rangle $ tells us whether
the cloning succeeded. If the cloning is successful we let the clones
go through the rest of the circuit, yielding output states $\left|
\phi_{i}\right\rangle=\frac{1}{4}\sum_{x=00}^{11}(-1)^{f_{0}(x)\oplus
f_{i}(x)}\left|x\right\rangle $, $(i=1,2)$. These states can be
measured in the basis defined by eqs. (\ref{h0000})-(\ref{h1001}) to
unambiguously decide which of the four sets $S_{0000},S_{0011},S_{0101}$ or $S_{1001}$ contains $f_{0}\oplus f_{i}$.}
\end{center}
\end{figure}

The quantum circuit that we apply to solve this problem is given in
figure \ref{fig probclo}.

Immediately after querying function $f_{0}$, we have one of three
possible linearly independent states (each corresponding to one of the possible $f_{0}$'s):

\begin{align}
\left|  h_{0010}\right\rangle  &  \equiv\frac{1}{2}\left[  \left|
00\right\rangle +\left|  01\right\rangle -\left|  10\right\rangle +\left|
11\right\rangle \right]  ,\label{f0state1}\\
\left|  h_{0101}\right\rangle  &  \equiv\frac{1}{2}\left[  \left|
00\right\rangle -\left|  01\right\rangle +\left|  10\right\rangle -\left|
11\right\rangle \right]  ,\label{f0state2}\\
\left|  h_{1001}\right\rangle  &  \equiv\frac{1}{2}\left[  -\left|
00\right\rangle +\left|  01\right\rangle +\left|  10\right\rangle -\left|
11\right\rangle \right]  . \label{f0state3}
\end{align}

As we have seen in section \ref{sec probclo}, we can build probabilistic cloning machines with different cloning
efficiencies for
each of the states (\ref{f0state1})-(\ref{f0state3}). Theorem 4.2 (due
to Duan and Guo and reviewed in section \ref{sec probclo}) provides us with inequalities that allow us to derive
achievable efficiencies for the probabilistic cloning process. I did
a numerical search that yielded the following achievable efficiencies
for probabilistically cloning the states in eqs. (\ref{f0state1})-(\ref{f0state3}):
\begin{align}
\gamma_{1}  &  \equiv\gamma(\left|  h_{0010}\right\rangle
)=0.14165,\label{gama1}\\
\gamma_{2}  &  \equiv\gamma(\left|  h_{0101}\right\rangle )=\gamma(\left|
h_{1001}\right\rangle )=0.57122. \label{gama2}
\end{align}

After the cloning process we can measure a `flag' subsystem and know
whether the cloning was successful or not. For this particular cloning
process, the probability of success is, on average,
$p_{\text{success}}=(\gamma_{1}+2\gamma_{2})/3\simeq0.4280$. Let us
suppose that it was successful. Then each of the cloning branches goes
through the second part of the circuit in fig. \ref{fig probclo}, to yield one of the
four orthogonal states:

\begin{align}
\left|  h_{0000}\right\rangle  &  \equiv\frac{1}{2}\left[  \left|
00\right\rangle +\left|  01\right\rangle +\left|  10\right\rangle +\left|
11\right\rangle \right]  ,\label{h0000}\\
\left|  h_{0011}\right\rangle  &  \equiv\frac{1}{2}\left[  \left|
00\right\rangle +\left|  01\right\rangle -\left|  10\right\rangle -\left|
11\right\rangle \right]  ,\label{h0011}\\
\left|  h_{0101}\right\rangle  &  \equiv\frac{1}{2}\left[  \left|
00\right\rangle -\left|  01\right\rangle +\left|  10\right\rangle -\left|
11\right\rangle \right]  ,\label{h0101}\\
\left|  h_{1001}\right\rangle  &  \equiv\frac{1}{2}\left[  -\left|
00\right\rangle +\left|  01\right\rangle +\left|  10\right\rangle -\left|
11\right\rangle \right]  ,\label{h1001}%
\end{align}
which can be discriminated unambiguously. We obtain state
$\left|h_{0000}\right\rangle $ if and only if the combined function
$f_{0}\oplus f_{i}$ is one of the two in set $S_{0000}$, as can be
checked by calculating the effect of the circuit in fig. \ref{fig probclo} for all
possible $f_{0},f_{1}$ and $f_{2}$. The situation is similar for the
other three states; the detection of each of them signals precisely
which one of the four sets $\{S_{0000},S_{0011},S_{0101},S_{1001}\}$
contains $f_{0}\oplus f_{i}$ . As a result, if the cloning process is
successful, we manage to accomplish our task.

However, the cloning process will fail with probability $(1-p_{\text{success}
})$. If this happens, a simple evaluation of the posterior
probabilities for function $f_{0}$ shows that it is more likely to be
$h_{0010}$ than the other two, thanks to the relatively low cloning
efficiency for \ the state in eq. (\ref{f0state1}), in relation to the
states in eqs. (\ref{f0state2}) and (\ref{f0state3}) [see eqs. (\ref{gama1}-\ref{gama2})]. If we then guess that $f_{0}=h_{0010}$, we will be right with probability

\[
p_{0010}=\frac{(1-\gamma_{1})}{(1-\gamma_{1})+2(1-\gamma_{2})}\simeq0.5002.
\]

What is more, we are still free to design quantum circuits to obtain
information about $f_{1}$ and $f_{2}$, since at this stage we still
have not queried them. Given our guess that $f_{0}=h_{0010}$, only the
four functions in $S_{1}$ can be candidates for $f_{1}$ and $f_{2}$,
because of the constraints given by eq. (\ref{constraints}). These four
possibilities can be discriminated unambiguously by running a circuit
like that of fig. \ref{fig discri} twice, once with $f_{1}$ and once with
$f_{2}$. The circuit produces one of four orthogonal states, each
corresponding to one of the four possibilities for $f_{i}$. Therefore,
if our guess that $f_{0}=h_{0010}$ was correct, we are able to find
the correct $f_{1}$ and $f_{2}$ and therefore accomplish our task. In
the case that $f_{0}\neq h_{0010}$ after all, we may still have
guessed the right sets by chance; a simple analysis shows that this
will happen with probability $1/16$.

The above considerations lead to an overall probability of success given by
\begin{eqnarray}
p_{2}&=&p_{\text{success}}+(1-p_{\text{success}})\left[  p_{0010}+(1-p_{0010}
)\frac{1}{16}\right]\nonumber\\
&\simeq&0.7320>p_{1}=0.6875,
\end{eqnarray}
thus showing that our cloning approach is more efficient than the
previous one, which does not use cloning. I have not proven that the
first approach is the most efficient among those that do not resort to cloning, but I conjecture that it is.

\subsection{Remarks}

Besides the larger probability of obtaining the correct result, our
cloning approach offers another advantage: the measurement of the
`flag' state allows us to be confident about having the correct result
in a larger fraction of our attempts.

For the probabilistic cloning
machines described above this fraction was $\simeq0.428$, but \ this
can be improved by choosing a different cloning machine, characterised
by $\gamma_{1}=0.3485,\gamma_{2}=0.5258$. This latter machine signals
a guaranteed correct result in a fraction
$(\gamma_{1}+2\gamma_{2})/3\simeq0.467$ of the runs.

The best no-cloning approach for obtaining these guaranteed correct results
would involve unambiguous discrimination of the function $f_{0}$,
followed by the distinction among the four possibilities for functions
$f_{1}$ and $f_{2}$ (this second step is simple if we know $f_{0}$ for
certain). Theorem 4.3 (see section \ref{sec probclo}), due to Duan and
Guo, provides us with a tool to
numerically determine the best efficiency for unambiguous
discrimination of $f_{0}$. A numerical search shows that this can
be done only with efficiency $\leq1/3$, and therefore this is the
limit for the fraction of runs for which we can obtain a guaranteed
correct result for the task at hand, if we do not resort to
cloning.

\section{Chapter summary}

I presented two examples of quantum computation tasks which
attain optimal performance by using an intermediate quantum cloning
step. In both examples, I showed that preserving and distributing
quantum information through a cloning process offers advantages over
extracting classical information mid-way in the quantum computation,
for example using optimal state estimation.

The first example models the situation in which we need to do $M$
quantum computations which have a first part in common. When this first part is lengthy or otherwise costly to compute, we can obtain
good approximations of all $M$ ideal output states by running the
first part only once, cloning the resulting state and feeding the
clones to the multiple subsequent computations. It was shown that the
average fidelity of the resulting approximate outputs is maximised
when we use an optimal, symmetric UQCM.

When performing quantum computations, we may often need to
distribute quantum information contained in a state about which we
have some partial information. In such cases, UQCM's may not be
the optimal solution. The second example I presented shows such a
case, in which we use a probabilistic, state-dependent cloning procedure
to distribute quantum information in the middle of a quantum
computation.

Of course, general quantum algorithms are already known to  manipulate
quantum information, distributing it among different parts of the
quantum memory during a computation. In this chapter I have shown that
quantum cloning can be taken as a natural tool to do this quantum information distribution, in order
to optimise our use of computational resources. It would be
interesting to find other tasks that could profit from cloning,
perhaps by combining already known quantum algorithms with some
intermediate cloning steps.


\chapter{On quantum random access codes\label{chap qcc}}

In section \ref{sec intro qracs} I described a quantum information
application called quantum random access codes (QRAC's). In this
chapter I will analyse some aspects of this application, stressing the
fundamental quantum characteristics behind its
better-than-classical performance for a type of information processing
task.

In section \ref{sec qraccontext} I will show the role quantum
contextuality plays in the simplest QRAC. In section \ref{sec
qracnonloc} I change the QRAC somewhat to show how it can make use of
quantum non-locality instead. Next, in section \ref{sec qracapplic} we
will see the relation between the $2\overset{p}{\rightarrow}1$ QRAC
and two quantum information protocols known as remote state
preparation and dense coding.

The simplicity of the $2\overset{p}{\rightarrow}1$ QRAC suggests it
may have been already derived in other contexts, with other interests
in view. In section \ref{sec twoqcc} I will show that the quantum
communication complexity protocol introduced by Buhrman, Cleve and van
Dam in \cite{BuhrmanCvD97} is
actually equivalent to this simple QRAC.  This shows the two
apparently dissimilar protocols are actually one and the same, and
rely on the same quantum characteristics for their
higher-than-classical performance.

Next I will derive some bounds on general QRAC's efficiency directly
from the invariant information of Brukner and Zeilinger, which we
overviewed in section \ref{sec invinfo}. 

Finally, in section \ref{sec 2qracfeasibility} I will briefly sketch
some of the experimental requirements for demonstrating the simplest
QRAC's in the laboratory.

\section{QRAC's and contextuality \label{sec qraccontext}}

In this section I will show how the higher-than-classical performance
of the simplest QRAC can be traced back to quantum
contextuality. Later we will see its relation to quantum non-locality
as well.

In section \ref{sec 21qrac} we have seen that the simplest classical
random access code consists of asking Alice to encode her two-bit
string $b_0b_1$ into a
single-bit message, which Bob must decode to read one of Alice's
bits, but not both at the same time. As we have seen, the optimal classical probability can be
attained through the obvious protocol consisting of just sending one
of the bits, which Bob can then always `decode'. If he is required to
read the unsent bit, then he just guesses it, getting it right half of
the time. This trivial code results in an average probability of success of $p_c=\tfrac{1}{2}\left(1+\tfrac{1}{2}\right)=\tfrac{3}{4}$.

We have also seen how quantum mechanics enables us to perform the task
with a higher probability of success by sending a qubit, instead of a
classical bit. Alice
prepares one of four non-orthogonal states of a qubit depending on her
two-bit string, and sends it to Bob. By appropriate measurements
on this single qubit, Bob manages to read out Alice's first bit or the
second, but not both, with a high probability of
success $p_q \simeq 0.85$ (see figure \ref{fig applic1}).

The preparation of the encoding state can be done in a
two-step process, in which Alice first creates an entangled state of
dimension $d \ge4$, and then prepares a qubit state through suitable
projective measurements on a subsystem.

Let us now investigate one possible way of
doing this, which will make apparent the role of quantum contextuality
in the higher-than-classical performance of this task. Alice starts by preparing
\begin{equation}
\left|\psi^{+}\right\rangle=\frac{1}{\sqrt{2}}\left(
\left|0_{A}0_{B}\right\rangle + \left|1_{A}1_{B}\right\rangle
\right).\label {psiplus}
\end{equation}

Let us use the same qubit parametrisation as in eq. (\ref{psithetaphi}):
\begin{equation}
\left|\psi(\theta,\phi)\right\rangle=\cos(\tfrac{\theta}{2})\left|0\right\rangle+\exp(i\phi)\sin(\tfrac{\theta}{2})\left|1\right\rangle.
\end{equation}

In order to prepare qubit $B$ in the desired state, she measures
qubit $A$ with one of the two observables used for maximal violation
of the CHSH inequality (see section \ref{sec chshineq}). As discussed in section \ref{sec chshineq}, these projectors must alternate on the same plane in the Bloch sphere, separated by angles of $\pi/4$ radians. For ease of comparison with fig. 4.1 depicting the encoding states, we can choose the following projectors. If Alice's
bit-string $b_0b_1=00$ or $11$ she will measure projector $\hat{P}_{a2}=\hat{P}(\pi/2,\pi/4)$;
if $b_0b_1=01$ or $10$ she measures $\hat{P}_{a1}=\hat{P}(\pi/2,7\pi/4)$. She
then performs a unitary rotation by $\pi$ radians around the $z$ axis
on qubit $B$, conditional on the outcome of her measurement. This
enables her to deterministically prepare any of the four states used
in the protocol.

As in the CHSH test, Bob's readout measurements are made using
observables different from those used by Alice to prepare the
states. He will measure qubit $B$ with projector
$\hat{P}_{b1}=\hat{P}(\pi/2,0)$ if he want to read bit $b_1$, and
$\hat{P}_{b2}=\hat{P}(\pi/2,\pi/2)$ if he want to read bit $b_2$. As we discussed in section \ref{sec 21qrac}, the measurement outcomes give him a
$2{\rightarrow}1$ QRAC a probability of success which is
higher than the optimal classical analogue.

This way of presenting the $2{\rightarrow}1$ QRAC shows
that its higher-than-classical performance has something to do with a
CHSH inequality test. What exactly is the relation? Since there are no
simultaneous measurements of different observables here, we cannot
claim here that the quantum advantage comes from non-locality.

Now I will show that the quantum advantage comes directly from
contextuality. First, let us note that when performing the protocol $N$ times on multiple pairs of qubits in
state (\ref{psiplus}), about $N/2$ times Alice will need to do the
$\pi$ rotation on Bob's qubit, and she will not in the other half of
the particles.

The subset of pairs not rotated by $\pi$ remain in
state (\ref{psiplus}), and the measurements performed are those that
maximally violate the CHSH inequality. Since a strict locality
condition was not imposed here, the CHSH inequality violation in this
case points towards violation of a weaker assumption, that of
non-contextuality of experiment outcomes.

It is easy to check that a $\pi$ rotation around the $z$-axis of qubit
$B$ takes state (\ref{psiplus}) into
\begin{equation}
\left|\psi^{-}\right\rangle=\frac{1}{\sqrt{2}}\left(
\left|0_{A}0_{B}\right\rangle - \left|1_{A}1_{B}\right\rangle
\right).\label {psiminusyep}
\end{equation}
It is true that Alice's measurement happens before Bob's, and
adherents of the `state collapse' interpretation of quantum theory can say the
state is no longer (\ref{psiminusyep}) when Bob measures it. Nevertheless,
the measurement statistics and correlations are the same as those
obtainable from state (\ref{psiminusyep}) under strict locality
conditions. This state also violates maximally the CHSH inequality
(\ref{CHSH ineq}).

Note that while each subset maximally violates the CHSH inequality,
taken together there is no violation. This, however, is no problem,
since Alice and Bob actually need to get together and compare notes in
order to verify the stronger-than-classical correlations -- then they
can check whether a rotation was performed or not, and accordingly
which state was effectively measured. This naturally leads to separate
analyses of the two subsets, resulting in a maximal violation of the
CHSH inequality.

We have just seen that the measurements violate the CHSH inequality,
but still are necessarily local, as they depend on Alice preparing
qubit $B$ and then sending it to Bob. As we have seen in section
\ref{sec nonlocality}, quantum non-locality tests are a special type
of contextuality tests. In particular, when any non-locality test is
performed with time-like separation between the measurements, we
cannot rule out a local explanation for the outcomes. But despite
being compatible with a local hidden-variable explanation, the
observed outcomes violate the non-contextuality hypothesis. In other words, the
time-like separated CHSH inequality measurements involved in this QRAC
protocol can be interpreted as a contextuality test. For further
discussion on experimental feasibility of contextuality tests which
exploit this relation between non-locality and contextuality see
\cite{SimonZWZ00,MichlerWZ00}.

Next we will have a look at a simple modification of the protocol,
which turns it into an equivalent of a CHSH non-locality test.

\section{QRAC's and non-locality \label{sec qracnonloc}}

Taking into consideration our discussion above of the
$2{\rightarrow}1$ QRAC, it is easy to change the protocol
somewhat and obtain the equivalent of a genuine CHSH inequality test.

The QRAC we presented above takes its advantage from using quantum
instead of classical communication. There is another way of obtaining
a quantum version of these codes, by keeping the  classical
communication, but using the resource represented by measurements on previously shared entangled states. By exploring the strong
correlations between measurements on these entangled pairs of
particles, the parties can save on classical communication use.

The simplest such entanglement-based QRAC is again a
$2{\rightarrow}1$ quantum code using classical
communication of one bit from Alice to Bob (the same amount as the classical
code), plus measurements on one previously shared maximally entangled
state (\ref{psiplus}). The measurements are those presented above,
i.e. the same as for a CHSH test. The difference from the QRAC
protocol described in the previous section is that here Alice never
performs the conditional $\pi$ rotation. This enables them to share the state beforehand, and even do
all measurements under strict locality conditions (i.e. with
space-like separation). As we have seen, however, half of the time Bob
gets exactly the opposite result of what he should get, for a
successful quantum protocol.

It is here that the classical communication part of the protocol comes in. With a single bit of communication, Alice can tell Bob
whether he should have performed the $\pi$ rotation she did not have
a chance to. While he cannot physically do this operation on his qubit after measurement, he can obtain the correct
outcome just by flipping whatever outcome he got, conditionally on
Alice's one-bit classical message. By doing this, he
recovers the outcomes he would have obtained using the four-state
encoding of the quantum-communication version of the
QRAC. This enables him to succeed with a higher-than-classical
probability of success $p_q=\cos^{2}(\pi/8)\simeq 0.85$.

We thus see that it is the non-locality of the measurement outcomes on
a maximally entangled state that allows for an entanglement-based QRAC
with higher-than-classical performance.

\section{Relations with other applications \label{sec qracapplic}}

We have just seen that by using a single bit of classical
communication plus entanglement, Alice and Bob could contrive a
protocol enabling Bob to deterministically obtain the precise state he
requires for a QRAC with high probability of success. This is a specific
example of a general process known as remote state preparation.

Remote state preparation is a protocol by which Bob is able to recover
an arbitrary quantum state \textit{known} to Alice, by suitable
measurements on some previously shared entangled states, plus
classical communication only. The simple protocol we have used above
to remotely prepare one of the four states necessary for the QRAC is
actually sufficient to prepare any state on the equator of the Bloch
sphere. This was pointed out by Pati in \cite{Pati01}.

Some more general remote state preparation protocols were developed by
Bennett and co-workers in \cite{BennettDSSTW01}. These include
protocols for general qubit states (not necessarily on the Bloch
sphere equator) and general higher-dimensional systems.

There is also a relation between the $2{\rightarrow}1$ QRAC
and a process known as quantum dense coding \cite{BennettW92}. Quantum
dense coding allows Alice to send Bob two bits of information with the
transmission of a single qubit, supplemented by previously shared
entanglement. Let us see how we can use dense coding in a type of
QRAC, and the caveats involved.

Dense coding is possible because a two-qubit Bell state can be locally
transformed into any of the four orthogonal Bell states at will. After
sharing a Bell state with Bob, Alice can encode a two-bit value by
locally transforming the two-qubit state into one out of the four
possible orthogonal Bell states. She can then send her qubit to Bob through a quantum channel. With both
qubits in his hands, he can discriminate the four orthogonal states
and read Alice's two-bit message.

In the previous sections we have seen two ways of implementing a
$2{\rightarrow}1$ QRAC: either by sending a single qubit
from Alice to Bob, or by sending a bit encoding the outcome of a local
measurement on previously shared entangled qubits. Dense coding offers
a third way of performing a QRAC, which combines the ingredients of
these two approaches. Using dense coding, Alice can
send a two-bit message to Bob using \textit{both} shared entanglement
and qubit communication. Interestingly, with both resources they can
do better than with either isolated resource, because now Bob can
decode both bits with probability $p_q=1$.

In order to do so, however, Bob's decoding procedure requires Bell
basis discrimination, and this is hard to do experimentally with
photons, as a coupling between them is required. For this reason, in
section \ref{sec 2qracfeasibility} below I will only consider the
experimental requirements for a realization of the
$2{\rightarrow}1$ QRAC using either entanglement or qubit
communication.
 
\section{QRAC's and quantum communication complexity \label{sec twoqcc}}

In this section we will see that an altogether different quantum
information processing protocol is actually equivalent to the simple
$2{\rightarrow}1$ QRAC we have been discussing. In \cite{BuhrmanCvD97}
 Buhrman, Cleve and van Dam introduced a two-party quantum
communication complexity protocol which was shown to be better than
its optimal classical counterpart. Let us now review this
communication complexity task, and show how the $2{\rightarrow}1$ QRAC
solves it.

\subsection{A simple two-party communication complexity task
\label{sec BCvD task}}

Let us now review the communication complexity problem proposed in
\cite{BuhrmanCvD97}. Let Alice and Bob each have two-bit strings
$x=x_1x_2$ and $y=y_1y_2$, respectively. Alice is allowed to send just
a single bit to Bob, who must then compute:
\begin{equation}
g(x,y)=x_1\oplus y_1\oplus(x_2\wedge y_2),
\end{equation}
where $\oplus$ denotes sum modulo 2, and $\wedge$ is just the AND
Boolean operator. The bit-strings $x$ and $y$ are drawn from a random
uniform distribution. A quantum protocol for this problem was
presented in \cite{BuhrmanCvD97} as the first two-party example of
entanglement-enhanced communication complexity.

It is clear that a single bit of communication from Alice to Bob is
not enough for him to always compute $g(x,y)$ successfully. What
interests us now is to find the probability of obtaining the correct
value of $g(x,y)$ in three communication complexity scenarios we have
sketched in section \ref{sec qccintro}. These involve using: a bit of
communication (classical model); a qubit of communication
(quantum-communication model); or a bit of communication plus
measurements on entangled pairs (entanglement-based model).

In the next section I will show that this communication complexity task is equivalent to a $2 \to 1$ random access code, and therefore also has an optimal classical probability of success $p_c=3/4$.

\subsection{Quantum protocols \label{sec 2qccqprot}}

Here we will see that the $2{\rightarrow}1$ QRAC we have
studied above can be used to solve this communication complexity
problem.

What exactly must Bob know in order to solve the task? He has some
partial information about function $g(x,y)$, given by his own data
$y$. For each value of $y$, the table below shows what $g(x,y)$ reduces
to, and lists one particular bit value that would enable Bob to compute $g(x,y)$:

\begin{center}
\begin{tabular}{|c|c|c|} \hline
y & $g(x,y)=x_1\oplus y_1\oplus(x_2\wedge y_2)$ & Bob needs to know \\
\hline\hline
00 & $x_1$ & $x_1$\\ \hline
01 & $x_1\oplus 1$ & $x_1$\\ \hline
10 & $x_0\oplus x_1$ & $x_0\oplus x_1$ \\ \hline
11 & $x_0\oplus x_1\oplus 1$ & $x_0\oplus x_1$ \\ \hline
\end{tabular}
\end{center}

We see that Bob can compute $g(x,y)$ if there is some way for him to
choose which bit to read: either $x_1$ or $x_0\oplus
x_1$. This can be achieved with average probability
$p_q=\cos^{2}(\pi/8)$ with the $2{\rightarrow}1$
QRAC I have discussed above. As we have seen in sections \ref{sec qraccontext} and \ref{sec qracnonloc}, this can be done either by
performing measurements on a maximally entangled pair of qubits (and
sending a bit of classical communication) or by directly sending a
qubit instead of a classical bit from Alice to Bob. These are the
entanglement-based and the quantum-communication protocols, respectively.

This solution using the $2{\rightarrow}1$
QRAC is equivalent to the original entanglement-based solution
presented in \cite{BuhrmanCvD97}. Its advantage is that it uncovers more clearly
the real nature of the problem. Now it is easier to see that this
quantum communication complexity protocol also derives its
higher-than-classical performance from quantum contextuality and
non-locality.

\section{Invariant information and QRAC's \label{sec invinfoqrac}}

In section \ref{sec invinfo} we have reviewed the definition and some
properties of the invariant quantum information $I_{total}$ introduced by
Brukner and Zeilinger. In that introduction, I stressed that
$I_{total}$ could
be motivated with a very simple application in mind, which involved
quantifying the entropy, or information, of mutually exclusive
measurements that can be performed on a single quantum system. The
measurements we considered were projective measurements on different
mutually unbiased bases (MUB's) \footnote{For the definition of MUB see section \ref{sec definvinfo}.}. Motivated by the appearance of two
MUB's for a qubit in the preceding sections of this chapter, here we
will investigate what $I_{total}$ can tell us about QRAC's.

The most symmetrical QRAC construction involves having Bob decode the
state sent by Alice through projection on a set of alternative
MUB's. The choice of MUB is made depending on which information Bob
wants to decode from the state sent by Alice. The simple $2 \to 1$
QRAC presented in section \ref{sec 21qrac} is an example of
that. Alice encodes her two bits of information into four states, each
of which has a high overlap one basis vector from each MUB. At the same time, each encoding state has a
small overlap with the other two vectors, representing the outcomes
that Alice does not want Bob to obtain. As we have seen, this simple
scheme can reach a probability of success $p_q=\cos^2(\pi/8) \simeq
0.85$, which is higher than the optimal classical protocol (which
achieves only $p_c=0.75$).

If we did not know about this $2 \to 1$ QRAC, however, it would
still be possible to obtain a tight bound for $p_q$ just by using
Brukner and Zeilinger's invariant information $I_{total}$ [see
eq. (\ref{Itotalddim})]. In this section we will see how $I_{total}$ can
be used to obtain bounds on the probability of success of a wide range
of QRAC's.

\subsection{Bounding the efficiency of simple QRAC's}

To obtain a bound on the quantum probability of success $p_q$ for the $2 \to 1$
qubit code we have analysed in section \ref{sec 21qrac}, let us
start by recalling the expression of $I_{total}$ for measurements over
two MUB's for a qubit:
\begin{equation}
I_{total}=\sum_{j=1}^{2}I(p_1^{(j)},p_2^{(j)}) =\sum_{j=1}^{2}\sum_{i=1}^{2} \left(p_i^{(j)} -\frac{1}{2} \right)^2, \label{I2b1qubit}
\end{equation}
where $p_i^{(j)}$ are the probabilities of the projection over MUB $j$
to have outcome $i$. Let us assume there exists an encoding which
maximises this information (achieving $I_{total}=1/2$), by equally
distributing the information over the two alternative
measurements. This would mean that the information for each
measurement would be half the maximal, or $I=1/4$. This is achieved
when the probability of success is 
\begin{equation}
I(p,1-p)=(p-1/2)^2+((1-p)-1/2)^2=1/4 \Rightarrow p=\frac{1}{2}+\frac{\sqrt{2}}{4} \simeq 0.85.
\end{equation}

This is actually achieved by the $2 \to 1$ QRAC we have reviewed
in section \ref{sec 21qrac}. We thus see that equally distributing the
invariant information over the two MUB's automatically allows us to
compute a tight bound for the efficiency of the $2 \to 1$
QRAC we have discussed previously. This also shows that this QRAC is
optimal.

We can also apply this simple reasoning to a $3 \to 1$ QRAC
which encodes three bits into a single qubit. This QRAC is a
straightforward extension of the $2 \to 1$ QRAC from the equator to
the whole surface of the Bloch sphere. It was attributed
to Chuang in reference \cite{AmbainisNT-SV99}.

\begin{figure}
\begin{center}
{\includegraphics[scale=0.50]{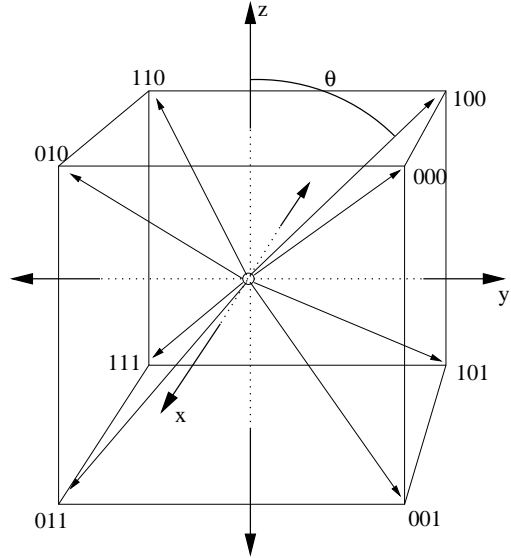}}
\caption[Encoding states for the qubit $3 \to 1$ QRAC. ]{\label{fig qcca}Encoding states for the $3 \to 1$ QRAC using a single
qubit. Alice prepares one out of eight states on the vertices of a
cube inscribed within the Bloch sphere, depending on her three-bit
string. The angle $\theta$ is such that
$\cos^2(\theta/2)=1/2+\sqrt(3)/6 \simeq 0.79$, which is the probability
of Bob correctly decoding a single bit out of the three.}
\end{center}
\end{figure}

With three bits, Alice has $2^3=8$ possible bit-strings $b_0b_1b_2$. For each
possibility she will prepare one particular state from the set
depicted in figure \ref{fig qcca}. These states lie on the vertices of a
cube inscribed within the Bloch sphere. If Bob wants to read out bit
$b_0$ he measures along the $x$-axis and associates a positive result
with $b_0=0$. To read bits $b_1$ [$b_2$] he measures along the
$y$-axis [$z$-axis] and again associates a positive result with
$b_1=0$ [$b_2=0$]. It is easy to see Bob's probability of success is
given by $p_q=\cos^2(\theta/2)=1/2+\sqrt(3)/6 \simeq 0.79$, where the
angle $\theta$ is given in figure \ref{fig qcca}. The optimal classical $3 \to 1$
random access code succeeds only with probability $p_c=0.75$, as can
be checked easily through a search over all deterministic protocols.

Also in this case, an equipartition of the invariant information
$I_{total}$ over the three MUB's bounds $p_q$ tightly. Dividing the total information of a qubit by three we get
that each two-outcome measurement can have at most
$I=\frac{1}{2}\frac{1}{3}=1/6$. To find explicitly the bound on
$p_1$ from the invariant information, we just need to solve:
\begin{equation}
I=(p-1/2)^2+((1-p)-1/2)^2=1/6 \Rightarrow p=\frac{1}{2}+\sqrt(3)/6 \simeq 0.79,
\end{equation}
which shows that equipartition of the invariant information over the
three MUB's again gives a tight bound on the efficiency of a $3 \to 1$
QRAC.

\subsection{More general bounds \label{sec 21trit}}

We can now ask whether $I_{total}$ can be used to obtain tight bounds
for more general QRAC's. In this section we will see that in the case of two and three MUB's for a
qubit, it is a mere geometrical coincidence that the maximum overlaps
between encoding and decoding states given by the invariant
information can be achieved. In higher-dimensional Hilbert spaces
there will not necessarily exist such symmetries allowing us to find
encoding states with the desired overlaps with the decoding projective
measurements. This difficulty can be illustrated with a simple
example.

Let us now work out a QRAC which encodes two trits (i.e. three-level
systems) into a single qutrit. Bob's measurements will consist of
three-outcome projective measurements over two MUB's. Alice's encoding
procedure involves preparing one of nine states, depending on the two
trits she has.

Let us use two MUB's for a qutrit presented in
\cite{BechmannPasquinucciP00}. They are given by:
\begin{equation}
\begin{array}{l}
|\alpha_1\rangle=(\omega|0\rangle+|1\rangle+|2\rangle)/\sqrt{3},\\ 
|\beta_1\rangle=
(|0\rangle+\omega|1\rangle+|2\rangle)/\sqrt{3},\\
|\gamma_1\rangle=
(|0\rangle+|1\rangle+\omega|2\rangle)/\sqrt{3},
\end{array}
\end{equation}
and
\begin{equation}
\begin{array}{l}
|\alpha_2\rangle=(\omega^{*}|0\rangle+|1\rangle+|2\rangle)/\sqrt{3},\\ 
|\beta_2\rangle=
(|0\rangle+\omega^{*}|1\rangle+|2\rangle)/\sqrt{3},\\
|\gamma_2\rangle=
(|0\rangle+|1\rangle+\omega^{*}|2\rangle)/\sqrt{3},
\end{array}
\end{equation}
where $\omega=\exp( 2\pi i/3)$.

We can search now for the encoding states Alice will use. To start
with, we must find a state which maximises the sum of the overlaps
with $|\alpha_1\rangle$ and $|\alpha_2\rangle$. A state with this
property is:
\begin{equation}
|\psi_{\alpha_1\alpha_2}\rangle= a|0\rangle +
 \sqrt{(1-a^2)/2}|1\rangle +\sqrt{(1-a^2)/2}|2\rangle,\label{psialphaalpha}
\end{equation}
with $a=-\left(\frac{1}{2}+\frac{\sqrt{3}}{6}\right)^{\frac{1}{2}}$.

For the decoding procedure, Bob will project the qutrit into the first
or second basis depending on which trit he wants to read. The state
(\ref{psialphaalpha}) above will result in an outcome $|\alpha_1\rangle$
or $|\alpha_2\rangle$ with probability
\begin{equation}
\left|\left\langle\psi_{\alpha_1\alpha_2}|\alpha_1
\right\rangle\right|^2=\left|\left\langle\psi_{\alpha_1\alpha_2}|\alpha_1
\right\rangle\right|^2=\frac{1}{2}+\frac{\sqrt{3}}{6} \simeq 0.789.
\end{equation}

Given the symmetries present in the two MUB's above, it is easy to
design unitaries which we can apply to state (\ref{psialphaalpha}) to obtain
the other eight encoding states, each of which has the same large
overlap with one basis state from each MUB. Let us now define two
unitaries which will be useful for that:
\begin{equation}
V:\left\{ 
\begin{array}{r@{}l}
&|0\rangle  \to |1\rangle \\
&|1\rangle  \to |2\rangle \\
&|2\rangle  \to |0\rangle \\
\end{array}\right. ,
 \mbox{ and } U:\left\{ 
\begin{array}{r@{}l}
&|0\rangle  \to |0\rangle \\
&|1\rangle  \to \omega|1\rangle \\
&|2\rangle  \to \omega^{*}|2\rangle \\
\end{array}\right. .\label{Vtrit}
\end{equation}
The unitaries $U$ and $V$ above transform the basis states in the two
MUB's in a useful way. We can apply some combinations of them on the
encoding state (\ref{psialphaalpha}) to obtain the other encoding
states, as we see in table \ref{table qractrit}.
\begin{table}\caption{\label{table qractrit}Encoding states for $2
\to 1$ QRAC with a qutrit}
\center
\begin{tabular}{|c|c|} \hline
Unitary applied to $|\psi_{\alpha_1\alpha_2}\rangle$ & Encoding state obtained\\
\hline\hline
\mbox{$1 \hspace{-1.0mm}  {\bf l}$} &
$|\psi_{\alpha_1\alpha_2}\rangle$\\ \hline
$V$ & $|\psi_{\beta_1\beta_2}\rangle$\\ \hline
$V^2$ & $|\psi_{\gamma_1\gamma_2}\rangle$\\ \hline
$U$ & $|\psi_{\gamma_1\beta_2}\rangle$\\ \hline
$VU$ & $|\psi_{\alpha_1\beta_2}\rangle$\\ \hline
$V^2U$ & $|\psi_{\beta_1\alpha_2}\rangle$\\ \hline
$U^2$ & $|\psi_{\beta_1\gamma_2}\rangle$\\ \hline
$VU^2$ & $|\psi_{\gamma_1\alpha_2}\rangle$\\ \hline
$V^2U^2$ & $|\psi_{\alpha_1\beta_2}\rangle$\\ \hline
\end{tabular}
\center
\end{table}

Together with the two MUB's above, these nine states form a $2 \to 1$
QRAC for trits which is successful with probability
$p_q=1/2+\sqrt{3}/6 \simeq 0.79$. We will now see that this $p_q$ is
smaller than the bound obtained from the invariant information.

If we partition the total information content of a qutrit ($I_{total}=2/3$)
into two equal parts corresponding to each MUB, we should have a
maximum of $I=1/3$ for each three-outcome measurement. The largest
probability $p$ compatible with $I=1/3$ is given by
\begin{equation}
\left(p-\frac{1}{3}\right)^2+2\left(\frac{(1-p)}{2}-1/3
\right)^2=\frac{1}{3} \Rightarrow p=\frac{1}{3}+\frac{\sqrt{2}}{3}
\simeq 0.805,
\end{equation}
but this is larger than $p_q=1/2+\sqrt{3}/6 \simeq
0.789$ found for the QRAC described above.

This shows that other considerations can get into the way of quantum
random access codes, other than the bounds imposed by the invariant
information of Brukner and Zeilinger. The geometry of the set of pure states
for $d \ge 3$ is quite complicated, and ultimately limits the
construction of QRAC's, as the example above shows. This difficulty
can show up in different ways. In general, it may
not be possible to share the invariant information equally among
just two MUB's. Even if that is possible, this may require two
high-probability outcomes together with a third low-probability
outcome. This is not favourable for the QRAC, as it ideally requires just one
high-probability outcome to guarantee a high decoding probability for
Bob. With just one high-probability outcome, the optimisation done
above shows it is impossible to find states which achieve the optimal
sharing of $I_{total}$ between two MUB's.


Much research has been reported recently on quantum information
protocols using systems of dimensionality $d \ge 3$ \cite{BuhrmanCWdW01,CerfBKG02,BruknerZZ02,Cabello02b}. It has been shown
that in some cases, it could be possible to implement such protocols
with low detection efficiency \cite{Massar02}. The $2 \to 1$ QRAC with
qutrits presented above represents an initial effort towards
developing QRAC's for higher-dimensional systems. It would be
interesting to find how the gap between quantum and classical
performance behaves as $d$ increases. Some results along these lines have already been reported in \cite{Nayak99}.

QRAC's offer a genuinely quantum way of encoding information, which
naturally imposes restrictions on how much of it can be read out. This
limitation of the quantum readout appears because of the
non-commutativity of the decoding measurements. Instead of considering
this as a problem, here it is actually used to obtain a
higher-than-classical efficiency for this task. This seems to be a
general idea behind the development of many quantum information
applications -- we start from
an apparent problem or curiosity of quantum theory, but turn the
understanding of these characteristics into useful quantum information
applications.

\section{Experimental feasibility \label{sec 2qracfeasibility}}

In previous sections we have seen that the simple
$2{\rightarrow}1$ QRAC is effectively a demonstration of
quantum contextuality, and if performed using entanglement, of
non-locality as well. Besides illustrating these fundamental
characteristics of quantum theory, we have seen that this QRAC is also
behind the higher-than-classical performance of another information
processing task, the communication complexity problem we discussed in
section \ref{sec twoqcc}. Taking all this into account, it is natural
for us to ask about the experimental difficulty of implementing the
$2{\rightarrow}1$ QRAC. In this section we will analyse the
experimental requirements necessary for implementing this
simple quantum protocol.

\subsection{Detection efficiency and noise rate \label{sec etamu}}

Our analysis of the experimental difficulties of implementing the QRAC
starts by discussing the two main limitations for these quantum
experiments: finite detection efficiency $\eta$ and non-zero
background rate $1-\mu$. In all the discussion on experimental
feasibility in this thesis, we will consider protocols involving one
or more detections of a single qubit only, which involve discrimination between
two given orthogonal states.

Let us now define more precisely what we mean by detection
efficiency $\eta$. The quantum protocols require the parties to project the
qubit along a certain basis of two orthogonal states. By detection
efficiency I mean the fraction of events for which the detecting
apparatus clicks, successfully telling apart the two orthogonal
states we are interested in. In a typical quantum
optical experiment, the detection efficiency depends basically on the
quantum efficiency of single-photon detectors, which are usually of
the order of a few percent only.

We should also be concerned about background
noise, i.e. spurious detector `clicks' arising from background
photons or detector noise. We can account for these by introducing a
factor $\mu$, defined such that $(1-\mu)$ is the fraction of
detections at each party due to the background or detector noise. We
assume that the noise is uniformly random, yielding each of
the two measurement outcomes with probability $p=1/2$. Note that if these outcomes are found to deviate from uniform
randomness in an actual experiment, then we would be able to use the
regularity to improve our protocol; in other words, a uniformly random
background is a worst-case assumption.

Imperfections in the state preparation procedure can also be
incorporated in our model through the parameter $\mu$. For example,
suppose Alice needs to send state $\left|0\right\rangle$ to Bob, but
only manages to prepare an imperfect state. Suppose further that they
perform many experiments to characterise exactly the state that she
does manage to produce, and they find out that with probability $p_0$
the outcome is what she intended (i.e $|0\rangle$), but with
probability $(1-p_0)$ it is the opposite. Then, for all practical purposes, we can model the
situation by assuming she actually prepares state
\begin{equation}
\rho=\mu\left|0\right\rangle\left\langle 0 \right|+(1-\mu) \frac{\mbox{$1
\hspace{-1.0mm}  {\bf l}$}}{2},
\end{equation}
where $\mu=2p_0-1$ is the random background rate. A similar procedure
can be used in the case of imperfect preparation of maximally
entangled states, used in the entanglement-based QRAC we have
discussed.

\subsection{Feasibility of qubit communication protocol}

In the original QRAC protocol Alice communicates a single qubit to
Bob, preparing it in one of four states depending on her two-bit string.
The quantum communication involved suggests that this protocol can be
implemented with photons. Alice sends Bob a photon whose polarisation
encodes the qubit. 

%

Sometimes the protocol will work and sometimes not. With probability
$\eta\mu$ Bob will successfully measure a signal photon, getting the
right result with probability $p_q\simeq0.854$, as we have seen in
section \ref{sec 2qccqprot}. Whenever this does not happen, he will
succeed with probability $1/2$, either because he had to guess after
a detector failure, or because he measured a background
photon. Taking all into account, for a higher-than-classical
performance of the $2{\rightarrow}1$ QRAC it is enough to
have:
\begin{equation}
\eta\mu p_{q}+(1-\eta\mu)\frac{1}{2}>p_{c},
\end{equation}
or, plugging in $p_q=\cos^{2}(\pi/8)$ and $p_c=3/4$,
\begin{equation}
\mu>\frac{\sqrt{2}}{2\eta}.\label{sufqubitqrac}
\end{equation}
If there is no background ($\mu=1$), we see that it is sufficient to
have a detection efficiency $\eta>\sqrt{2}/2\simeq0.71$. The analysis
above was first made by van Dam in \cite{vDam99} for the communication
complexity protocol presented in section \ref{sec twoqcc}, but
restricted to the cases $\eta=0$ or $\mu=1$.

If an experiment fulfilling condition (\ref{sufqubitqrac}) is performed,
then we will have a one-qubit quantum protocol which outperforms the
optimal classical protocol using just one classical bit. These
conditions, however, cannot yet be reached in existing quantum optical
experiments (for a state-of-the-art experiment see \cite{KurtsieferOW01}). In the next chapter we will analyse a similar problem
whose quantum protocol can achieve a  higher-than-classical
performance in a currently feasible experiment.

\subsection{Feasibility of entanglement-based protocol}

As we have seen in section
\ref{sec qracnonloc}, the entanglement-based protocol is equivalent to a CHSH inequality
test. When analysing the feasibility of a CHSH test, some rather
complex arguments are necessary to find the conditions on $\eta$ and
$\mu$ that exclude any hidden-variable theory explanation (see
\cite{Larsson98b}). Interestingly, I will show these conditions can be
obtained directly from the analysis of the entanglement-based QRAC,
by requiring that it must have a higher-than-classical probability of
success.

First I present the case of limited efficiency $\eta<1$ but $\mu=1$
(no noise but limited detection efficiency), as discussed by van Dam
\cite{vDam99} in the context of the communication complexity problem
reviewed in section \ref{sec BCvD task}.

Before trying the quantum protocol, Alice and Bob
must agree on a procedure for the case when their detectors do not
fire. Of course, they are not
allowed to communicate the failure, as this would constitute some
further communication beyond the allowed one bit. The most effective
procedure is for Alice and Bob to use the optimal classical protocol
whenever their detectors fail. In the case of a single detector
failure, Bob's guess will still be correct with $p=1/2$. When
\textit{both} detectors fail, however, they will still obtain an
optimal classical probability of success
$p_c=3/4$. Assuming independent errors at the two
detectors, we can calculate the minimum detector efficiency needed:
\begin{align}
&
\eta^{2}p_{q}+(1-\eta)^{2}p_{c}+\eta(1-\eta)\frac{1}{2}+(1-\eta)\eta \frac{1}{2}>p_{c}\nonumber\\
&  \Rightarrow\eta>\frac{2p_{c}-1}{p_{q}+p_{c}-1}.
\end{align}
Our $2{\rightarrow}1$ QRAC has $p_{q}=\cos^{2}(\pi/8)$ and $p_{c}=3/4$, which results in a minimum necessary detector efficiency
\begin{equation}
\eta_{\min}=2(\sqrt{2}-1)\simeq0.828,
\end{equation}
which is also the minimum detector efficiency necessary for a
loophole-free CHSH inequality test \cite{GargM87}.

To take into account the probability $(1-\mu)$ of each detector firing
due to a background event, we can classify each run of the experiment
into one of three classes:

1) Alice and Bob measure their particles accurately with probability
$\eta^{2}\mu^{2}$, in which case they apply the quantum protocol and
succeed with probability $p_{q}=\cos^{2}(\pi/8)$.

2) Both their detectors fail to detect anything, which will happen in
a fraction $(1-\eta)^{2}$ of the runs. In this situation they will
both use the classical protocol with a probability of success $p_{c}=3/4$.

3) In all the other situations Bob will use either the quantum or the
classical protocol, depending on whether he detects a photon or
not. However, due to either background or lack of detection at Alice's
side his guesses will be random, yielding a success rate of only
$1/2$.

\begin{figure}
\begin{center}
\includegraphics[scale=0.55
]{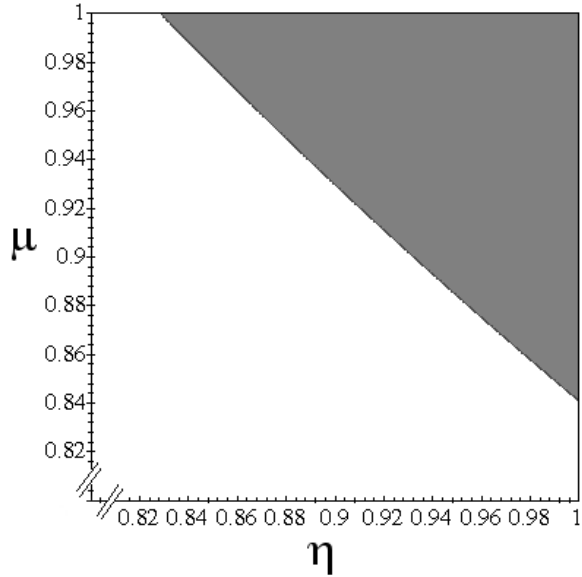}
\caption[Experimental conditions for a better-than-classical
entanglement-based $2 \to 1$ QRAC.]{\label{fig 21qrac}The shaded area indicates the region where the background level
$(1-\mu)$ and detector efficiency $\eta$ allow for an
entanglement-based $2 \to 1$ QRAC with better-than-classical performance. The area corresponds to that given by inequality
(\ref{bound visibility}).}
\end{center}
\end{figure}

Taking all this into account, the condition for a quantum protocol
which is better than the optimal classical one is:
\begin{equation}
\eta^{2}\mu^{2}p_{q}+(1-\eta)^{2}p_{c}+(1-\eta^{2}\mu^{2}-(1-\eta)^{2})\frac{1}{2}>p_{c}. \label{mu and eta}
\end{equation}
Plugging in the values for $p_q$ and $p_c$ we obtain
\begin{equation}
\mu>\sqrt[4]{\frac{1}{2}}\sqrt{\frac{2}{\eta}-1}.\label{bound visibility}
\end{equation}
In particular, for perfect detectors ($\eta=1$), we need
$\mu>2^{-1/4}\simeq0.841$. In figure \ref{fig 21qrac} I represent the
region in the $\eta\times\mu$ parameter space that guarantees a
higher-than-classical success rate.

The favourable $\eta\times\mu$ region coincides exactly with the
conditions necessary for a violation of the CHSH inequality (as can be
gathered from \cite{Larsson99}, for example). This is not a surprise,
given the equivalence between the quantum protocol and a CHSH test
pointed out in section \ref{sec qracnonloc}.

Interestingly, we have obtained information on how hard it is to
experimentally establish quantum non-locality by a simple analysis of
a  quantum information processing application. It is important to
remember that to beat the optimal classical protocol it is
\textit{not} necessary to enforce strict locality for the
measurements. It is enough to obtain the stronger-than-classical
correlations with time-like separation only. Under these conditions,
the QRAC becomes a quantum contextuality test, and derives its
higher-than-classical performance from this quantum characteristic.

A first candidate for experimental implementation of the protocol
would be entangled photons generated by a parametric down conversion
crystal. The CHSH test with entangled photons which achieved the
highest detection efficiency to date seems to be the one reported by
Weinfurter's group \cite{KurtsieferOW01}, the detection efficiency $\eta$ being slightly over $28\%$.

Maximally entangled pairs of photons can be generated with high
fidelity and measured with precision, which guarantees the equivalent
of a very high $\mu$. However, detector efficiencies $\eta$ are still
short of those necessary. Ion traps techniques, on the other hand, can
reach high $\eta$ and $\mu$ \cite{RoweKMSIMW01}. This makes ion traps a
natural choice for the experimental demonstration of the simple quantum random access code presented here.

\section{Chapter summary}

In this chapter I have analysed different aspects of some of the
simplest quantum random access codes. We have seen that the simplest of
them, the $2\to 1$ QRAC with a qubit, can be shown to derive its
performance either from quantum contextuality or from non-locality,
depending on how the protocol is implemented.

I have shown how the simplest QRAC solves a two-party communication
complexity problem proposed in \cite{BuhrmanCvD97}. This shows that
the higher-than-classical performance of that protocol can also be
attributed to contextuality and non-locality.

We have also seen that the invariant information of Brukner and
Zeilinger, which I presented in section \ref{sec invinfo}, can be used to
obtain upper bounds for the efficiency of a class of QRAC's. By building
a QRAC with three-level systems, or qutrits, we have seen that those
bounds are not tight in general, despite being tight for the QRAC's I
presented using a qubit.

Then I presented an analysis of experimental conditions on detector
efficiency and background counts, which could enable a
better-than-classical protocol. From a simple analysis of the quantum and the
optimal classical protocol I showed that these conditions are
equivalent to those required for a CHSH non-locality test, except that
locality conditions are not necessary. By relaxing this requirement
the protocol becomes equivalent to a quantum contextuality test, and
is amenable to implementation using current ion trap technology.



\chapter{On multi-party quantum communication complexity \label{chap feasible}}

In this chapter I will analyse different
ways of solving a simple multi-party communication complexity problem
first proposed by Buhrman and co-authors in \cite{BuhrmanCvD97,BuhrmanvDHT99}. As we have seen
in section \ref{sec ccintro}, quantum mechanics can help in reducing
the communication complexity of certain problems. In the context of
the problem we will analyse, references
\cite{BuhrmanCvD97,BuhrmanvDHT99} showed that multi-party entanglement
can reduce the required amount of communication for a flawless protocol. In this chapter I approach the problem somewhat
differently, in a way that allows a fairer comparison with the
classical protocols. For that, I will restrain the amount of
communication allowed and explicitly compute the probability of
success for both the classical and quantum protocols. This will enable
us to consider realistic implementations, involving limited detection
efficiencies and non-zero noise.

We will see that the entanglement-based quantum protocol is better
than classical protocols due to multipartite
non-locality. Accordingly, the protocol can be interpreted also as a
multi-particle non-locality test. I will find a set of sufficient
conditions on detection efficiency and noise for such tests, directly
from the quantum protocol.

Remarkably, the quantum protocol involving qubit-communication can be
demonstrated with realistically low detection efficiencies. This is a
practically unique situation in quantum information experiments, which
usually have low detection efficiencies, which requires post-selection of experimental runs or a fair sampling assumption. This is simply not an option if we are interested in
practical quantum applications, as opposed to proof-of-principle
demonstrations. We will see that the qubit-communication protocol can be
shown to be better than any equivalent protocol using classical
communication with no need for such artificial performance-enhancing
procedures.

This qubit-communication solution to the communication complexity
problem suggests that entanglement is not the only non-classical
resource giving quantum information processing its power. We will see
that this quantum protocol relies on the continuous nature of the set
of quantum pure states for its higher-than-classical performance. In
the next chapter I will analyse this issue in more depth by discussing
the quantum solution to a similar problem. We will see that the
quantum protocols tell us many things about the foundations of
quantum theory and computation capabilities of quantum systems.

This chapter is organised as follows. After a review of the
communication complexity task definition (section \ref{sec mqcc def}),
I will discuss different ways to compare quantum and classical
protocols, and the reasons for the particular choice I make. In
section \ref{sec optclascc} I determine the optimal classical
probability of success for the task, for a fixed amount of classical
communication. Then in section \ref{sec qcc prot} I present quantum
protocols that solve the task ideally with unit probability, resorting
either to entanglement or to qubit communication. In section \ref{sec
expqcc} I discuss how these protocols are affected by experimental
limitations, and I conclude with a chapter summary in section \ref{sec
qccconcl}.

\section{A multi-party communication complexity task \label{sec mqcc
def}}

In this section I review the \textit{Modulo-4 Sum} communication
complexity problem defined for three parties by Buhrman, Cleve and van
Dam \cite{BuhrmanCvD97}, and later generalised to $N$ parties
($N\geq3$) in \cite{BuhrmanvDHT99}. An updated version of
\cite{BuhrmanCvD97} was published in \cite{BuhrmanCvD01}.

The problem can be stated as follows. Each party $P_{i}$ receives a
two-bit number input $x_{i}$, subject to the constraint:
\begin{equation}
\left( \sum_{i=1}^{N}x_{i} \right) \operatorname{mod}2=0. \label{cond
gmn}
\end{equation}
The strings are chosen randomly with a uniform probability
distribution among those combinations that satisfy
eq. (\ref{cond gmn}) above. After possibly some distributed
computation, involving some  communication between the
parties, one of them (say the last one $P_{N}$) must announce the value
of the Boolean function
\begin{equation}
f(\overrightarrow{x})=\frac{1}{2}\left[ \left(   \sum_{i=1}^{N}x_{i}\right)
\operatorname{mod}4\right]  .
\end{equation}
In other words, each party is given a number $x_{i}\in\{0,1,2,3\}$,
subject to the constraint that the sum of all $x_{i}$ is even. After
some communication the last party must decide whether the sum
modulo-$4$ is equal to $0$ or $2$.

The type (and amount) of communication allowed will be important when
we compare quantum and classical protocols. In the next section I
present two different sets of constraints on the communication
allowed, and the reasons that motivated me to consider the differences
between quantum and classical in these two settings.

\subsection{Restraining the amount of communication}

There are different approaches one can take when studying this
problem. The original papers \cite{BuhrmanCvD97, BuhrmanvDHT99} proved
bounds on the amount of communication necessary for solving this
problem classically in the public communication setting, where each
party broadcasts to all others. Then an entanglement-based quantum
communication complexity protocol was shown to be possible, which
reduced the necessary amount of communication for a perfect protocol.

Instead of calculating the minimal amount of communication sufficient
to solve the task, another thing one might want to do is to limit the
amount of communication, and then find the optimal probability of
success achieved by quantum and classical protocols. This
is the approach I adopt here, for the reasons I sketch below.

First, it could be the case that exact protocols exhibit a large
gap between quantum and classical, which however disappears when we
consider more realistic, probabilistic protocols. Directly comparing
the probability of success of general probabilistic classical
protocols with quantum protocols (which are sometimes intrinsically
probabilistic) seems to be a fairer comparison.

A second motivation for this approach are experiments: knowing the
optimal probability of success of ideal quantum and classical
protocols, we will be able to determine sufficient experimental
requirements for implementing these protocols in practice. Factors
such as noise and finite quantum detector efficiency  tend to lower the
efficiency of a realistic quantum protocol, and these can be easily
included in my analysis. I will derive sufficient conditions on these
factors for practical quantum protocols that are still better than the
optimal classical ones.

Still keeping an eye on feasible experimental implementations, in what
comes below we will require the communication to be
sequential. This decision is related to the
fact that the sequential quantum communication necessary to solve this
problem can be conveniently realized by sending a single photon
through a series of optical elements representing the parties. We will
soon see that such quantum communication can outperform the equivalent
classical communication setting even with realistically low quantum
detection efficiencies.

In common with the two-party communication complexity problem I described in the last chapter, the quantum
protocols I will consider here are of two types, involving either
using quantum entanglement or using quantum communication instead of
classical communication. For each type of communication I will compare
the quantum protocol with an equivalent classical protocol, and find
the conditions under which the quantum protocol presents an
advantage. For the reasons I explained in the last
paragraphs, I will impose two different restrictions on the amount of communication allowed:

\vspace{ 2 mm}
A- \textbf{Entanglement-based protocol.} Here we will compare quantum
and classical performances when each party $P_j$ is allowed to send
just a single bit of classical communication to the last party
$P_N$. We compare the optimal classical protocol with the
entanglement-based quantum protocol which supplements this
communication with measurements on multi-party entangled states.

\vspace{ 2mm}
B- \textbf{Sequential qubit communication protocol.} Here the allowed
one (qu)bit of communication must be \textit{sequential}: party $P_1$
sends one (qu)bit to party $P_2$, who sends one (qu)bit to party $P_3$
and so forth, until the last party $P_N$ who has to announce the value
of $f(\overrightarrow{x})$. The comparison between quantum and
classical is made by comparing the probability of success of the
optimal one-bit message classical protocol with that of a one-qubit
message quantum protocol.
\vspace{2 mm}

The entanglement-based protocol can be interpreted as a
non-locality test for three or more entangled qubits. Accordingly, the sufficient
experimental conditions I will derive will also be sufficient for
genuine multi-particle non-locality tests. This represents a simple
example in which a quantum information application brings results
which are relevant to foundational tests of quantum theory.

We will see that the qubit-communication protocol  derives
its higher-than-classical probability of success from another quantum
characteristic. I will briefly discuss this at the end of section
\ref{sec expqubitgmn}, and at considerably more length in the next
chapter. The suggestion of a qubit-communication protocol for the
tripartite Modulo-4 Sum problem was made independently also by
Buhrman, Cleve and van Dam \cite{BuhrmanCvD01}, in an update of the
earlier paper \cite{BuhrmanCvD97}.

\section{Determining the optimal classical protocols \label{sec
optclascc}}

\subsection{Sequential communication \label{sec qccseqcomm}}

First, let us obtain the optimal classical success rate for the
Modulo-4 Sum problem, when we allow only one bit of sequential
communication from one party to the next.

For the moment let us consider only deterministic protocols. The first
party $P_{1}$ has access only to her two-bit string $x_{1}$, and so
can choose between $2^{4}$ protocols. These can be represented by the
four-bit string $prot_{1}$, whose $n^{th}$ ($n=0,1,2,3$) bit encodes
the message $m_{1}$ to be sent to $P_{2}$ if $x_{1}=n$. The other
parties $P_{j}$ ($j=2,...,N-1$) can choose among $2^{8}$ protocols
that take into consideration both $x_{j}$ and the message $m_{j-1}$
received from the previous party. Each of these protocols can be
represented by an 8-bit string $prot_{j}$, whose $n^{th}$
($n=0,1,...,7$) bit encodes the message to be sent when
$2x_{j}+m_{j-1}=n$.

Each possible deterministic protocol can then be represented by the
$(N-1)$-tuple
\[
\overrightarrow{prot}= (prot_{1}, prot_{2}, ..., prot_{N-1}).
\]
Finding the probability of success of a given protocol $\overrightarrow{prot}$
is a straightforward computation. It must be done in parts, each part
corresponding to one of the possible values for the two-bit string
$x_N$ belonging to the last party. Starting with $x_N=0$, we produce a list of all possible input data $\{x_{1},x_{2},...,x_{N-1}\}$ compatible with
$x_{N}=0$. Then we compute the messages $m_{N-1}$ corresponding to
each (according to protocol $\overrightarrow{prot}$), and find the
fraction of cases in which $P_{N}$'s most likely guess about the value
of function $f$ would in fact be correct. This is repeated for $x_{N}=1,2$
and $3$, and the results averaged to obtain the overall probability of
success $p_{c}$. The optimal deterministic protocol can then be found
by a computer search over all $2^{4}(2^{8})^{N-2}=2^{(8N-12)}$
protocols.

I wrote a C computer program that finds the optimal classical
probability of success for $N=3,4,5$ and $6$. Moreover, a limited
search over protocols for larger number of parties yields some lower
bounds for $p_c$. The results are summarised in table \ref{table pcseq}.

\begin{table}\caption{\label{table pcseq}Optimal $p_c$ for $N$ parties and sequential communication}
\begin{center}
\begin{tabular}{|c|c|c|c|c|c|c|} \hline
$N$ & 3 & 4 & 5 & 6 & 7 & 8 \\
\hline
$p_c$ & $3/4$ & $3/4$ & $5/8$ & $5/8$ & $\ge 9/16$ & $\ge 9/16$ \\
\hline
\end{tabular}
\end{center}
\end{table}

The optimal $p_{c}$
for $N=3,4,5$ and $6$ is attained by many protocols, for example the
one consisting of $prot_{1}=0011$ and all the other
$prot_{j}=01011010$. The same protocol yields the lower bounds for the
optimal probabilities of success presented above for $N=7$ and
$8$. Checking that these lower bounds are tight would involve a very
long exhaustive search over all protocols. For the purpose of
comparison with the quantum protocol given below, it would be
desirable to obtain at least an analytical upper bound for $p_{c}^{N}$
that decreases with $N$. A rigorous proof has escaped me, but in the
next section I make a conjecture about $p_c$ for general $N$.

Up to now we have been computing the probability of success for
deterministic protocols. In a probabilistic protocol, each party
$P_{j}$ implements his/her own protocol by probabilistically picking a
deterministic protocol $prot_{j}$ from some set of protocols,
according to probabilities obtained from a list of random
numbers. Since this list of numbers could have been shared beforehand
between the parties, the last party $P_{N}$ can know exactly which
protocols were chosen by each of the other parties for each run of the
probabilistic protocol. This means that each run of the probabilistic
protocol is effectively a deterministic one, with a probability of
success bounded by the optimal deterministic $p_{c}$ derived
above. The relation between deterministic and probabilistic protocols
for communication complexity tasks is further discussed in chapter 3
of the book by Kushilevitz and Nisan \cite{KushilevitzN97}.

\subsection{Non-sequential communication}

Here we discuss how to obtain the optimal probabilities of success for
classical protocols with one bit of communication from each of the
first $(N-1)$ parties to the last party $P_N$. As has been done
a couple of times already in this thesis, the optimal
protocols are obtained through an exhaustive search over all
deterministic $N$-party protocols.

How many protocols are there? The two-bit string $x_i$ belonging to
party $P_j$ can only assume four values. Each party's protocol can
involve a distinct binary choice of message for each of these values;
there are therefore $2^{4}=16$ possible protocols for each
party. Since a total of $(N-1)$ parties enter in the communication
protocol, the total number of protocols is $2^{4N-4}$.

Again, we see that the number of protocols increases
exponentially. Despite that, an exhaustive search is feasible for
small $N$. I wrote a computer program in C that computes the optimal
probability of success for $N=3$ to $7$ parties. The
results are summarised in table \ref{table pcnonseq}.

\begin{table}\caption{\label{table pcnonseq}Optimal $p_c$ for $N$ parties and non-sequential communication}
\begin{center}
\begin{tabular}{|c|c|c|c|c|c|} \hline
N & 3 & 4 & 5 & 6 & 7 \\
\hline
$p_c$ & 3/4 & 3/4 & 5/8 & 5/8 & 9/16 \\
\hline
\end{tabular}
\end{center}
\end{table}

As we see, the task gets harder and harder as the number of parties
increases, as happened in the sequential communication model. At least
for small number of parties, the optimal probabilities of success are
the same for the two communication models.

Based on the structure of the problem and on the values for $p_c$
calculated above and in the previous section, I conjecture that the optimal classical probability of success is given by:
\begin{equation}
\mbox{Conjecture: } p_c=\frac{1}{2}+\frac{1}{2^{\frac{N+1}{2}}},\label{conj pc}
\end{equation}
for odd $N$, being the same for both models of computation (sequential
and non-sequential). This would approach the random guess success
($p=1/2$) exponentially with the number of parties $N$.

\section{Quantum protocols \label{sec qcc prot}}
\subsection{Entanglement-based quantum protocol \label{sec entgmn}}

In this section I will review the entanglement-based protocol for
this task, first presented by Buhrman, Cleve and van Dam in
\cite{BuhrmanCvD97}, and further elaborated in \cite{BuhrmanvDHT99}. My presentation here will stress
the fact that it is a quantum (non-local) phase that enables a
higher-than-classical performance for this task. This suggested the
use of the phase in a single qubit for the quantum communication
protocol we will discuss in the next section.

The quantum resource used will be a $N$-party highly entangled GHZ
state:
\begin{equation}
\left|  \mbox{GHZ($\phi$)}\right\rangle =\frac{1}{\sqrt{2}}\left(  \left|  0_{1}0_{2}%
\cdots0_{N}\right\rangle +e^{i\phi}\left|  1_{1}1_{2}\cdots1_{N}\right\rangle
\right)  , \label{GHZ state}
\end{equation}
where $\phi$ is a phase which we will use in the protocol. 

To simplify the discussion, let us present the entanglement-based
protocol for three parties only. The results generalise in a trivial
way for more parties.

The parties initially share a three-party
$\left|\mbox{GHZ(0)}\right\rangle$ state. The quantum protocol starts by having each party $P_i$ apply the following unitary to his/her qubit:
\begin{equation}
U:\left\{
\begin{tabular}{c} $\left|0\right\rangle\to|0\rangle$\\
$\left|1\right\rangle\to\exp(i\frac{\pi x_i}{2})|1\rangle$ 
\end{tabular}
\right. ,\label{unitgmn}
\end{equation}
which effectively just rotates the phase $\phi$ in the GHZ state
(\ref{GHZ state}) by an amount proportional to $P_i$'s number $x_i$.

After the three parties have performed their rotations, because of the
promise (\ref{cond gmn}) the initial state will be transformed to:
\begin{eqnarray}
\left|\mbox{GHZ(0)}\right\rangle \to
\left\{
\begin{tabular}{c}
$\left|\mbox{GHZ($0$)}\right\rangle
\mbox{ if $f(x_A,x_B,x_C)=0$ or}$\\
$\left|\mbox{GHZ($\pi$)}\right\rangle
\mbox{ if $f(x_A,x_B,x_C)=1$.}$
\end{tabular}
\right. \label{ghzt} 
\end{eqnarray}

It turns out
that these two orthogonal states can be distinguished [and $f(\overrightarrow{x})$
computed] with local operations and one bit of classical
communication from Alice to Clare, and one from Bob to Clare. The local
operation needed at each party consists of applying a Hadamard gate
\begin{equation}
H:\left\{ \begin{tabular} {c}
$\left|0\right\rangle 
\to \frac{1}{\sqrt{2}}\left(\left|0\right\rangle +
\left|1\right\rangle \right)$ \\
$\left|1\right\rangle 
\to \frac{1}{\sqrt{2}}\left(\left|0\right\rangle - \left|1\right\rangle\right)$
\end{tabular}\right. ,\label{hadamard}
\end{equation}
followed by a measurement on the computational basis. It is easy to
check that the three Hadamard gates transform state (\ref{ghzt}) as:
\begin{eqnarray}
 \left|\mbox{GHZ(0)}\right\rangle &\to&
\left|000\right\rangle+\left|011\right\rangle+\left|101\right\rangle+\left|110\right\rangle
,\\
\left|\mbox{GHZ($\pi$)}\right\rangle &\to&
\left|001\right\rangle+\left|010\right\rangle+\left|100\right\rangle+\left|111\right\rangle .
\end{eqnarray}
Alice and Bob then measure their qubits in the computational basis,
and send the measurement outcomes to Clare. It is clear from the two equations above that Clare can distinguish perfectly between
$\left|\mbox{GHZ($0$)}\right\rangle$ and
$\left|\mbox{GHZ($\pi$)}\right\rangle$, solving the task with certainty.

The above protocol for three parties generalises for $N$ parties in a
straightforward way. Each party $P_i$ performs the unitary
(\ref{unitgmn}) depending on his/her data $x_i$, followed by a
Hadamard gate and a measurement on the computational basis. Each of the first
$(N-1)$ parties then broadcasts the result to the last party $P_N$,
who can then compute $f(\overrightarrow{x})$ unambiguously.

\subsubsection{Relation with multi-party non-locality and contextuality}

The communication complexity task was devised
explicitly to take advantage from the strong quantum correlations
present in the $\left|\mbox{GHZ}\right\rangle$ state. These
correlations allow for the inequality-free proof of quantum
non-locality proposed in  \cite{GreenbergerHZ89, Mermin90}. The higher
probability of success of the quantum protocol with respect to the
optimal classical protocol relies heavily on this form of
non-locality.

In the discussion above I obtained the optimal
probability of success for classical protocols, when the information
is restricted to one bit from each party. This had not been done in
previous discussions of this communication complexity problem, and
enables us to directly compare the quantum and classical
performances. In section \ref{sec expqcc} I will do this, taking into account finite detection efficiencies
and imperfect state preparation. If, even with these imperfections, one
can implement a better-than-classical quantum protocol with space-like
separated measurements, this would constitute perfectly valid
experimental evidence of multi-particle non-locality. This is because
our calculation of the optimal classical probability of success
imposes a constraint on any local hidden-variable theory: it must not
enable a protocol that is better than the optimal classical one. Any
protocol with space-like separated measurements that performs better
than the optimal classical protocol must rely on non-local
correlations inconsistent with local hidden-variable theories.

In a more realistic setting, the measurements in a practical
implementation would not be space-like separated. In this case we
cannot claim that the measurements contradict the locality assumption;
they do, however, contradict the non-contextuality assumption. Again
we see that contextuality counts as a non-classical resource to be
explored in quantum information protocols, as I have discussed in
section \ref{sec qraccontext}.

This approach stresses the non-classicality of quantum information
processing protocols as a way to investigate multipartite quantum
non-locality and contextuality. It is interesting to see that simple
arguments and computations allow us to obtain upper
bounds to what is possible classically, and show that quantum
mechanics violates these bounds even with mixed states, or with
inefficient detectors. While Bell-type bipartite non-locality
inequalities are simple and compelling, they get unsatisfyingly
cumbersome when we increase the number of parties, or the number of
settings at each party. My approach here also has its conceptual
attractiveness, as it links non-locality and contextuality to an
enhanced computational capability. Moreover, when we try to quantify
multipartite entanglement of high-dimensional systems, this approach
can be more fruitful and intuitively satisfying than deriving complex
Bell-type inequalities which do not suggest any practical applications
to the non-locality they uncover.

\subsection{Sequential qubit communication protocol}

In section \ref{sec qccseqcomm} we have seen that the Modulo-4 Sum
problem gets harder and harder to solve classically with sequential communication, as the number of parties increases. There is,
however, a simple quantum protocol with sequential qubit communication
that has a probability of success $p_{q}=1$ \textit{independently} of
the number of parties involved. We will see that this protocol is a
simple adaptation of the entanglement-based protocol presented in
\cite{BuhrmanvDHT99} and reviewed above. 

The idea is to start with the qubit in state

\begin{equation}
\left|  \psi_{i}\right\rangle =\frac{1}{\sqrt{2}}\left(  \left|
0\right\rangle +\left|  1\right\rangle \right)
\end{equation}
and send it flying by all the parties, from first to last. Each party
needs only act upon the qubit with unitary $U$ given by eq. (\ref{unitgmn}).
After going through the $N$ unitaries the qubit state will be
\[
\left|  \psi_{f}\right\rangle =\frac{1}{\sqrt{2}}\left(  \left|
0\right\rangle +(-1)^{f(\overrightarrow{x})}\left|  1\right\rangle \right)  ,
\]
due to the constraint (\ref{constraint}) on the possible inputs
$x_{j}$. The last party can then measure $\left|
\psi_{f}\right\rangle $ in the $\{\frac{1}{\sqrt{2}}(\left|  0\right\rangle +\left|  1\right\rangle
),\frac{1}{\sqrt{2}}(\left|  0\right\rangle -\left|  1\right\rangle
\}$ basis, obtaining $f(\overrightarrow{x})$ with probability $p_{q}=1$.

The protocol above is an adaptation of the entanglement-based protocol
presented in \cite{BuhrmanvDHT99} (and reviewed in section \ref{sec
entgmn}) to the qubit-communication setting. While the entanglement-based protocol uses a non-local
quantum phase which can be operated on locally by each party, the
qubit communication protocol above uses a quantum phase of a single
qubit to acquire information on $f(\overrightarrow{x})$ as it flies by
the parties towards the last party $P_{N}$, where a single detection
reveals the result. The fact that this protocol involves only one
detection (as opposed to the $N$ detections for the entanglement-based
protocol) leads to an experimental implementation which is robust
against low detection efficiencies, as we will see in section \ref{sec expqubitgmn}.

\section{Experimental feasibility \label{sec expqcc}}

\subsection{Feasibility of entanglement-based protocol}

Creating entangled states of more than two qubits is
experimentally much more challenging than creating bipartite
states. Having said that, let us assume, for a moment, that the technical difficulties
of creating GHZ states have been surpassed. In
this section we will see that, in principle, the multi-party quantum
protocols require lower detection efficiency \footnote{For the definitions of detection efficiency $\eta$
and background noise $(1-\mu)$ see section \ref{sec etamu}.} $\eta$ and tolerates higher
background noise $(1-\mu)$  than the two-party protocol of section \ref{sec
2qracfeasibility}. As we discussed in section \ref{sec entgmn}, these results
translate also as sufficient conditions for experimental
violation of multi-particle non-locality.

The analysis is similar to that presented in section \ref{sec 2qracfeasibility}, with the difference that here we have $N$ detections
instead of just two. Whenever we detect a valid signal from all the
$N$ detectors (which happens with probability $\eta^{N}\mu^{N}$), the
quantum protocol works with probability $p_{q}=1$. If the last party's
detector does not click, she will rely on the possibility of the other
parties' detectors having failed as well, and on them using the
optimal classical protocol; this will happen with probability $(1-\eta)^{N}$ and result
in a probability of success $p_{c}$ given by table \ref{table pcnonseq}. In all other
cases, the last party will only be able to make random guesses,
succeeding with probability $1/2$ only. Assuming independent errors at
each detector, for a higher-than-classical probability of success we
need

\begin{equation}
\eta^{N}\mu^{N}p_{q}+(1-\eta)^{N}p_{c}+[1-\eta^{N}\mu^{N}-(1-\eta)^{N}
]\frac{1}{2}>p_{c}. \label{ineq tri}
\end{equation}

Substituting $p_{c}$ from table \ref{table pcnonseq} into the
inequality above we obtain the results summarised in table \ref{table
ent7prot}, for number of parties $N= 3$ to 7.

\begin{table}\caption{\label{table ent7prot}Conditions for better-than-classical performance} \begin{tabular} {|c|c|c|c|} \hline

N & $\mu_{min}$ & $\eta_{min}$ & favourable $\mu$ x $\eta$ region \\
\hline \hline
3 & $2^{-\frac{1}{3}} \simeq 0.794$ & $\simeq 0.791$  & $\mu >
\frac{1}{2\eta}\left(
4\eta^{3}-12\eta^{2}+12\eta\right)^{\frac{1}{3}}$ \\ \hline

4 & $2^{-\frac{1}{4}} \simeq 0.841$ & $\simeq 0.841$ &  $\mu >
\frac{1}{2\eta} 2^{\frac{3}{4}}\left(
-\eta^{4}+4\eta^{3}-6\eta^{2}+4\eta\right)^{\frac{1}{4}}$ \\
\hline

5 & $2^{-\frac{2}{5}} \simeq 0.758$ & $\simeq 0.758$ &  $\mu >
\frac{1}{2\eta} 2^{\frac{3}{5}}\left(
\eta^{5}-5\eta^{4}+10\eta^{3}-10\eta^{2}+5\eta\right)^{\frac{1}{5}}$ \\
\hline

6 & $2^{-\frac{1}{3}} \simeq 0.794$ & $\simeq 0.794$ &  $\mu >
\frac{1}{2\eta} 2^{\frac{2}{3}}\left(
-\eta^{6}+6\eta^{5}-15\eta^{4}+20\eta^3 -15\eta^2 +6 \eta \right)^{\frac{1}{6}}$ \\ \hline

7 & $2^{-\frac{3}{7}} \simeq 0.743$ & $\simeq 0.743$ &  $\mu >
\frac{1}{2\eta} 2^{\frac{4}{7}}\left(
\eta^{7}-7\eta^{6}+21\eta^{5}-35\eta^{4}+35\eta^3 -21\eta^2 +7\eta \right)^{\frac{1}{7}}$ \\ \hline
\end{tabular}
\end{table}

Table \ref{table ent7prot} shows the minimum $\eta$ (when $\mu=1$),
the minimum $\mu$ (when $\eta=1$), and the compromises between these
two factors which still result in a quantum protocol with
higher-than-classical probability of success.

\begin{figure}
\begin{center}
\includegraphics[scale=0.45,angle=270
]{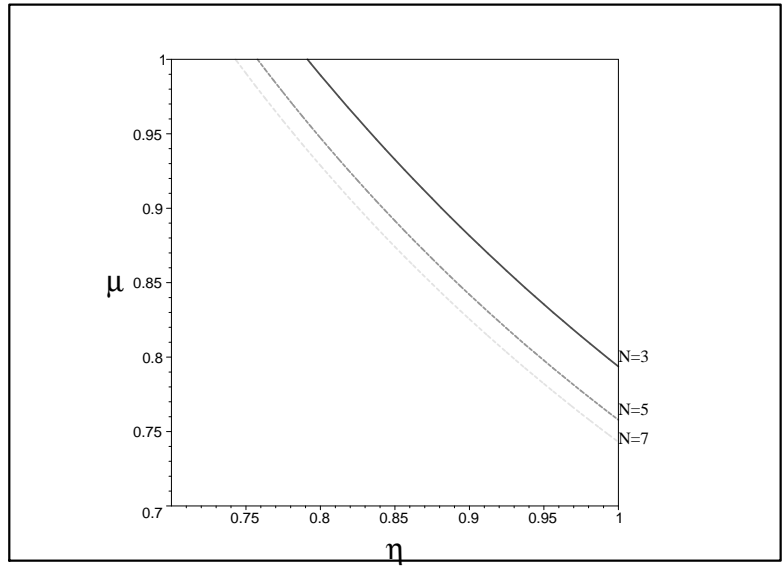}
\caption[Experimental conditions for a better-than-classical
entanglement-based $N$-party quantum communication complexity protocol.]{\label{fig feasible1}Here we plot the lines delimiting the
region where the background level $(1-\mu)$ and detector efficiency
$\eta$ allow for a $N$-party entanglement-based quantum communication
complexity protocol which is more efficient than the classical
counterpart. The favourable regions correspond to the area above the
lines indicated for $N=3,5$ and $7$. Analytical
expressions for these favourable regions are given in table \ref{table ent7prot}.}
\end{center}
\end{figure}

In figure \ref{fig feasible1} I plotted the favourable $\eta\times\mu$ region for
$N=3,5$ and $7$. Note that the region monotonically increases for
increasing odd $N$. Also, the region is larger than that required for
the two-party QRAC we investigated in section \ref{sec
2qracfeasibility} [see inequality  (\ref{bound visibility})]. This is
related to the fact that quantum non-locality tests for multi-particle
states can be done with lower detector efficiencies than for the
two-particle case (see for example \cite{Larsson98, LarssonS01}).

As we discussed in section \ref{sec entgmn}, the requirements in table
\ref{table ent7prot} also represent sufficient conditions for a multi-particle non-locality
test with $N=3$ to $7$ qubits. Besides limited detector efficiency,
here I also take into account limited visibility (due to background
counts); this possibility has been addressed with respect to the
particular experiment described in \cite{BouwmeesterPDWZ99} in
references \cite{deBarrosS00} and \cite{SzaboF02}. As far as I know, it is the
first time simultaneous bounds both on detector efficiency and noise
rate are obtained for more than three parties.

This quantum protocol can also be used to obtain bounds on the
amount of mixedness we can add to entangled multi-particle states and
still obtain an entangled density matrix. This has been used as a
measure of `strength' of locality violation for large-dimensional
states in \cite{CollinsGLMP02, KaszlikowskiGZMZ00}, even though there
has been some criticism to this criterion \cite{AcinDGL02}. To see how to
apply this idea, imagine that instead of performing the protocol with a
perfect $\left|  \mbox{GHZ(0)}\right\rangle$ [see equation (\ref{GHZ state})], we instead use a
mixture of it with the separable, maximally mixed state:
\begin{equation}
\rho=\epsilon\left|\mbox{GHZ}\right\rangle\left\langle \mbox{GHZ} \right|+(1-\epsilon) \frac{\mbox{$1
\hspace{-1.0mm}  {\bf l}$}}{2^{N}},
\end{equation}
where $(1-\epsilon)$ is the amount of mixedness added. The mixed state
thus obtained is sometimes referred to as a Werner state. For large
enough $\epsilon$ it is clear (from continuity) that the quantum
protocol should still go through with better-than-classical
performance, but there must be an intermediary value for which this is
not true anymore. This can be worked out in a straightforward way
using an argument like the one I used above for detection efficiencies
and noise rates. For a better-than-classical performance we need:
\begin{equation}
\epsilon +(1-\epsilon)\frac{1}{2}>p_c \Rightarrow \epsilon > 2p_c -1.
\end{equation}
With the conjectured $p_c$ from equation (\ref{conj pc}), the equation
above would lead to the requirement that 
\begin{equation}
\epsilon>2^{-\frac{N-1}{2}}
\end{equation}
for odd $N$. This would imply that the  amount of mixedness that
multi-qubit entangled states can tolerate and still remain entangled
increases exponentially with the number of qubits. That this is indeed
the case has been shown through other methods by a number of authors,
see for example \cite{DeuarMN00}.

\subsection{Feasibility of qubit communication protocol \label{sec expqubitgmn}}

In this section we will work out the detector efficiencies and noise
rates which allow for a qubit-communication protocol which is
better than its optimal classical counterpart. We will see that the
sufficient detector efficiency decreases monotonically with the number
of parties. This means that a better-than-classical quantum protocol
can be implemented in a feasible experiment.

For the moment, let us assume that the only limitation in implementing the
protocol is $\eta<1$ (we will deal with the more realistic case
below). In case of a successful detection (which occurs with
probability $\eta$) the probability of success is equal to one. In
case the detection fails (probability $1-\eta$), the last party
$P_{N}$ has to make a random guess about the value of $F$, succeeding
only with probability $1/2$. Thus for a higher-than-classical
performance we need to implement the quantum protocol with a detection
efficiency $\eta$ such that

\begin{equation}
\eta+(1-\eta)\frac{1}{2}>p_{c}.
\end{equation}
Thus, it is sufficient to have $\eta>2p_{c}-1$. In section \ref{sec
qccseqcomm} we have seen that the
optimal classical protocol for $N=5$ parties has a success rate
$p_{c}^{N=5}=5/8$, and therefore can be beaten by the quantum protocol
if the detection efficiency $\eta>0.25$, in the absence of other
experimental losses.

For a more realistic grasp of the experimental difficulties, let us
examine a simple quantum optical setup that implements the quantum
protocol for this problem. The flying qubit is encoded in the
polarisation state of a single photon. For a fair comparison with the
classical protocol, it is important to allow only a single photon per
run to pass by the parties and arrive at $P_{N}$. One way to achieve
this is to use a parametric down conversion crystal pumped by a
laser. Detection of one of the twin photons generated can then be used
as a trigger to let the second photon go towards the parties. For the
triggering mechanism to work we need to introduce a delay for the
second photon, which can be easily achieved by coupling it to a few
meters of optical fibre. Upon detection of the first photon, the
second photon is allowed to come through the $N$ parties. Each party
consists of an optical element using birefringent materials to perform
the unitary phase shift given by eq. (\ref{unitgmn}). In the end, the
last party $P_{N}$ must also detect the photon in the proper basis.

Such a setup has other imperfections that must be considered, besides
the limited detection efficiency $\eta$. The first is the finite
transmissivity $t$ of the combination of $N$ birefringent plates used
to perform the unitaries $U_j$. Another problem is the
fraction $(1-\mu)$ of detected events which are due to detector dark
counts. Finally, even if the detected photon is a signal photon, the
success rate $s$ of the quantum protocol can be less than perfect,
because of imperfections and misalignment of the optical elements that produce the initial state, introduce the phase shifts and measure the
final polarisation. Taking all these limitations into account, for a
higher-than-classical probability of success we would need:
\begin{equation}
p_{q}^{eff}=\mu\eta ts+[1-\mu\eta t]\frac{1}{2}>p_{c}.
\label{big ineq}
\end{equation}

Now let us make some realistic estimates for these parameters for the
protocol with $N=5$ parties. By using quartz plates with
anti-reflection coating, it is possible to obtain transmission of a
fraction $0.995$ of the incident photons per plate, which in the case
of five parties would result in $t=(0.995)^{5}\simeq0.975$. It is
relatively straightforward to bring dark count rates below the $1\%$
level \cite{KurtsieferOW01}, so let us take $\mu=0.99$. Good alignment
of the optical elements should enable a success rate of $s\simeq0.90$ whenever a signal photon is detected; for example, visibilities
of up to $96\%$ in simple Bell tests using entangled photons have been
reported \cite{KurtsieferOW01}.

Plugging these estimates for $t,\mu$
and $s$ in inequality (\ref{big ineq}), we see that in order to obtain a
better-than-classical probability of success it is sufficient to have
a detection efficiency $\eta\gtrsim0.33$, which is within reach of
current technology \cite{KurtsieferOW01}. It is reasonable to
conjecture that the optimal $p_{c}^{N}$ continues to decrease for
$N\geq7$, in which case the sufficient detection efficiency could be
dramatically lower. In principle, one way to calculate $p_{c}^{N}$ for
$N\geq7$ is through an exhaustive search over all deterministic
protocols, as was done in section \ref{sec qccseqcomm} for $N=3,4$ and $5$.

It is clear that essentially the same setup can be used to solve the
Modulo-4 Sum problem using classical polarised light. In common with a
qubit, classical light has a continuous variable (the phase) that can
be manipulated, as opposed to classical bits that can only assume two
discrete values. The counter-intuitive quantum feature that helps in
communication complexity is the fact that even single photons still
retain the continuous description of the classical electromagnetic
field. More generally, a $d$-dimensional pure quantum state is
characterised by $2(d-1)$ real parameters that can be used for communication purposes, as opposed to the $d$ discrete states available to
a classical system of same dimensionality. Defining exactly for which
communication tasks such a different resource can be used to advantage
is a central research problem in quantum information theory.

In the next chapter I will present a task in which the contrast
between quantum and classical information is even more pronounced. We
will see that the continuous nature of the set of quantum pure states
enables a single qubit to encode more information that any finite
number of classical bits.

\section{Chapter summary \label{sec qccconcl}}

In this chapter I have analysed a simple multi-party communication
complexity task, and different quantum protocols to solve it for the
cases of $N=3$ to $7$ parties. Simple arguments and an exhaustive
computer search enabled me to obtain the probability of success of the
optimal classical protocol for two different situations: when the
parties communicate directly with the last one, and when they
communicate sequentially to one another until the last.

In the first situation, we have seen that multi-party quantum
entanglement improves the probability of success. I also found the
detector efficiency and tolerable noise rates which are sufficient for
the protocol to work with better-than-classical performance. These
requirements are equivalent to those for a multi-qubit non-locality
test, and were shown to be less stringent than those needed for a
bipartite CHSH non-locality test.

In the second situation I compared the optimal
classical protocol with just one bit of sequential communication with
the quantum protocol in which a single qubit is communicated from
party to party through a quantum channel. I showed that the required
detection efficiency also decreases with the number of parties, and
allows for a realistic quantum optical setup with a feasible detection
efficiency $\eta=0.33$ to beat the optimal classical protocol. This is
a unique case among quantum information applications, where the
quantum performance is better than classical even in realistic
experimental conditions. The higher-than-classical performance of the
quantum communication protocol arises directly from the use of a
quantum phase to encode information.

If implemented, this would be the first experiment
to demonstrate the superiority of quantum communication over classical
communication for distributed computation tasks.

In the next chapter we will analyse a similar information processing
task, and show how a single qubit can be more efficient that an
arbitrarily large number of classical bits in some situations. We will
also see other motivations for implementing the feasible
qubit-communication protocol discussed here.



\chapter{Encoding information in a single qubit \label{chap qubit}}

A general pure state of a two-level quantum system -- a qubit -- can
be written as
\begin{equation}
\left|\psi \right\rangle=a|0\rangle+b|1\rangle, \label{qubit state}
\end{equation}
where $a$ and $b$ are complex numbers satisfying $|a|^{2}+|b|^{2}=1$. Given

that the overall complex phase has no physical significance, we can
describe state (\ref{qubit state}) with two real numbers. This means
that the full specification of a single qubit's quantum state in general requires an infinite
number of bits of information, necessary to specify these real numbers
with arbitrary accuracy.

This is in marked contrast with a classical two-level system, whose
complete information content can be encoded in a single classical
bit. Given this fundamental difference between a bit and a qubit, it is natural
to ask whether we can use a qubit to communicate an arbitrarily
large amount of classical information.

Of course, it turns out that a single qubit of communication can at
most be used to send one bit of information between two previously
unentangled parties, Alice and Bob. This is the essence of Holevo's
theorem \cite{Holevo73}, which was further refined by Nielsen
\cite{Nielsen98,ClevevDNT99}. Even in the case where Alice and Bob are
allowed to use some previously shared entangled state, a single qubit
of communication can result in a transmission of only two classical
bits (using quantum dense coding \cite{BennettW92}). This is still a
long way from the arbitrarily large amount of information we might
have hoped for.

The fact that we cannot use a qubit to communicate an unlimited amount
of information raises doubts about the reality (i.e. objective existence) of the infinite amount of information encoded in a single
qubit state. In this chapter I will show how the real nature of the
parameters in eq. (\ref{qubit state}) can be used in an information processing application. The trick
is not to require that this information can be actually read out, but
rather to use it to substitute for an unlimited amount of classical
communication during a distributed computation task. I illustrate this with a very
simple information processing task requiring an arbitrarily large
amount of classical communication, but which can be done perfectly
using a single qubit instead.

In section \ref{sec qccintro} we have seen that quantum communication
complexity results have already established that quantum communication
can be more efficient than classical communication for many
distributed computation problems. The task I discuss in this chapter
was inspired by the Sum Modulo-4 communication complexity problem
originally proposed in \cite{BuhrmanCvD97,BuhrmanvDHT99} and discussed
in chapter \ref{chap feasible}. In a way, the task we discuss in this
chapter is a continuous version of the qubit-communication protocol
presented in chapter \ref{chap feasible}. This generalisation enables
us to prove an unbounded separation in communication power between a qubit and any
finite number of bits. In sections \ref{sec whybetter} - \ref{sec
simentcla} we explore a number of consequences of this result,
relating to issues in computational complexity theory and to the
foundations of quantum theory itself.

From a computational perspective, in section
\ref{sec interprcomput} I will show the relevance of this task in
investigating the trade-offs between different types of resources (such
as memory, time, accuracy in operations) available in the quantum and
classical models of computation. We will see that in some instances
the quantum advantage arises from certain characteristics quantum
computing shares with analog computing, and not from quantum
entanglement, for example.

From a more foundational point of view, in section \ref{sec
whybetter} I will argue that the
unbounded quantum advantage can be traced back to a crucial axiom that
differentiates between classical probability theory and quantum
theory, as proposed by Hardy \cite{Hardy01, Hardy01b}. It also implies
some constraints on hidden-variable theories capable of reproducing
quantum-mechanical predictions for a single qubit.

A third way to interpret this result sheds light on an altogether
different quantum resource: quantum entanglement. In section \ref{sec
simentcla} I will show that no finite amount of one-way classical
communication can perfectly simulate the communication advantage
provided by entanglement.

\section{The task}

Let us now define an information processing task, whose classical and quantum solutions we will analyse in the next sections.

We have a one dimensional real field $\varphi(x)$ defined on the
straight line between points $A$ and $B$ in the $x$-axis. The
integrated value of the field is guaranteed to be equal to an unknown
integer $m$, times a known real constant, $\alpha$:

\begin{equation}
\int_{A}^{B}\varphi(x) dx= m\alpha. \label{constraint}
\end{equation}

\begin{figure}
\begin{center}
{\includegraphics[scale=0.65]{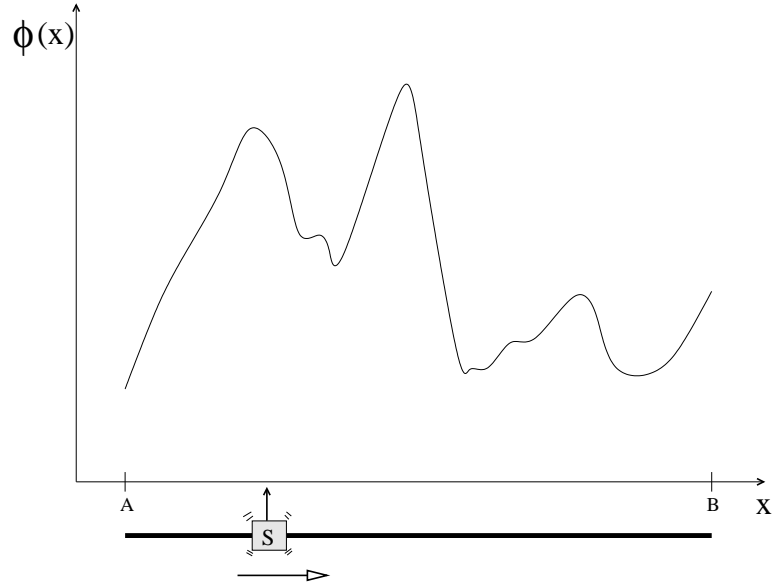}}
\caption[An information processing task.]{\label{fig qubit1}We are allowed to take system $S$ from points $A$ to $B$ in
the $x$-axis, coupling it in any local way with the field
$\varphi(x)$. The task is to obtain one bit of information about
$\int_A^B \varphi(x) dx$, as described in the main text. }
\end{center}
\end{figure}

The task is simply to find out whether $m$ is odd or even by sending
some physical system $S$ directly from point $A$ to point $B$ (it is
not allowed to move backwards). We are allowed to couple the system in
any local way with the field $\varphi$ (see figure \ref{fig qubit1}), and we are searching for a
deterministic protocol that solves the task with certainty.

In the next section I will present two simple protocols that solve
this problem, one using a classical system $S$ and one using a quantum
$S$. The classical $S$ we use is a system with an infinite number
of distinguishable states, such as a classical pointer. Communication with such a system effectively represents the
transmission of an unlimited amount of classical information.

The quantum protocol I will consider also solves the task, but using
only a single qubit of quantum communication. After presenting these
two contrasting protocols, in section \ref{sec noclassprot} I will
show rigorously that this huge difference in communication resources holds between the quantum protocol and \textit{any} classical protocol. The consequences
of this will be explored in sections \ref{sec whybetter}, \ref{sec
interprcomput}, and \ref{sec simentcla}.

\section{Two simple solutions}
\subsection{A quantum protocol \label{sec Faraday}}

Let the quantum system $S$ be a spin-half particle initially prepared
in the spin up (along the $z$ direction) state. Now let us couple the
qubit to the field so that the spin orientation will rotate in the
$y-z$ plane by an amount proportional to the strength of the
field. This constant of proportionality can be chosen to be such that, by the time $S$ reaches $B$, the spin has been
rotated through $m/2$ full rotations. Thus, if $m$ is even then the
spin will be up, and if $m$ is odd it will be down. The task can then
be solved by a simple measurement of the spin along the $z$ direction
at $B$.

This protocol can be realized experimentally in a simple way using the
following quantum optical setup. We encode the classical field
$\varphi(x)$ in the magnetic field
$B_{x}=\overrightarrow{B}\cdot\hat{x}$ in the centre of a solenoid. In
principle, arbitrary functions $\varphi(x)$ can be generated by using
a non-uniform current density; the field $B_{x}$ will be proportional
to the local current density.

In the centre of the solenoid we have a transparent rod. If a
photon with vertical polarisation is sent through the rod, its
polarisation vector at the other end will have been rotated by an
amount proportional to the integrated value of the field (due to the
Faraday effect). The constant of proportionality associated with the
interaction can be altered, for example by changing the physical properties of the
transparent rod. The polarisation can be analysed at $B$ along the
vertical-horizontal directions. If the constant of proportionality is
correctly adjusted, the polarisation at $B$ will be vertical in the
case where $m$ is even and horizontal if $m$ is odd, solving the task.

\subsection{A classical protocol \label{sec classanalog}}

To solve the problem with a classical system, let $S$ be a rod pivoted
at one end and free to rotate in the $y-z$ plane. The position of the
rod is given by a real number $\theta$ whose specification requires an
infinite number of classical bits. In other words, the rod is a
classical system with an infinite number of distinguishable
states. From the computational point of view we will develop in
section \ref{sec interprcomput}, we will see that the rod can be thought
of as a classical analog computer memory, or alternatively as an infinite
digital memory.

To solve the above information processing problem, we start with the
rod in the up direction. We couple the rod $S$ to the field so that as
it moves through distance $dx$ we rotate it by $\eta\varphi(x)dx$. By
choosing the coupling constant $\eta=\pi/\alpha$, the rod will be
rotated by $m/2$ turns. At the end the rod will be pointing up if $m$
is even and down if $m$ is odd.

At this point it is natural to ask whether any classical solution
really requires a system $S$ with an infinite number of
distinguishable states. After all, the final position of the rod is
giving us much more information about $\varphi(x)$ than we actually
need; we can actually read out the value of the integral of
$\varphi(x)$ modulo $2\alpha$ (given by the rod's final position) and
not just the answer to our yes/no question. This indicates that a more
careful analysis is required before we conclude that it is impossible
to solve the task with a classical $S$ having only a finite number of
distinguishable states. In the next section I will prove that this is
indeed the case.

\section{Necessity of unlimited amount of classical communication
\label{sec noclassprot}}

The classical protocol I presented above requires communication using
a classical continuous-variable system with an infinite number of
distinguishable states. By using such a system, we are effectively
communicating an unlimited amount of information, encoded in
the continuous parameter characterising the rod's position. In this
section I will prove that the stated task cannot be solved by any
classical protocol using only a finite amount of classical
communication, encoded in a system with any finite number of
distinguishable states.

We start by dividing the interval $AB$ into $N$ equal sections, each
of length $1/N$, and label them sequentially from $1$ to $N$, starting
from $A$. Now define $\phi_{n}$ to be the integrated value of the
field $\varphi(x)$ over the $n$th section. Constraint (\ref{constraint}) now reads:
\begin{equation}
\sum_{n=1}^{N}\phi_{n}=m\alpha. \label{summalpha}
\end{equation}

We do not need to impose any restrictions on the nature of the
function $\varphi(x)$ except that it can be integrated. If we want
$\varphi(x)$ to correspond to a physical field it must be continuous,
but that is not required for our proof to work.

Even with continuous $\varphi(x)$, there is no constraint on the numbers
$\phi_{n}$ except that they sum to $m\alpha$. Hence, we can find
fields $\varphi(x)$ which correspond to any point in the coordinate
space $\{\phi_{n}\}$ for $n=1$ to $N-1$, with $\phi_{N}$ being chosen
so that eq. (\ref{summalpha}) holds.

By breaking the problem in $N$ parts, we effectively have $N$ parties
in line, who can communicate sequentially from the first to the last
one, with the objective of enabling the last one to solve the
problem. After this simplification we see the similarity with the
sequential communication version of the communication complexity
problem discussed in chapter \ref{chap feasible}.

If this communication task can be
solved for an arbitrary field $\varphi(x)$, then it can be solved for
any subset of points in the coordinate space $\{\phi_{n}\}$. Hence, we can further simplify by
choosing a discrete set of values that the $\phi_{n}$ can take, to
enable us to prove the result in a simple way. We allow each $\phi_{n}$ to take only discrete values ${\alpha
k_{n}/K}$, where $k_{n}\in\{0,1,...,2K-1\}$. We will also require the
integer $K$ to be a power of two, for a technical reason
that will become apparent later. Note
that $\phi_{n}$ takes values in the range $0$ to $\alpha(2-1/K)$. This
makes sense as the sum is only important modulo $2\alpha$ (since we
are only interested in whether $m$ is odd or even).

From now on we will refer to the $k_{n}$'s instead to the $\phi_{n}$'s. Condition (\ref{summalpha}) becomes
\begin{equation}
\sum_{n=1}^{N}k_{n}=mK.
\end{equation}
Again, since we are only interested in whether $m$ is odd or even we
will only be concerned with sums over the $k_{n}$'s modulo $2K$.

\subsection{Notation}

Now, let us establish a notation that enables us to discuss all
possible classical protocols involving communication with a
finite-dimensional classical system $S$. Let $L$ denote the number of
distinguishable states of $S$. Communication with a classical system
with $L$ distinguishable states is the equivalent of sending a message
of size $\log_{2}L$ classical bits.

Each party has a number $k_{n}$ (or equivalently $\phi_{n}$). The
$n$th party receives $S$ from the ($n-1$)th party. This will be in a certain state
$l_{n-1}\in\{1,2,...,L\}$. Since the first party receives no
information we will put $l_{0}=1$ so we can still use this
notation.

The most general protocol the $n$th party can follow consists of
selecting a function $f_{l_{n-1}}^{n}(k_{n})$ which outputs a number
$l_{n}\in\{1,2,...,L\}$. He then prepares system $S$ in this state
$l_{n}$, sending it on to the ($n+1$)th party. The protocols adopted
by all the parties must enable the last party ($N$th) to observe
system $S$ in state $l_{N-1}$, and use this together with his local
information $k_{N}$ to determine whether $m$ is odd or even.

In the next section I will show that for a perfect classical protocol
it is necessary to have $L\geq2N-1$. Since $N$ and $K$ are arbitrary
and can be chosen to be as large as we want, the number of bits
encoded in system $S$ must also be arbitrarily large.

\subsection{Proof \label{sec qubitinfproof}}

We start by choosing the number of distinguishable states $L$ to be
less than what is necessary to encode each party's number; so let us
set $L=2K-1$. The first party sends message
$l_{1}=f_{0}^{1}(k_{1})$ to the second party. However, this function
cannot be one to one since $L<2K$. Thus, for some $l_{1}$, there must
exist $a \ne b$ such that
$f_{0}^{1}(a)=f_{0}^{1}(b)=l_{1}$. Because of that, when the second
party receives the message $l_{1}$ he does not know whether $k_{1}=a$
or $k_{1}=b$.

The $n$th party will try to enable the next one to learn the value of the
partial sum $(k_{1}+k_{2}+\dots+k_{n})\mbox{ mod }2K$, as this is
the data that the last party needs to know in order to solve the
problem. However, we have just seen that after just one communication step, there are already two different values of the
first party's data $k_1$ assigned to a particular message $l_1$. The
idea of the proof is to show that this uncertainty increases with each
communication step, until there exists a message $l_{N-1}$
that leads the last party to error.

Consider the $n$th party. Let $A_{n}$ be the set of distinct values of
the partial sum $(k_{1}+k_{2}+\dots+k_{n-1})\mbox{ mod }2K$ which are
consistent with the message $l_{n-1}$ he has received. In the case
of the last party, we identify here a possible cause for a flawed
protocol. If $A_{N-1}$ has elements that differ by $K$, there will
exist one value of $k_N$ for which the last party will be unable to
solve the problem with certainty.

The $n$th party will send $l_{n}=f_{l_{n-1}}^{n}(k_{n})$ to the
($n+1$)th party. For some $l_{n}$ there must exist distinct $a$ and
$b$ such that $l_{n}=f_{l_{n-1}}^{n}(a)=f_{l_{n-1}}^{n}(b)$ (since
$L<2K$). If the $(n+1)$th party receives this $l_{n}$ then
\begin{equation}
|A_{n+1}|\geq|A_{n}\oplus\{a,b\}|,\label{anplus1}
\end{equation}
where $|A|$ denotes the number of elements of the set $A$ and $A\oplus
B$ is the set of all distinct sums, modulo $2K$, of one element from
$A$ and one element from $B$. Already we see that the number of
elements of $|A_{n}|$ cannot decrease as $n$ increases.

Let us now prove that any successful protocol must have
\begin{equation}
|A_{n}\oplus\{a,b\}|\geq|A_{n}|+1 \hspace{2 mm}, \label{gthana}
\end{equation}
which together with (\ref{anplus1}) means that $|A_{n+1}|$ is strictly
larger than $|A_n|$.

The proof will be by contradiction. Let us thus assume that there is a
successful protocol in which eq. (\ref{gthana}) is false. This is equivalent to saying that $|A_{n}\oplus\{a,b\}|=|A_{n}|$, as $|A_{n}\oplus\{a,b\}|$ cannot be
less than $|A_{n}|$. Note also that
$|A_{n}|=|A_{n}\oplus\{a\}|=|A_{n}\oplus\{b\}|$. From our assumption and from writing
\begin{equation}
A_{n}\oplus\{a,b\}=(A_{n}\oplus\{a\})\cup(A_{n}\oplus\{b\}), \label{aunionb}
\end{equation}
it follows that $A_{n}\oplus\{a\}=A_{n}\oplus\{b\}$ (if they were not
the same, their union would have more elements than each of them separately).

Since we are doing modulo arithmetic we can think of the members of
the set $A_{n}$ as being arranged around a circle. Then the effect of
the $\oplus\{a\}$ operation is to rotate them all forward by $a$ and
similarly for the $b$ case. Hence, $A_{n}\oplus\{a\}=A_{n}\oplus\{b\}$
implies that $A_{n}=A_{n}\oplus\{\left|  b-a\right|  \}$. Put
$|b-a|=\Delta_{1}$. We can apply this shift as many times as we
like. Hence, $A_{n}=A_{n}\oplus \{i\Delta_{1}\}$ where
$i=1,2,\cdots$. It is possible that $\Delta_{1}$ divides $2K$ in which
case it is a period of the set $A_{n}$. If it is not period then we
can still prove that $A_{n}$ must have a period. To see this note that
either $\Delta_{1}$ is a period or there exists an integer $i_{1}$
such that $0< i_{1}\Delta_{1}\mod2K<\Delta_{1}$. Put
$\Delta_{2}=i_{1} \Delta_{1}\mod 2K$. Now,
$A_{n}=A_{n}\oplus\{i\Delta_{2}\}$ and hence either  $\Delta_{2}$ is a
period or there exists an integer $i_{2}$ such that $0<
i_{2}\Delta_{2} \mod2K<\Delta_{2}$. We can continue in this way
generating a sequence of $\Delta_{j}$'s. At some point this sequence
must terminate since $\Delta_{j+1}<\Delta_{j}$ and
$\Delta_{j}\geq1$. The last member of the sequence must be a period of
$A_{n}$. Let this period be $v$.

As we have chosen $K$ to be a power of two, this divisor $v$ must be
either equal to one, or to some power of two as well. These two possibilities entail the existence of elements of $A_n$ which differ
by $K$, leading to an error. Since we assumed from the beginning that
there was a successful protocol in which eq. (\ref{gthana}) was false,
to avoid a contradiction we are forced to conclude that
eq. (\ref{gthana}) must be true.

Thus, in  any successful protocol eq. (\ref{gthana}) must hold for all
$n<N$. The last party then may receive a message for which
$|A_{N-1}|=N$. If this number is larger than $K$ then there must exist
elements in $A_{N-1}$ separated by $K$, leading to an error. So we need $K \ge N$. Now let us remember
that we fixed the number of distinguishable states in system $S$ to be
$L=2K-1$. This means that any flawless protocol requires that $L
\ge2N-1$. As $K$ and $N$ can be chosen at will, this proves that any
classical protocol for the task proposed requires a system $S$ with an
arbitrarily large number of distinguishable states.

\section{Why is quantum better?\label{sec whybetter}}

It is instructive to revisit the classical protocol and spot the
reason why the rod could be replaced by a qubit. The rod's coupling
with the field $\phi(x)$ takes it through its continuous set of
distinguishable states in such a way that its final state encodes the
property of $\phi(x)$ that we want to know. The key point to note is
that the protocol works just as well if we never acquire information
about the rod's state until the final point $B$. There, constraint
(\ref{constraint}) guarantees that acquiring a single bit of
information about the rod's state is enough to solve the task. The rod
offers us more than we actually need: it is enough to have just two
distinguishable states for the read-out, plus the continuous set of
pure (but non-distinguishable) states for the encoding. This is
precisely what a qubit offers us.

It is clear that the task could be solved by using a classical
electromagnetic field instead of a single polarised photon, as we have
discussed in section \ref{sec Faraday}. The protocol using a single
photon, however, is as economical as possible, working perfectly
with a communication consisting of the smallest possible physical system.
Besides this practical point, we have to remember that a classical
electromagnetic field, with its infinite number of distinguishable
states, is only an approximation to the actual physical properties of
the quantum electromagnetic field.

By quantising the classical electromagnetic field, we obtain a
more physically accurate description of the system, and find out that
the single building blocks (the quanta) retain a curious mix of
characteristics. Unlike the classical field, they have only a finite number of
distinguishable states (hence the term `quanta' to describe them and
the whole theory). On the other hand, they somehow still retain the
continuous classical description of the classical field they constitute, by
having a continuous set of pure, but non-distinguishable, states.

Interestingly, the existence of this continuous set of pure states
turns out to be the distinguishing trait between quantum theory and
classical probability theory. We have seen this in section \ref{sec
hardyaxioms}, where we briefly presented Hardy's axioms for quantum
mechanics.

In Hardy's approach, the crucial difference between quantum and classical
physics is the continuous nature of the set of quantum pure states, as
opposed to the discrete nature of the set of classical pure states. As
we have seen, it is also this that allows for an
unbounded separation in communication power between a qubit and any
finite-dimensional classical system.

We thus see that the unbounded separation we have noted is not simply
a curious event arising from some peripheric
characteristic of the single qubit case. It actually results from the
unavoidable, fundamental trait that distinguishes quantum theory from
classical probability theory.

The result I have presented in this chapter places a significant
constraint on any hidden-variable theory (HVT) describing a single
qubit. It implies that the HVT must have at least one continuous
parameter, capable of taking on an arbitrarily large number of
values. While this is indeed the case for Bell's deterministic
hidden-variable model for a single spin-half particle \cite{Bell66},
my result shows that this is an unavoidable trait of any HVT
consistent with quantum theory.

\section{A computational perspective \label{sec interprcomput}}

In this section we will analyse the information processing task I
proposed from a computational perspective. I will start by re-stating
the bound on communication resources as a bound on the sizes of
classical and quantum memories necessary for some computations. This
is a result in computation space complexity, which can be motivated
using the `ROM-based' computation model
of Travaglione \textit{et al.} \cite{TravaglioneNWA01}, which I also
present in the next section.

Then I will show that this quantum advantage in memory size comes at
the expense of an exponentially increasing accuracy required for the
quantum operations. Fortunately, we will see that this accuracy can
be reached with polynomial resources at each party.

I will argue that this is an example of the more general problem of
computational resource trade-offs between different models of
computation (in this case, classical and quantum).

In the remainder of the chapter we will often be analysing how
different resources scale with the problem size. It is natural to take $n=\log_2{N}$ as the
problem size, where $N$ is the number of parties in the discretized
version of the communication protocol introduced earlier. While there
are other ways to define problem size, for our purposes it is
sufficient to stick to a consistent definition, as we will be only
concerned with the asymptotic \textit{differences} between the quantum
and classical models.

\subsection{Time- versus space-complexity}

When computer scientists analyse the complexity of an algorithm, they
usually put more emphasis on the \textit{time} complexity, which deals
with the expected number of computational operations needed
for performing an algorithm. It has been shown that quantum computing
offers a decrease in the time complexity of important algorithms: a
quadratic speed-up for database search (Grover \cite{Grover97b})
and a conjectured exponential speed-up for integer factoring (Shor
\cite{Shor94, Shor97}).

Time, however, represents only one of the resources necessary for
computing. \textit{Space} complexity analyses how much writable memory
is necessary to perform different algorithms, and how the size of this
memory scales with problem size. This problem has been addressed in a
few recent papers \cite{AmbainisF98,Watrous99,TravaglioneNWA01,
SypherBWHT02}, but in general it has drawn much less interest than
time-complexity.

In \cite{TravaglioneNWA01} Travaglione \textit{et al.} discussed what
they called the `ROM-based model' of computation, and were able to
obtain a simple analytical result comparing the classical and the
quantum space-complexity. Since this is important for understanding my
result, let us have a look at their findings.

\subsubsection{ROM-based computation \label{sec rombased}}

If we want to analyse space efficiency of computers, it is convenient
to distinguish between that part of the memory that is used for the
read-only data, and the part of the memory that is actually operated upon during
the computation. This motivated Travaglione and co-workers
\cite{TravaglioneNWA01} to define a deterministic, reversible model
that enabled them to compare the space-efficiency of quantum and
classical computers.

Their model can be presented as follows. All the input data necessary
for the computation is available as a Read-Only Memory (ROM) of $j$
bits $u_1 u_2 u_3,...,u_j$. Besides this ROM, we have also $m$ (qu)bits as
writable (qu)bits, which can be operated upon one at a time by
reversible gates conditionally on the value of one of the ROM
bits. The gates are required to be reversible for a fair comparison
between the quantum and the classical models, since the quantum memory
necessarily undergoes a unitary (hence reversible) evolution. After
many such computational steps, in the quantum case we require the
quantum writable memory to be in a disentangled state, consisting of
the chosen computational basis states. This results in a computation which
is deterministic and reversible, in both cases, classical or quantum.

A ROM-based computer is called universal if it is capable of
transforming the $m$ writable (qu)bits to any one of the $2^{\wedge}
(m2^{j})$ possible boolean outputs of the $j$ ROM-bits. Travaglione \textit{et al.} proved
that universal computers in this definition require only a single
qubit of quantum memory, or two bits of classical memory. Their proof,
however, does not state anything about the difference in
time-complexity between the quantum and the classical models.

Next we will examine the relation between this model of memory-bounded
computation and the task we are studying in this chapter. We will see
that our task corresponds to a specific ROM-based computation, in
which we first constrain the number of times the ROM bits can be read,
and then ask what the size of the minimal computer memory must be for
a certain deterministic computation.

\subsubsection{Task as a ROM-based computation}

We can view the discrete version of the protocol we have studied in
this chapter as such a ROM-based computation. We start by defining a
function $g:m\rightarrow k$, where $m\in\{1,2,...,N\}$ represents each
of the parties and $k\in\{0,1,...,2K-1\}$ represents the numbers each party
has. The constraint on the integral of $\phi(x)$ translates as a
constraint on the function:
\begin{equation}
\sum_{j=1}^{N}g(j)\mbox{ mod }2K=0 \mbox{ or }K. 
\end{equation}
The Boolean function we want to evaluate is one which distinguishes
between these two possibilities.

The role of the ROM bits is played here by each party's data, which
they can use to conditionally apply unitaries on the quantum
register, or to change the state of the classical digital
register. We have a `ROM call' each time a party uses his data to
operate on the quantum memory.

The universality result for ROM-based computation means here
that this task can be solved either with a classical memory
of just two writable bits, or with a quantum memory of a single
qubit. It is possible that there exists a huge gap in time-efficiency
between these two space-efficient computations; however, as far as space
complexity goes, the separation between classical and quantum in this case
is very small.

The problem we solved in this chapter imposes a further constraint:
each party has access to the memory only once, before passing it to
the next party in line. This corresponds to a constraint in time: the protocol must end after exactly $N$ operations on the
memory, one by each party. So here we are actually imposing
constraints both on time- and space-efficiency. We start by fixing the
allowed time-complexity (in number of ROM calls) to be $N$, and then
we ask what the minimal memory size is, for classical or quantum memories.

By fixing the time of the computation to be $N$ function calls, we are
effectively allowing the parties the exact minimum number of ROM calls
necessary for the computation. This is true because a change in a
single party's number can flip the overall sum from $0$ modulo $2K$ to
$K$ modulo $2K$, changing the result of the computation. This means
that if we miss any single party's number, we could be led to error in
the final result. Therefore, the minimal number of ROM calls necessary for this task is exactly one per party, or $N$ in total.

Now let us re-interpret our result: we have shown that with this
minimal time complexity, any classical ROM-based computer needs a
writable memory of at least $\log_{2}(2N-1)$ bits. In terms of the
problem size $n=\log_2{N}$, the classical memory is $O(n)$,
i.e. linear in $n$. A quantum computer, on the other hand, succeeds
with a fixed memory of the smallest possible size, consisting of
a single qubit. This is a striking separation between the capabilities
of these two models of computation.

Here we must add a word of caution about our time-complexity
definition. As is common in such discussions, we adopted a time
complexity which is measured by the number of function calls,
sometimes called in the jargon \textit{oracle} calls. This, however,
can be misleading, as each oracle call can involve a complex quantum operation
which requires a large number of elementary gates to be performed.

For a more physically motivated definition of the time complexity, we
need to actually address the complexity of implementing each oracle
call. As each party has a classical number, our oracle already
consists of a classical oracle in both the quantum and classical
cases, which is good. But depending on the oracle result, each party
has to implement a unitary on the qubit, whose complexity we must
evaluate.

In the next sections we tackle this problem. We will start by showing
that the quantum algorithm with small error $\delta$ requires an exponential accuracy for each
party's unitary operation, which is bad news. Surprisingly, however,
we will see that this exponential accuracy in each unitary can be
reached with a number of gates which is only polynomial, which is good
news.

This will enable us to compare the quantum and classical computations
in a fairer way. We will see that, with the same number of function
evaluations, quantum computation offers a polynomial space advantage
over classical computation, while requiring a polynomial time
overhead.

\subsection{On the accuracy needed \label{sec accrequ}}

Even in classical computational complexity theory, there are more
resources to account for, besides time and space. For example, the
work of Landauer \cite{Landauer61} has shown that there is a
lower bound on the energy consumption of irreversible
computation. This motivated the study of classical reversible
computation, which can be done essentially with no energy consumption \cite{Bennett73}.
In today's computers energy use is
not a major concern, but this may change in future computer
architectures.

When we consider quantum computation, it is not yet completely clear
which resources will prove important to consider, from a
physical point of view. Suppose, for example, that in the next decades
we obtain a better understanding of quantum theory, or maybe a more
successful theory that supersedes it. It may be the case that such a
theory imposes restrictions on the type of entanglement that can be
physically produced, through decoherence, for example. This
possibility is already being considered in different contexts, for
example in Penrose's idea for gravitationally-induced decoherence
\cite{Penrose96}. Then it would become important to quantify how much
of this `difficult' entanglement is required by quantum algorithms,
and count that as a computing resource to be minimised.

It is already clear, however, that the \textit{accuracy} with which we can
implement single qubit gates is limited. Much work has been done to
investigate what is the number of gates required to achieve the
necessary accuracy for quantum unitaries during a quantum
computation. For an introductory treatment of this problem
and references, see Chapter 4 of the book by Nielsen and Chuang
\cite{NielsenC00}.

This issue is important in the context of the computational problem I
have presented in this chapter. The first thing to note is that for an
absolutely perfect quantum protocol, the accuracy needed is infinite,
as any small inaccuracy in the implementation of a single unitary in
general leads to some probability of error at the end. This, however,
can be also said of the digital memory involved in the classical
protocol, as its performance also assumes perfect circuit elements.

What we can do from a theoretical point of view is to calculate the
accuracy needed to implement each quantum unitary in our problem, so
that the quantum protocol succeeds with high probability
$1-\delta$. We will see that the accuracy needed increases
exponentially with the problem size $n$.

The feasibility of quantum computation networks depends on some
fundamental results. One of the first things we need to know is
whether it is possible to compose many imperfect unitary operations, in
such a way that the final probability of error does not increase too
fast with the number of unitaries. Fortunately, it was shown (see
section 4.5.3 of \cite{NielsenC00}) that this is indeed the case. More
precisely, in order to approximate $N$ successive
unitaries with a final probability of error smaller than $\delta$, it
is enough to approximate each unitary with an error bounded by
$\epsilon \le \delta/(2N)$. In other words, the errors in the
individual unitaries compose only linearly with the number of
unitaries applied.

If we go back to our problem, this means that each party must be able
to implement his unitary with an error of at most
$\epsilon=\delta/(2N)$. This may not seem too bad, but we must
remember that the problem size is given by $n=\log_2{N}$, where $N$ is
the number of parties. This means that the accuracy necessary for each
party's unitary operation increases exponentially with problem size.

\subsection{Achieving the accuracy}

It may come as a surprise that this exponential accuracy can be
achieved with just polynomial resources at each party. In order to
evaluate the resources needed, we need to assume that each party can
implement a universal set of only two single-qubit gates with
perfection. These can be the $\pi/8$-rotation and Hadamard gates, for
example. This assumption may seem too idealistic, but represents the
same kind of assumption we made to derive the classical digital
protocol -- that we can perform classical bit operations perfectly.

Going back to the quantum case, each party then needs to apply a
certain sequence of these gates in order to approximate the desired
unitary to accuracy $\epsilon$. How many times does he need to apply
the gates?

The Solovay-Kitaev
theorem gives a near-optimal bound answering this question
\cite{Kitaev97}. It was shown that to approximate a unitary with an
error of at most $\epsilon$, we need to apply the gates from our
discrete set only O($\log_2^{c}(1/\epsilon))$ times, where $c$ is a
constant approximately equal to two \footnote{More precisely, a proof
can be given for $2<c<2+\eta$, with $\eta$ arbitrarily small. See \cite[Appendix 3]{NielsenC00}).}. As we have seen above, for a
computation with error bounded by $\delta$, we require $\epsilon
\le\delta/(2^{n+1})$. Using the Solovay-Kitaev result, we see that
this requires each party to use $O(\log_2^{c}{2^{n}/\delta})$ gates,
which is only polynomial in $n$.

So we see that the exponential accuracy at each party can be reached
with a number of gates which is polynomial in $n$. Since the number of
parties is exponential in $n$, so is the total number of gates
necessary for a quantum computation with high probability of
success. We must remember, however, that this is also the case for the
classical computation.

To summarise what we found in the last two sections, we have seen that
our task can be computed either with a digital classical memory or
with a quantum memory. The same exponential number $N=2^n$ of function
calls is necessary in both cases, and each function call requires a
manipulation of the memory involving a polynomial number of
gates. In the classical case the number of gates at each party is
$O(n)$, that necessary for adding two $n$-bit numbers. In the quantum
case this number is $O(n^c)$, where $c$ is approximately equal to
two. Interestingly, with the same number of function calls a quantum
computer performs the computation using only a single qubit memory, as
opposed to the necessary $O(n)$-bit classical memory.

This is just one example of a trade-off between computation resources
in different models of computation. In our case, we increase the
number of gates used by a polynomial amount, while reducing the memory
used also by a polynomial amount (from $O(n)$ to 1,
i.e. constant). Studying trade-off compromises improves our
understanding of the advantages provided by different models of
computation, such as analog or quantum. It is also important for
experiments involving small-scale quantum computers, as
it helps quantify the advantage one can get with limited resources of
different types. Some related work has been presented by Sypher
\textit{et al.} in \cite{SypherBWHT02}, where some simple computations
using a quantum computer with one and two qubits were experimentally performed.

\subsection{Quantum versus analog computation}

In the sections above we have been comparing a perfect digital
classical computation with a quantum computation using a qubit that
succeeds with high probability $1-\delta$. As we have noted already in
section \ref{sec classanalog}, however, there is a simple way of
solving the task with an analog classical memory (a rod). In this
section I compare the quantum computation using a qubit with this
classical analog computation.

It is important to note that the classical analog computation model is not physically sound -- at some point quantum
phenomena and the atomic structure of any analog device will make the
system deviate from our classical expectations. This is already a
good argument for consideration of quantum systems instead, because
they reflect the computational capacity of Nature to the best of our
current knowledge.

Unfortunately, there is relatively little reported research work on the complexity of analog computation. It has been noted that it
is possible to make some analog models which solve factoring and the
satisfiability problem in polynomial time \cite{Vazirani98}. In
particular, there has even been a proposal for a prototype of a fast
factoring device using light-emitting diodes (LED's) and photodetectors
\cite{Shamir79,LenstraS00,Shamir99}. These models, however, in general
have have limited scalability due to exponentially increasing
resources of some kind. Defining which resources have to be considered
in alternative analog models of computation is an open research
problem. Some aspects of analog computation complexity were discussed
by Vergis \textit{et al.} in \cite{VergisSD86}.

This notwithstanding, we can at least compare some characteristics of
the ideal quantum and (analog) classical computation of our task. Let us start by presenting some of the
advantages of the analog classical rod over our qubit-based
computation. First, the rod is more powerful because it enables us to
read out $\int \varphi(x)dx \mbox{ mod }2\alpha$, instead of the just
the single bit necessary for our task. Moreover, the analog
computation is robust against small imperfections in the rotations, as
the rod still ends up close enough to the ideal final position for us
to read out the correct answer.

The quantum computation with a qubit is not robust in that sense, as any imperfection in one unitary generally results
in some probability of error at the end. This is not much of a problem
if we accept a small probability of error $\delta$ at the end. The
demands for such a probabilistic quantum computation were evaluated in
the last two section, in terms of accuracy and number of gates. Despite the fact that it is not
clear how to weigh computational resources in analog computation, it
is reasonable to claim that the quantum computation with a qubit is
enormously more economical in space than the analog
computation. Any analog system used as a memory for our task must be
physically constituted by a great number of truly fundamental quantum
systems, each of which can be compared with the qubit.

A fairer comparison of the quantum and analog solutions to our task
would demand further work to be done. First, we would need to
establish a physically reasonable model for errors in the rod's
rotations, and evaluate the computations' tolerance to them. Then we
could compare that to the analysis we did in the previous two
sections, for an error-bounded quantum computation.

\section{Simulating entanglement with classical communication
\label{sec simentcla}}

Entanglement figures as a resource in a number of communication
tasks. It provides space-like separated parties with a means of
obtaining a string of bits which are correlated more strongly than is
possible with classical, local physics. If the parties are
time-like separated, however, the same strongly correlated string of
bits can be obtained using some classical communication between the
parties. By
studying how much classical communication is necessary to simulate the
presence of entanglement in this sense, we can get a better
understanding of the communication advantage provided by entanglement.

\begin{figure}
\begin{center}
{\includegraphics[scale=0.70]{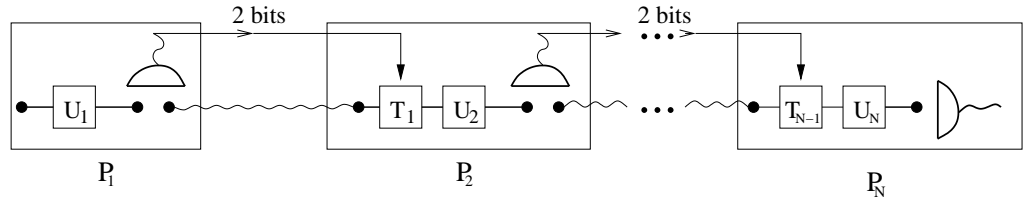}}
\caption[Solution to the task based on teleportation.]{\label{fig qubit2} Instead of sending the qubit through a
quantum channel, we can teleport it from party to party. Each party
 needs to transmit two bits to the next party $P_j$, who then
 implements a suitable unitary $T$ to complete the
 teleportation. Then the unitary $U_j$ is applied, depending on
 $P_j$'s data. After that the party teleports the qubit to
 the next one, through a Bell-basis projection of the two qubit system
 consisting of the resulting qubit, plus the qubit entangled with the
 next party.}
\end{center}
\end{figure}

The task we have been studying in this chapter can be used to shed
some light on the cost of simulating entanglement with classical communication. In order to see
how, let us go back to the discrete version of the quantum protocol
(with $N$ parties), and imagine that the parties do not have a quantum
channel to send the qubits through. Instead of offering the parties
the advantage of quantum communication, we give them quantum
entanglement: each neighbouring pair of parties starts with a maximally entangled
pair of qubits. Now they can overcome the lack of a quantum channel
and solve the task using $(N-1)$ teleportation steps (see figure
\ref{fig qubit2}). Each
teleportation step requires only two bits of one-way communication,
plus the previously shared entanglement.

In section \ref{sec qubitinfproof} we proved that no finite amount of
one-way classical communication can perfectly solve the task. The
procedure outlined above, however, succeeds with only two bits of
one-way classical communication, plus entanglement. We are thus forced
to conclude that no finite amount of one-way communication can
perfectly simulate the communication advantage provided by entanglement.

This contrasts with previous results obtained by various authors,
which show that entanglement simulation can be achieved with a finite
amount of communication only
\cite{BrassardCT99,CerfGM00,MassarBCC01}. Next we will have a brief
look at these results, and show how they can be reconciled with our
proof.

\subsection{Some results on entanglement simulation}

Brassard \textit{et al.} \cite{BrassardCT99} start by defining exactly
which quantum
measurements we are trying to simulate, and the means allowed. They define a \textit{quantum measurement scenario} with a triple of
the form ($\left|\Psi\right\rangle_{AB}, M_A, M_B$), where
$\left|\Psi\right\rangle_{AB}$ is the bipartite quantum state in consideration, and $M_A$ and $M_B$ are the sets of measurements at $A$
and $B$ whose results we will try to simulate.

We are allowed to use any local hidden variable scheme, augmented by a
certain amount of classical communication between the parties (which
we will try to quantify). Their main result can be stated in a theorem:

\begin{description}
\item[Theorem (Brassard \textit{et al.} \cite{BrassardCT99}):] For the quantum measurement
scenario ($\left|\Phi\right\rangle_{AB},M_A,M_B$), with
$\left|\Phi\right\rangle_{AB}=\frac{1}{\sqrt{2}}(\left|00\right\rangle
+ \left|11\right\rangle)$ and $M_A,M_B$ consisting of any von Neumann
measurement on each qubit, there exists a local hidden variable scheme augmented
with eight bits of communication (from Alice to Bob) that exactly
simulates it.
\end{description}

Another protocol related to entanglement simulation was presented in
\cite{CerfGM00}, where Cerf, Gisin and Massar discussed what they called
`classical teleportation'. They considered the situation in which Alice
needs to send information to Bob about a \textit{known} state she has,
in such a way that Bob can reproduce any measurement outcome on that
state. This is related to Brassard \textit{et al.}'s protocol, but
there are some significant differences. Unlike them, Cerf \textit{et
al.} consider more general POVM's, and not only projective
measurements.

Their main result is that on average, only a finite amount of
bi-directional communication is enough to do this `classical
teleportation'. Note, however, that our task demands only one-way
communication, which means that their result need not apply in our
case.

In the next section we will discuss the relation between our task and
the type of entanglement simulation discussed in the two papers
briefly reviewed here.

\subsection{Simulating our task}

The results reviewed in the previous sections imply that a finite
amount of communication is sufficient to simulate entanglement. In
this section we will qualify this statement, explaining why it does
not apply to the simulation of entanglement-based version of our task.

The theorem by Brassard \textit{et al.} implies that it is easy to
simulate one teleportation step with classical communication. This may
not be intuitive but, given their results, we must accept that some
operations (involving a few real-valued parameters) on a single
entangled pair can be perfectly simulated with classical
communication. Despite the communication being of just a few bits, we
must remember that the parties can previously share an unlimited
amount of instructions, i.e. local hidden variables.

The situation proposed in our task is different, as the outcomes of the measurement made at the last party
actually depend on a (potentially unlimited) \textit{chain} of
real-valued processes affecting the quantum systems used, such as
unitaries and POVM measurements. To see this, let
us consider what each party needs to do (see figure \ref{fig qubit2}). Each party $P_k$ has two
qubits, one entangled with party $P_{k-1}$ (qubit 1) and another entangled
with party $P_{k+1}$ (qubit 2). Upon receiving the two classical bits
from $P_{k-1}$, party $P_{k}$  applies the
corresponding unitary $T_{k-1}$ on qubit 1 to complete the teleportation
process. Then, $P_k$ will do a special projective measurement on the
joint system of qubits 1 and 2. This measurement consists of first
applying a unitary to qubit 1, which is dependent on $P_k$'s data; and
then projecting the qubits on the Bell basis, broadcasting the result
to party $P_{k+1}$. These two steps can be
viewed as resulting in a POVM measurement on each of the
qubits. Because the promise of the problem does not impose any
restriction on each party's unitary, each such POVM measurement is
effectively random.

In order to perform the task with classical communication only, we
would need to simulate perfectly the projective measurement performed
on the qubit by the last party. However, this requires information about all the POVM's performed by all the previous
parties. The bound we obtained shows that this amount of information,
if conveyed by classical means, is unbounded, increasing with the
number of parties.

It is interesting to view our entanglement-based task from a slightly
different perspective. We can consider the $(N-1)$ entangled pairs of qubits as one long
quantum system, whose constituent qubits we can measure at will. The particular sequence of measurements that implement our
entanglement-based protocol tells us something interesting about
information flow in quantum systems. We have proven that quantum mechanics (or any
hidden-variable theory compatible with it) necessarily requires an
unbounded amount of one-way information flow between the ends of this
entangled chain.

This result is related to the necessity of real-valued variables in
any hidden-variable theory for quantum mechanics, which we discussed
in section \ref{sec whybetter}.

\section{Chapter summary}

The description of quantum systems and operations involve real
numbers, and as such require an unbounded amount of classical
information. However, we can only send one bit of classical
communication per qubit of quantum communication.

In this chapter we explored a simple communication task which shows
that quantum theory, and indeed any hidden-variable theory consistent
with it, necessarily requires real-valued variables. What is more, I
have shown how these real-valued parameters can be used to advantage
in distributed computation applications involving either qubit
communication or use of quantum entanglement.

I have shown that in this application the quantum-based protocols
outperform any classical protocols using only a finite amount of
classical communication. In particular, I have shown that in order to
account for the information flow required by measurements on entangled
states, we necessarily need an unbounded amount of one-way classical
communication.

From a computational perspective, we have seen that the simple task we
analysed clarifies possible computational resources trade-offs we can
obtain when changing from classical to quantum computation. In
particular, I have discussed how quantum computers can offer a memory
size advantage, while requiring more computational steps for the same
task. Consideration of this kind of resource trade-offs is important
for appreciating and maximising the computational capabilities of
quantum or analog computers.


\chapter{Conclusion}

In this thesis I have tried to establish connections between simple
quantum information protocols and some foundational issues in quantum
theory. In this last chapter I comment on my main results and suggest
some open research problems related to them.

In chapters \ref{chap qcc}, \ref{chap feasible} and \ref{chap qubit} I examined a few simple
applications and identified the fundamental quantum traits behind
their better-than-classical performance. A generally acknowledged
quantum resource is quantum non-locality, which is indeed essential
in some of the applications reviewed. In this thesis I have identified two other quantum resources giving better-than-classical performance for some information processing
tasks: quantum contextuality, and the continuous nature of the set of
pure quantum states.

Regarding contextuality, it may have come as as surprise that in chapter
\ref{chap found} it was  presented as a more fundamental quantum
property, from which non-locality follows. There are two main
reasons for presenting contextuality the way I did.

The first reason is simplicity: I think it is conceptually easier to
first understand the impossibility of assigning (non-contextual)
values to experimental outcomes, and then impose the locality
condition and try to do the same. This way of putting things reveals
non-locality as a manifestation of contextuality with a `twist',
i.e. contextuality of space-like separated measurements.

This brings us to the second, and more important, reason for
discussing contextuality. My preoccupation with experimental
feasibility of quantum information applications led me to
realize that in most circumstances the quantum measurements involved are
\textit{not} space-like separated. Strictly speaking, this rules out
quantum non-locality as the characteristic responsible for their
higher-than-classical performance.

Some researchers have been aware of that for a long time, and prefer
to use more accurate terms such as `quantum correlations' instead of non-locality. At the same time, from
a more fundamental perspective, there has been a drive towards
understanding the nature of the `loopholes' that remain open in
quantum non-locality testing, such as inefficiencies in detection,
imperfect state preparation and the lack of strict locality
conditions for the measurements.

In chapters  \ref{chap qcc} and \ref{chap feasible} I revisited these
ideas to stress the point that non-locality tests without strict
locality conditions (i.e. without closing the locality `loophole') are
in fact equivalent to contextuality tests. Some relations between
contextuality and non-locality have been pointed out
by other authors \cite{Mermin90b,MichlerWZ00,SimonZWZ00}. In this
thesis I make the connection explicit, and show how contextuality can
be \textit{used} in experimental implementations of quantum
information applications.  It is not just
the case that feasible implementations fail to impose locality
conditions -- many times these \textit{cannot} be imposed even in
principle. That is the case of the $2 \to 1$ QRAC using qubit
communication, in which Alice prepares a state and sends it to Bob.

In my opinion quantum contextuality should be given more importance as
a potential resource in quantum information applications. This point
has been implicitly addressed also by Vaidman \cite{Vaidman01}, who
argued for the importance of using what he calls `quantum games' to
perform information processing and communication tasks which are
impossible classically. These applications are of the type that I
describe in this thesis, relying also on the stronger-than-classical
correlations offered by quantum contextuality.

Contextuality and non-locality can be considered as resources to be
understood and quantified for use in quantum information
applications. This leads us to the subject I discussed in chapters
\ref{chap ent} and \ref{chap tri}, the characterisation of tripartite
quantum entanglement.

An interesting feature of multipartite entanglement is the fact that
it appears in a number of qualitatively different types. Classifying these
types is quite challenging, but may help identify states which may be potentially useful
due to their unusual entanglement structure. Two examples of such unusual states are the four-party cluster states (\ref{wuzhangstate}) used in an
alternative model of quantum computation \cite{BriegelR01,
RaussendorfB01}, and the $W$ states (\ref{Wstatedef}) use in non-locality and contextuality proofs
\cite{Cabello02,Zheng02}, and in an application involving
quantum games \cite{HanZG02}.

In chapter \ref{chap tri} I identified several good candidates for
displaying a new type of tripartite entanglement under asymptotic
transformations.  At least one such family of states has since been
rigorously shown to be inequivalent to the previously known tripartite
entanglement classes \cite{AcinVC02}. It would be
interesting to prove the same for the other states I singled
out in chapter \ref{chap tri}.

More generally, one may ask whether there exists a finite Minimal
Entanglement Generating Set (MREGS) for tripartite entanglement. It
may actually be the case that multipartite entanglement
manipulations are genuinely irreversible, as happens with bound
entanglement in the bipartite setting. Identifying the causes of this
could be very interesting from a theoretical point of view.

In section \ref{sec wvsghz} I gave an argument showing that in at least one
well-motivated sense, the $W$ state of three qubits is more entangled
than the GHZ state. In the near future I intend to investigate this
question more closely, with the aim of checking whether it is possible
to explicitly describe non-locality tests supporting this claim.

Changing direction a bit, we may ask what fundamental laws govern the
distribution of quantum information, and how to make optimal use of
the manipulations that are allowed. The no-cloning theorem offered an
initial negative result, which motivated further research into the
allowed processes of quantum information distribution. In chapter
\ref{chap cloning} I investigated how quantum cloning could be
useful in quantum computing. This resulted in the two examples of
tasks for which cloning offers some advantages, using two different
types of cloning protocols.

An interesting research question on cloning consists in trying to find
a more fundamental justification of the optimal achievable cloning
fidelities. A step in that direction was provided by Gisin's work
bounding the fidelities from the requirement that there should be no
faster-than-light signalling \cite{Gisin98}. I suspect approaches deriving from quantum
information may prove useful for this problem. For example, one
can establish bounds on the cloning fidelities using the invariant
information of Brukner and Zeilinger, in a similar way as I did with QRAC's in section \ref{sec invinfoqrac}. Another suggestive
result is the curious fidelity balance I found in section \ref{sec
curfbr}. Clearly, there is much still to be understood about quantum
information distribution.

Physical theories and theoretical applications have one thing in
common: their justification comes only through successful
experiments. Some quantum information applications may be
experimentally feasible already, or may become so in the near
future. In chapters \ref{chap qcc} and \ref{chap feasible} I
demonstrated the feasibility of a couple of applications, and also
argued that their implementation would be equivalent to tests of
properties such as non-locality and contextuality.

Given these first results, it is clear that more feasibility studies
are necessary at this point. There are a number of quantum information
applications which seem promising and possibly interesting from a
foundational aspect, among them quantum cryptographic protocols,
absorption-reduced imaging \cite{Kwiat98,Jang99,InoueB00,MassarMP01},
and quantum games (see \cite{FlitneyA02} and references therein).

In chapter \ref{chap qubit} I proposed an information processing task
in which a single qubit is used to encode information which would
require an unlimited classical digital memory to store. Surprisingly,
this has bearings on different problems: space complexity theory,
general hidden-variable theories, and on the classical communication
cost of simulating entanglement.

In this case, the advantage comes not from the exponentially large
Hilbert space (as in Shor's algorithm) or from entanglement (as in
teleportation), but from the continuous nature of the set of pure
states available to quantum systems. This quantum characteristic is
central in Hardy's axiomatic formulation of quantum theory, and had
not been explicitly identified as an important quantum information
resource up to now.

This suggests that in some senses quantum computation is similar to
classical \textit{analog} computation. An open research question
consists in clarifying the similarities and differences between these
two models, with a view to understanding better the quantum advantage
provided by entanglement, for example. As we have seen, it is also
important to understand quantitatively the requirements on
computational resources such as space, time and accuracy. Regarding
the task studied in chapter \ref{chap qubit}, it would be
interesting to quantify the trade-offs possible between these
resources, in the three models of computation:  classical digital,
classical analog, and quantum.

The general approach I took in this thesis suggests different
motivations for experimental implementations of quantum information
protocols. This comes from the realization that simple experiments may
actually correspond to a number of different applications. For
example, in chapter \ref{chap qcc} I showed that one
simple experiment can be interpreted alternatively as: a locality or
contextuality test; or as protocol for remote state preparation; or as
a  quantum communication complexity protocol; or as a $2 \to 1$
quantum random access code. Another example is the feasible one-qubit
communication complexity protocol of chapter \ref{chap feasible},
which can also be interpreted as an experiment to demonstrate that a
one-qubit quantum computer is better than a one-bit classical computer
for some computations. This follows from the computational
interpretation of the generalization of that task developed in chapter
\ref{chap qubit}.

The applications I analysed help clarify some of the requirements for
experiments to establish some fundamental quantum characteristics. For example, the multi-party entanglement-based
communication complexity protocol of chapter \ref{chap feasible} automatically
gives us sufficient conditions for establishing multipartite
non-locality or contextuality. In the same chapter I also analysed the experimental requirements for a demonstration that
quantum communication can be better than classical communication for
some tasks.

Of course, there are many other approaches one may use to gain
insights on the foundations of quantum theory. Most attempts to unify
quantum theory and gravitation, for example, seem to acknowledge the
importance of recovering the information-theoretic description of
black hole entropy obtained by Beckenstein \cite{Bekenstein73}. Other
authors have uncovered interesting analogies between
entanglement and thermodynamics
\cite{PopescuR97,VedralK02,HorodeckiOH02}. Also, there are results
suggesting that quantum phenomena may bring surprising consequences to
thermodynamics, for example by improving the efficiencies of some
types of heat engines \cite{Scully02}. On the experimental front,
recent experiments revealing entanglement between macroscopic objects
\cite{JulsgaardKP01} suggest that the robustness of large-scale
entanglement may be investigated experimentally. Needless to say,
these techniques may also be useful for practical quantum information
applications.

These different approaches seem to be complementary, pointing towards
an understanding of quantum theory which is being shaped around the
idea of information as a foundational concept. From a more practical
point of view, quantum information applications are finally bringing quantum
theory's most counter-intuitive characteristics to within reach of
experiments. These developments engage the interest of the general
public, and provide challenges to an expanding, interdisciplinary
research community. It is a privilege to be doing research at a time
when exciting quantum theoretical concepts are finally maturing into
useful applications.

\addcontentsline{toc}{chapter}{Bibliography}
\bibliographystyle{plain}
\bibliography{thesis}

\end{document}